\documentclass[preprint,12pt]{elsarticle}

\usepackage{amssymb}
\usepackage{amsmath,amsthm,amsfonts,amscd,amsmath,amsfonts} % Some packages to write mathematics.
\usepackage{upgreek}
\usepackage{mdframed}
\usepackage{multicol}
\usepackage{graphicx}
\usepackage{nomencl}
\usepackage[nottoc]{tocbibind}
\usepackage{setspace}
\usepackage{subfig}
\usepackage{verbatim}   	% Allows quoting source with commands.
\usepackage{makeidx}    	% Package to make an index.
\usepackage{color}
\usepackage{xcolor}
\usepackage{float}
\usepackage{enumerate}
\usepackage{url}	
\usepackage{algpseudocode}
\usepackage{tcolorbox}
\usepackage{xcolor}
\usepackage{multirow}
\usepackage[linesnumbered,ruled,vlined]{algorithm2e}
\usepackage{ulem}
\usepackage[pdftex,bookmarks]{hyperref}
\usepackage{multirow}
\usepackage{bbm, dsfont}
\usepackage[percent]{overpic}
\newcommand{\blue}[1]{\textcolor{black}{#1}}

\newcommand{\bs}{\boldsymbol}

\usepackage[margin=1.2in]{geometry} % Adjust margins here

\usepackage{etoolbox} % noindent after the sections/subsections
\makeatletter
\patchcmd{\@startsection}
  {\@afterindenttrue}
  {\@afterindentfalse}
  {}{}
\makeatother

\journal{Computational Materials Science}

\begin{document}

%========================================================================
% First Page
%========================================================================
\begin{frontmatter}



\title{Stochastic Deep Learning Surrogate Models for Uncertainty Propagation in Microstructure-Properties of Ceramic Aerogels}


\author[inst1]{Md Azharul Islam}
\author[inst2]{Dwyer Deighan}
\author[inst1]{Shayan Bhattacharjee}
\author[inst1]{Daniel Tantalo}
\author[inst1]{Pratyush Kumar Singh}
\author[inst1]{David Salac}
\author[inst1]{Danial Faghihi\corref{cor1}}

\affiliation[inst1]{organization={Department of Mechanical and Aerospace Engineering, \\University at Buffalo},%Department and Organization
            %addressline={Address One}, 
            city={Buffalo},
            %postcode={00000}, 
            state={NY},
            country={USA}}

\affiliation[inst2]{organization={Computational and Data-Enabled Sciences, \\University at Buffalo},%Department and Organization
            %addressline={Address One}, 
            city={Buffalo},
            %postcode={00000}, 
            state={NY},
            country={USA}}

\cortext[cor1]{Corresponding Author, \texttt{danialfa@buffalo.edu} (D. Faghihi)}            

%========================================================================
% Abstract
%========================================================================
\begin{abstract}
\blue{
This study presents an integrated computational framework that, given synthesis parameters, predicts the resulting microstructural morphology and mechanical response of ceramic aerogel porous materials by combining physics-based simulations with deep learning surrogate models.
} 
Lattice Boltzmann simulations are \blue{employed} to model microstructure formation during material synthesis process, while a finite element \blue{model is used to compute the corresponding mechanical properties}.
To overcome the prohibitive computational demands of repeated physics-based simulations required for characterizing the impact of microstructure randomness on mechanical properties, surrogate models are developed using Convolutional Neural Networks (CNNs) for both microstructure generation and microstructure-property mapping. CNN training is formulated as a Bayesian inference problem to enable uncertainty quantification and provide confidence estimates in surrogate model predictions, under limited training data furnished by physics-based simulations.
Numerical results demonstrate that the microstructure surrogate model effectively generates microstructural images consistent with the morphology of training data across larger domains. 
The Bayesian CNN surrogate accurately predicts strain energy for in-distribution microstructures and its generalization capability to interpolated morphologies are further investigated. 
Finally, the surrogate models are employed for efficient uncertainty propagation, quantifying the influence of microstructural variability on macroscopic mechanical property.

\end{abstract}


\end{frontmatter}

%%\linenumbers

%%\tableofcontents

%\newpage
%========================================================================
% Document body
%========================================================================
\section{Introduction}
\label{sec:introduction}
%------------------------------------------------------------------------------------------------
% 1. Identify the problem and say why it is important
%------------------------------------------------------------------------------------------------
% Aerogel importance and issue with mechanical properties
Ceramic aerogels, composed of interconnected amorphous ceramic nanoparticles, are highly porous materials with exceptional thermal insulation properties, making them attractive for aerospace, microelectronics, fire-resistant wearables, and energy-efficient buildings \cite{nasa, lu2020wearable, wang2017graphene, choi2022ultralow, tan2022predictive}. However, their widespread adoption is limited by poor mechanical robustness. Physics-based synthesis-microstructure-property models offer a pathway to understanding and optimizing material performance, e.g., \cite{rajan2013informatics, national2008integrated, xu2022computational} to achieve mechanically robust aerogels without compromising insulation performance.
% need for the surrogate models for uncertainty propagation
Due to the inherent randomness (stochasticity) in microstructural features, such as pore size, shape, and spatial distribution, uncertainty propagation is essential for quantifying variability in mechanical properties. However, physics-based models are often computationally prohibitive for this task, as it demands large number of samples involving repeated model evaluations. To overcome this challenge, data-driven surrogate models \cite{mojumder2025adaptive, senthilnathan2025surrogate}, trained on high-fidelity data from physics-based simulations, can efficiently approximate the underlying models, enabling uncertainty propagation at significantly reduced computational cost.

%------------------------------------------------------------------------------------------------
% 2. The current state of the field and current limitation(s) in the field
%------------------------------------------------------------------------------------------------
% CNN surrogate models and uncertainty in data
Convolutional Neural Networks (CNNs) have become increasingly prominent in computational materials science \cite{zeng2019atom, choudhary2022recent, xiong2024indexing, zhang2024named, qian2024permeability, qian2023biomimetic}, particularly for generating realistic microstructural images and predicting macroscopic properties. However, deploying CNNs as surrogate models often requires training on sparse and limited datasets generated from computationally expensive physics-based simulations. While maximum likelihood estimation is the standard training approach, its performance can degrade in small-data or high-complexity regimes without appropriate safeguards. Overfitting and overconfident predictions may occur in such cases, but these risks are commonly mitigated through techniques such as regularization, cross-validation, transfer learning, and self-supervised learning \cite{gunduz2023self, dawson2023impact, yan2021applying, zong2024recent}.
% Bayesian and literature review
An alternative and complementary strategy is to formulate CNN training within a Bayesian inference framework \cite{neal2012bayesian, blundell2015weight}, where model parameters are treated as random variables with posterior distributions conditioned on the training data.
In principle, Bayesian inference enables uncertainty quantification and provides confidence estimates in surrogate model predictions \cite{singh2024opal, olivier2021bayesian}.
However, the practical application of Bayesian CNNs in surrogate modeling remains limited due to the computational challenges posed by high-dimensional parameter spaces and the sensitivity to prior and likelihood specifications, both of which can significantly influence accuracy and reliability of model prediction. Zhu et al. \cite{zhu2018bayesian} addressed the computational challenges by employing a variational gradient descent algorithm based on Stein’s method, extending Bayesian inference to convolutional encoder-decoder networks. Their approach captured uncertainty in flow through heterogeneous media, even in scenarios where the input (permeability) and output (flow and pressure) fields lacked shared underlying structures. Similarly, Shridhar et al. \cite{shridhar2019comprehensive} applied variational inference to estimate posterior weight distributions in CNNs, effectively propagating uncertainty in classification tasks across datasets like MNIST and CIFAR-100.

%------------------------------------------------------------------------------------------------
% 3. How do you plan on filling this hole with a novel approach
%------------------------------------------------------------------------------------------------
%
\begin{figure}[ht]
\centering
    \includegraphics[trim = 0mm 0mm 0mm 0mm, clip, width=1\textwidth]{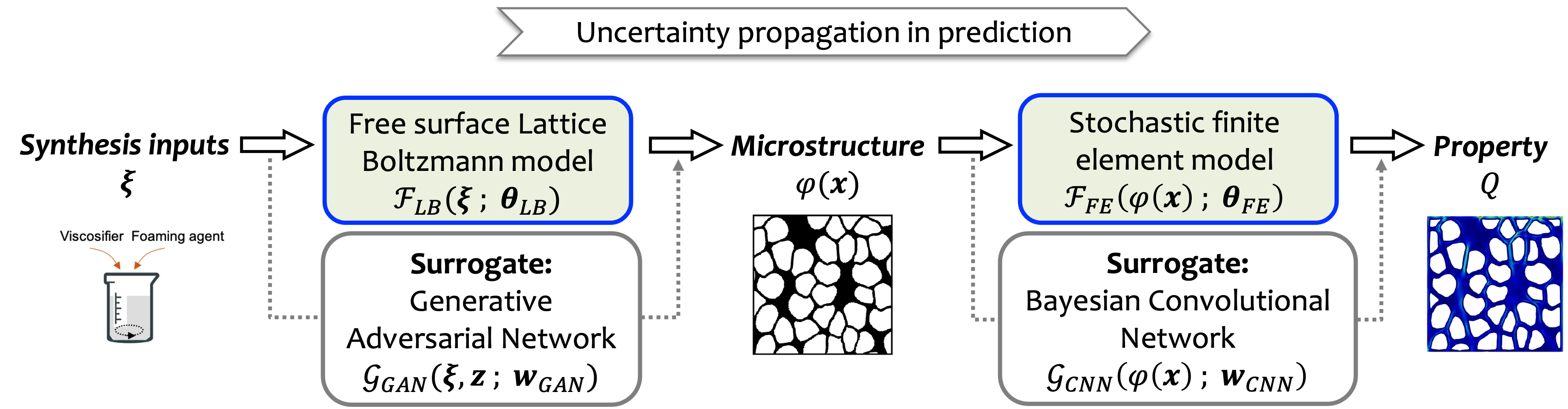}
    \vspace{-0.1in}
    \caption{
    Flowchart of the integrated modeling framework developed in this work for propagating uncertainty through synthesis inputs $\bs \xi$, microstructure spatial field $\varphi(\mathbf{x})$, and mechanical property $Q$ of ceramic aerogels.
    The physics-based models, $\mathcal{F}_{LB}(\cdot)$ and $\mathcal{F}_{FE}(\cdot)$, are parameterized by $\bs \theta_{LB}$ and $\bs \theta_{FE}$, and utilize synthesis inputs and microstructural patterns to predict material properties. 
    Computational efficiency is achieved through surrogate models $\mathcal{G}_{GAN}(\cdot)$ and $\mathcal{G}_{CNN}(\cdot)$, constructed using high-fidelity training data from physics-based simulations. The parameter of the surrogate models include the stochastic latent vector $\bs z$,
    deterministic weights of the generative network $\bs w_{GAN}$, and posterior probability distributions of the wights $\bs w_{CNN}$ from the Bayesian inference.
    }
    \vspace{-0.12in}
        \label{fig:framework}
\end{figure}

This work introduces an integrated modeling framework (Figure \ref{fig:framework}) for simulating the synthesis, microstructure, and properties of ceramic aerogels. Specifically, it focuses on a novel synthesis method utilizing in situ bubble-supported pore formation with foaming agents \cite{Ren2019hierarchical, lu2020wearable}, which addresses the limitations of conventional supercritical drying by preserving the porous structure without collapsing the solid matrix. This synthesis enables ambient-pressure, room-temperature drying, significantly reducing processing time while facilitating scalable and cost-effective production of ceramic aerogels.
The synthesis-microstructure physics-based model in Figure \ref{fig:framework}, relies on Lattice Boltzmann simulations of the pre-aerogel foaming process, capturing stochastic bubble nucleation, bubble dynamics, and coalescence to predict microstructure formation. The microstructure-property model employs finite element simulations to analyze the elastic deformation of the solid phase of the aerogel. Using the microstructural images, this model computes the displacement and strain fields to estimate the mechanical properties of the aerogel, i.e., the strain energy as the quantity of interest (QoI) for the modeling framework.
To enable efficient uncertainty propagation from microstructure to mechanical property, we develop surrogate models of the expensive physics-based simulations. The synthesis–microstructure surrogate employs a scalable generative adversarial network trained on Lattice Boltzmann-generated microstructures, allowing the generation of morphologically realistic microstructures across a range of domain sizes. The microstructure-property surrogate employs a CNN trained to map microstructural images to strain energy.
A key innovation of this framework is the formulation of CNN training as Bayesian inference, enabling the estimation of the probability distribution of weight parameters using high-fidelity training data from finite element simulations. To address the computational challenges of Bayesian inference in high-dimensional parameter spaces, we extend the variational inference approach commonly used in Bayesian Neural Networks, e.g., \cite{olivier2021bayesian} to accommodate the complexity of CNN architectures. The resulting Bayesian CNN (BayesCNN) enables not only prediction of the strain energy but also quantification of the associated uncertainty, providing a measure of confidence in the surrogate model's predictions.
These surrogates are then used to perform efficient uncertainty propagation analyses, that is, to quantify how microstructural randomness influences variability in mechanical properties. The generative model produces ensembles of statistically consistent microstructures, while the BayesCNN evaluates the corresponding strain energy to estimate the probability distribution and credible intervals of the QoI (strain energy) across  microstructures with different average pore sizes.

%------------------------------------------------------------------------------------------------
% 4. The final paragraph: summary of sections
%------------------------------------------------------------------------------------------------
The remainder of this manuscript is structured as follows. 
Section 2 provides an overview of the ceramic aerogel synthesis process, including its modeling using Lattice Boltzmann simulations and the development of surrogate models to represent microstructural patterns.
Section 3 describes the finite element model used to establish microstructure-property relationships and presents the comprehensive formulation and efficient solution algorithm of BayesCNN for surrogate modeling.
Section 4 outlines the construction of individual surrogate models, detailing the creation of training datasets from high-fidelity physics-based simulations and the validation of these models.
\blue{Additionally, this section illustrates the use of the surrogate models to efficiently propagate uncertainty from stochastic microstructure to the predicted QoI.}
Finally, Section 5 presents a discussion of the findings and the concluding remarks.

%++++++++++++++++++++++++++++++++++++++++++++++++++++++++++++++++++++++++
%++++++++++++++++++++++++++++++++++++++++++++++++++++++++++++++++++++++++
\section{Physical and surrogate models of aerogel microstructure }\label{sec:surrogate}
The exceptional thermal insulation of silica aerogels arises from their high porosity and nanoscale pore structure, which limits gas molecule movement. However, widespread use in engineering applications is hindered by the high production costs of current methods, such as supercritical or freeze-drying, e.g., \cite{dorcheh2008silica}.
To overcome these limitations, a novel synthesis process has been developed, utilizing in-situ bubble-supported pore formation for ambient pressure, room-temperature drying \cite{Ren2019hierarchical, lu2020wearable, wang2020transparent}. This method employs a silica precursor solution with cetrimonium bromide (CTAB) and an aqueous urea-based foaming agent. Hydrolysis of tetraethyl orthosilicate (TEOS) at CTAB micelle interfaces, combined with urea decomposition and ammonia and carbon dioxide release, generates in-situ bubbles, producing aerogels with randomly distributed pore sizes and morphologies.
The primary synthesis inputs $\bs \xi$ include the concentration of TEOS ($C_{TEOS}$), which regulates the solid phase volume fraction and thereby affects the porosity and pore size of the aerogel. The concentration of CTAB ($C_{CTAB}$) plays a critical role in bubble formation, influencing the resulting pore size distribution. The concentration of urea ($C_{urea}$), acting as a foaming agent, reduces surface tension between the precursor solution and gaseous bubbles, thereby impacting the pore distribution. Additionally, the concentration of viscosifier ($C_{vis}$) determines the stability and morphology of the pore structure during the foaming process. These parameters collectively control the microstructural characteristics and properties of the ceramic aerogel.

%++++++++++++++++++++++++++++++++++++++++++++++++++++++++++++++++++++++++
\subsection{Lattice Boltzmann simulation of microstructure formation}
The computational modeling of the ceramic aerogel synthesis process, as discussed in the previous section, requires the integration of key physical phenomena involved in pre-aerogel foaming. These include bubble nucleation, gas diffusion into bubbles, bubble dynamics and coalescence, and the rheological behavior of the precursor, which often exhibits non-Newtonian characteristics. In this study, we utilize the Free Surface Lattice Boltzmann Method (FSLBM) for foaming simulations, e.g., \cite{korner2002,korner2005,anderl2014,donath2009}. Specifically, we employ the open-source software package LBfoam \cite{ataei2020lbfoam}, which is implemented using the Parallel Lattice Boltzmann (Palabos) library \cite{palabos} for fluid dynamics and integrates a Volume-of-Fluid technique for accurate interface tracking. A brief overview of the fundamental equations is provided here for context, while detailed equations and solution algorithms can be found in \cite{ataei2020lbfoam}.

The core of FSLBM is the Lattice Boltzmann simulation of complex fluid systems, e.g., \cite{chen1998, do2000}. the computational domain is discretized into a regular lattice, where the particle distribution function $f_i(\mathbf{x},t)$ represents the expected number of particles at lattice site $\mathbf{x}$ and time $t$ in the state $i=0, 1, \cdots, N$ defined by a set of discrete velocities $\mathbf{v}_i$. 
The discretized Lattice Boltzmann equation is
\begin{equation}\label{eq:LBM}
f_i(\mathbf{x} + \mathbf{v}_i \Delta t, t+\Delta t) = f_i(\mathbf{x},t) + B(\mathbf{x},t),
\end{equation}
with the Bhatnagar-Gross-Krook (BGK) collision operator \cite{do2000} as,
\begin{equation} 
B(\mathbf{x},t) = - \frac{\Delta t}{\tau} \left(f_i(\mathbf{x},t) - f_i^{eq}(\mathbf{x},t) \right),
\end{equation}
where $\Delta t$ is the time step, $\tau$ is a relaxation time, and $f_i^{eq}$ is the distribution function at equilibrium \cite{do2000}.
The solution of Eq. (\ref{eq:LBM}) involves two iterative steps: streaming, where the distribution functions $f_i$  move to neighboring lattice sites, and collision, where $f_i$  is locally updated at each site. Macroscopic fluid properties, such as density and velocity, are then derived from the moments of $f_i$.
Building on this foundation, FSLBM simulations utilize a fast mass-tracking algorithm \cite{korner2005} to monitor the interfaces between gas and liquid phases throughout the computational domain. The liquid-gas interface is modeled as a free surface, with \blue{the Volume-of-Fluid (VOF)} method employed to track the interface using a level function, which represents the liquid fraction within each computational cell. Surface tension $\sigma_{LB}$ are incorporated by calculating the interface curvature from the level function and applying the resulting pressure jump across the interface.
In this work, to capture the highly viscous and shear-thinning behavior of the aerogel precursor, governed by the concentration of viscosifier $C_{vis}$, the Carreau-Yasuda model \cite{boyd2007analysis} is adopted to describe the fluid's rheological properties. The model provides an accurate representation of the non-Newtonian characteristics essential for simulating the foaming process.
The influence of TEOS concentration on bubble growth is incorporated by modeling gas diffusion into the bubbles and within the liquid phase using the advection-diffusion equation for gas concentration $C$,
\begin{equation}\label{eq:adv_diff}
\frac{\partial C}{\partial t} + \mathbf{v} \cdot \nabla C = D \nabla^2 C + S,
\end{equation}
where $\nabla$ represents the spatial gradient operator, $D$ denotes the diffusion coefficient, directly linked to $C_{TEOS}$, and $S$ is a source term representing gas generation or absorption as a function of $C_{urea}$, which regulates the rate of bubble growth.

\begin{figure}[h!]
\centering
    \includegraphics[trim = 0mm 0mm 0mm 0mm, clip, width=0.14\textwidth]{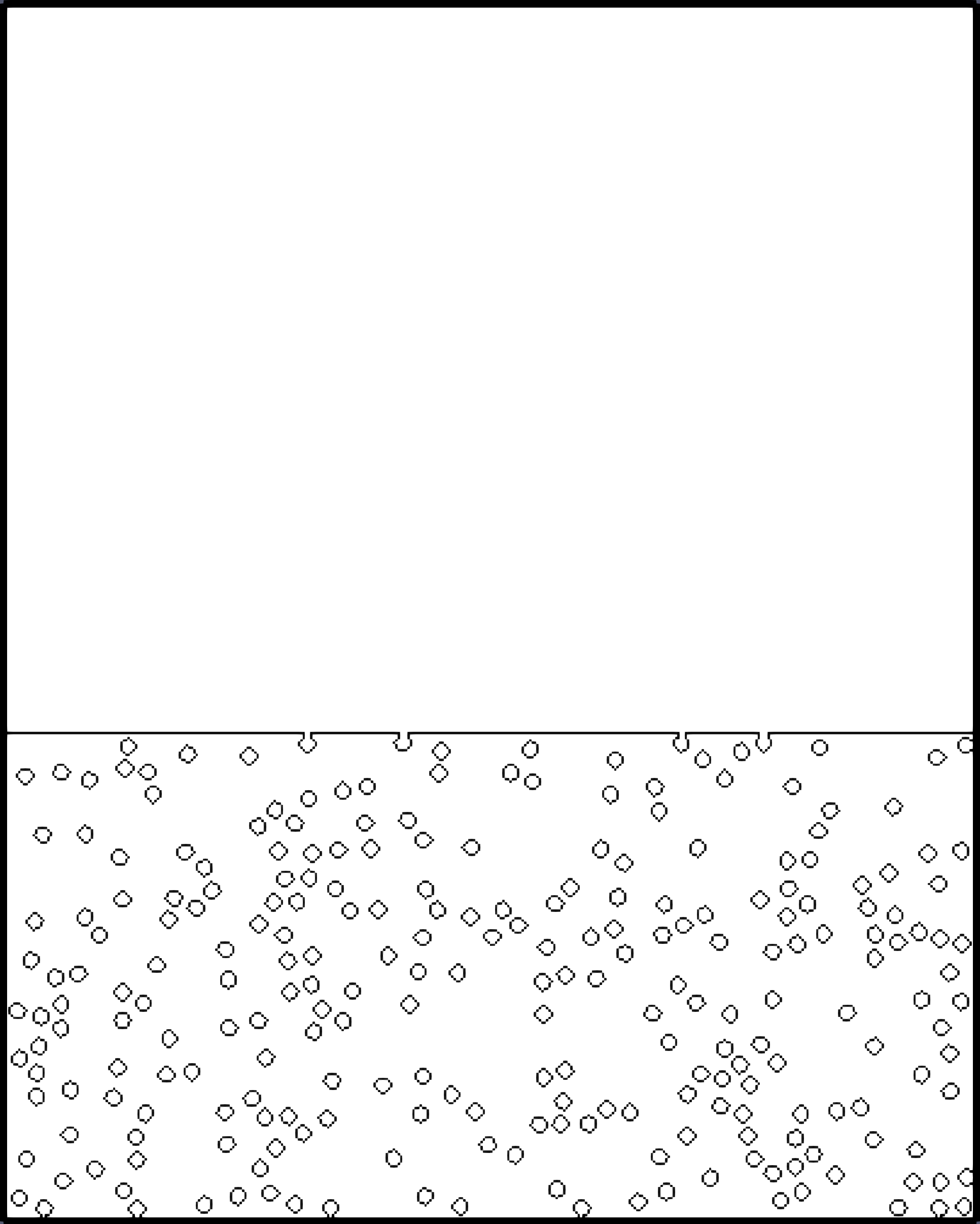}
    ~
    \includegraphics[trim = 0mm 0mm 0mm 0mm, clip, width=0.14\textwidth]{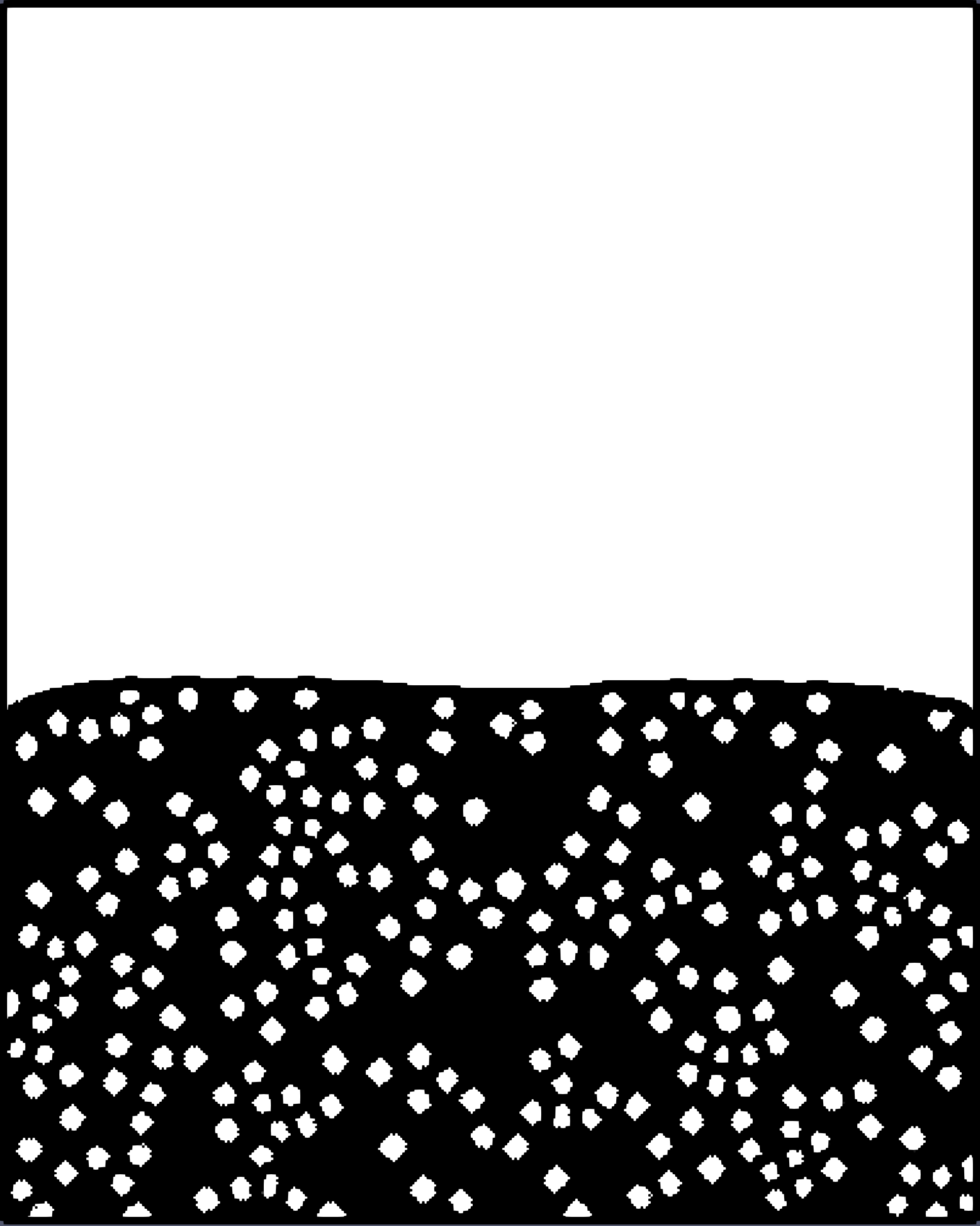} 
    ~
    \includegraphics[trim = 0mm 0mm 0mm 0mm, clip, width=0.14\textwidth]{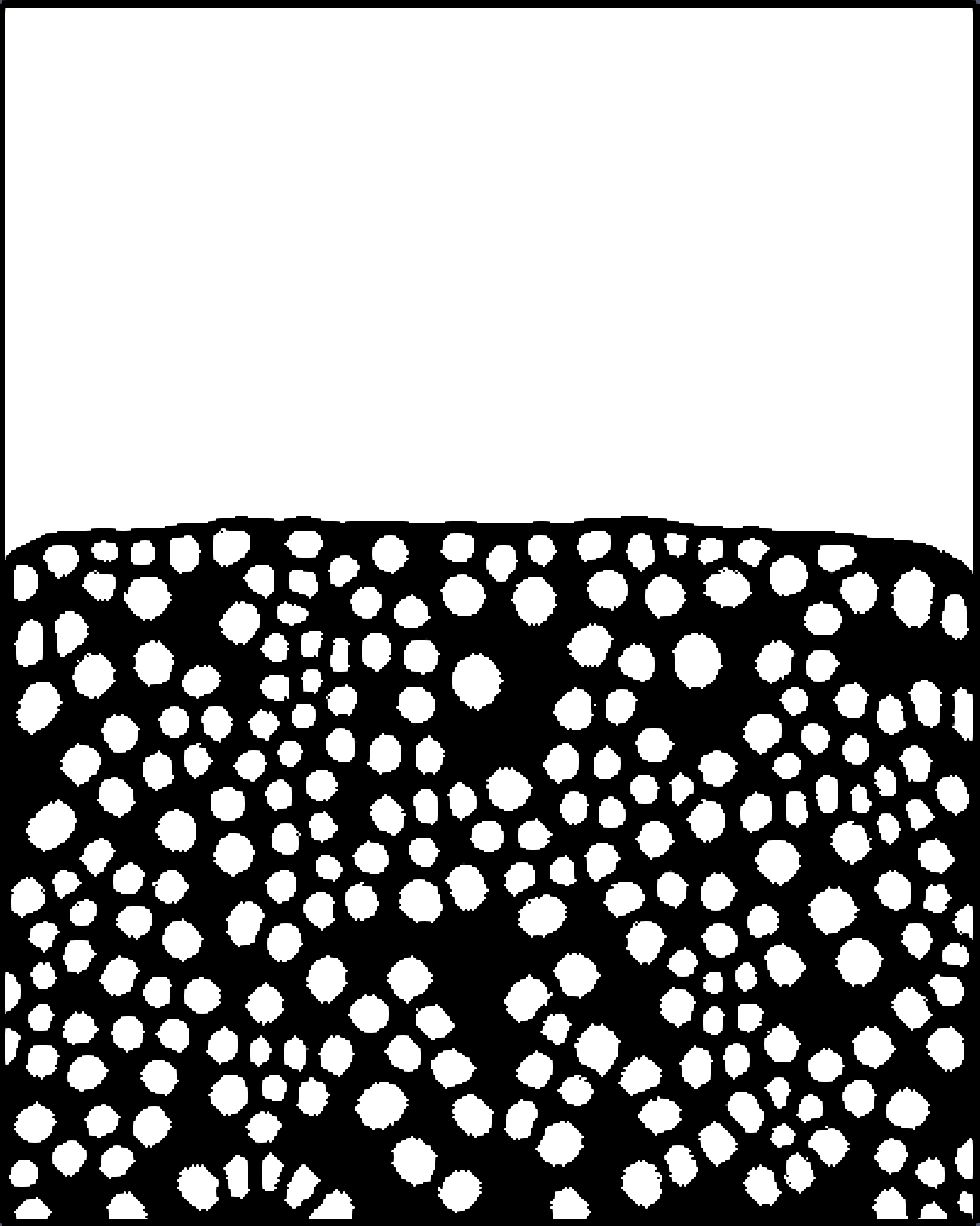} 
    ~
    \includegraphics[trim = 0mm 0mm 0mm 0mm, clip, width=0.14\textwidth]{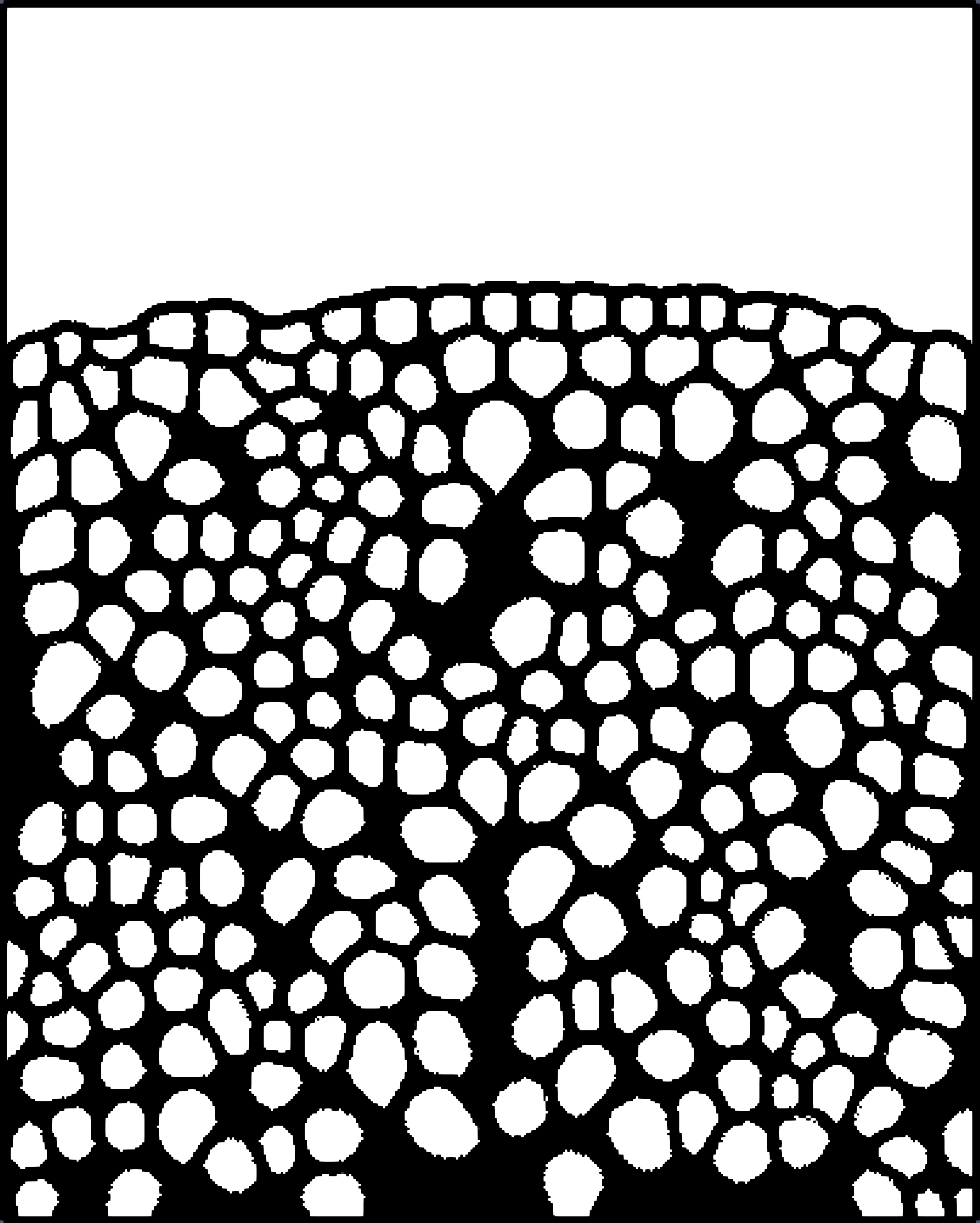} 
    ~
    \includegraphics[trim = 0mm 0mm 0mm 0mm, clip, width=0.14\textwidth]{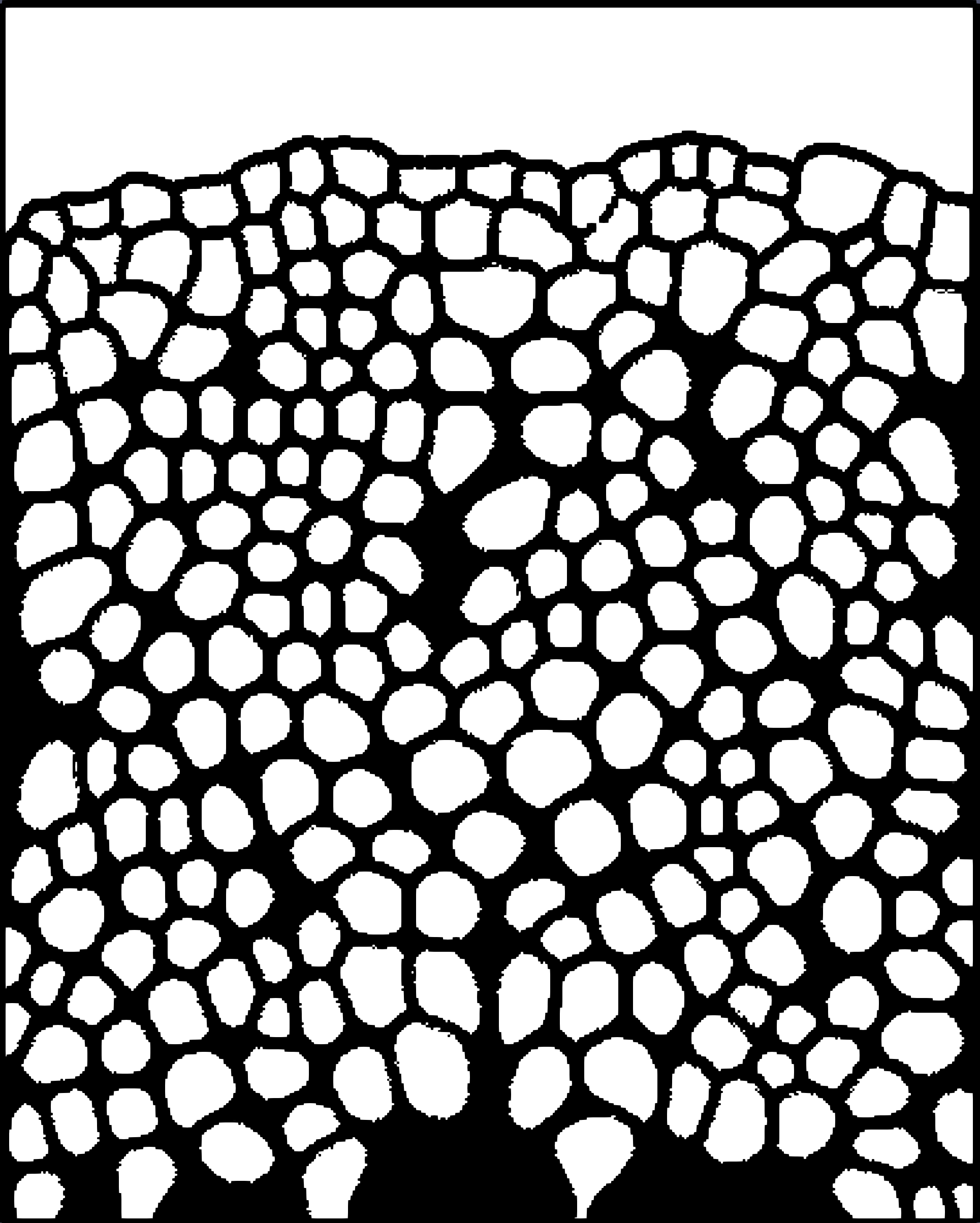} 
    ~
    \includegraphics[trim = 0mm 0mm 0mm 0mm, clip, width=0.14\textwidth]{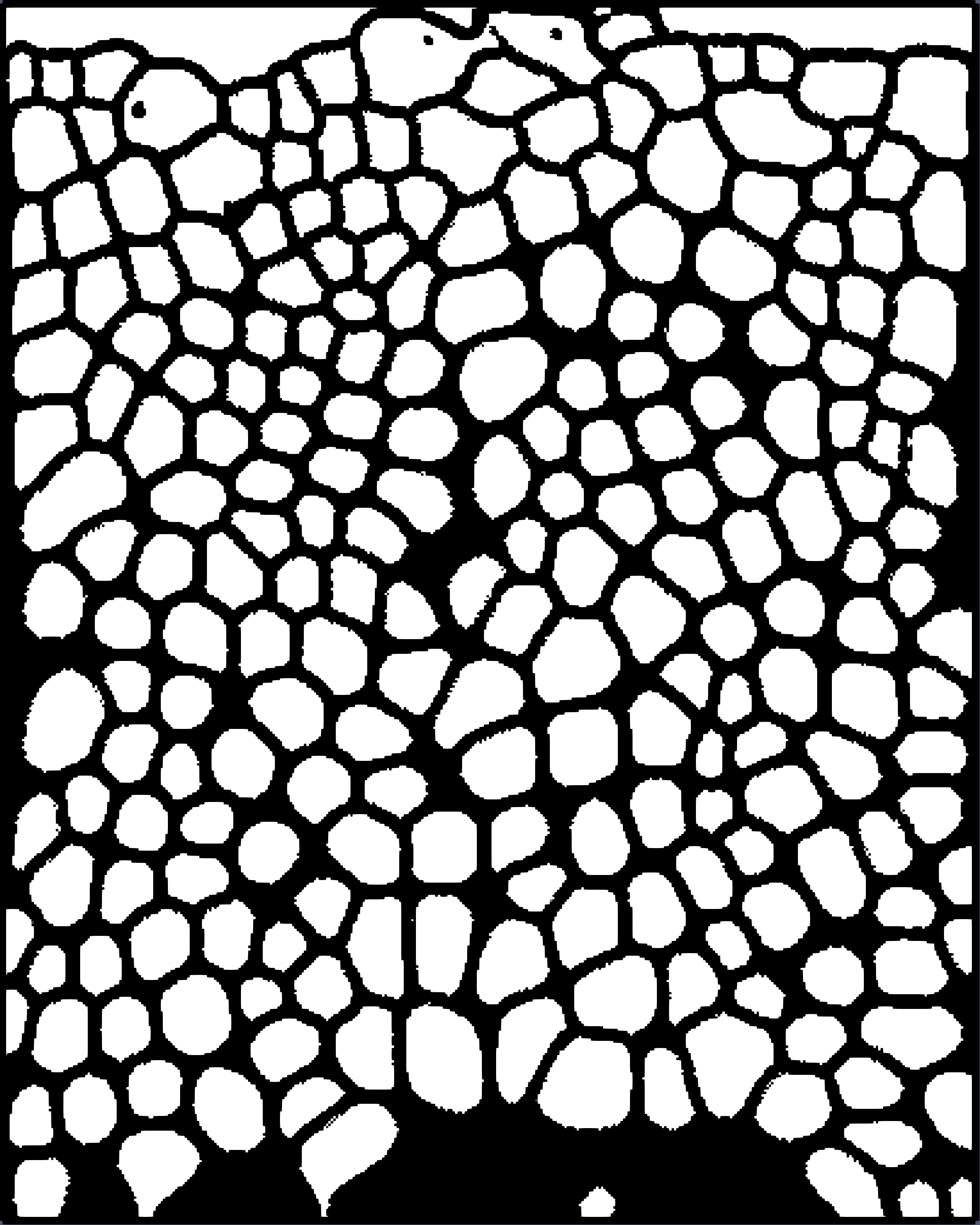}
        \\      (a)     \\
    \includegraphics[trim = 0mm 0mm 0mm 0mm, clip, width=0.14\textwidth]{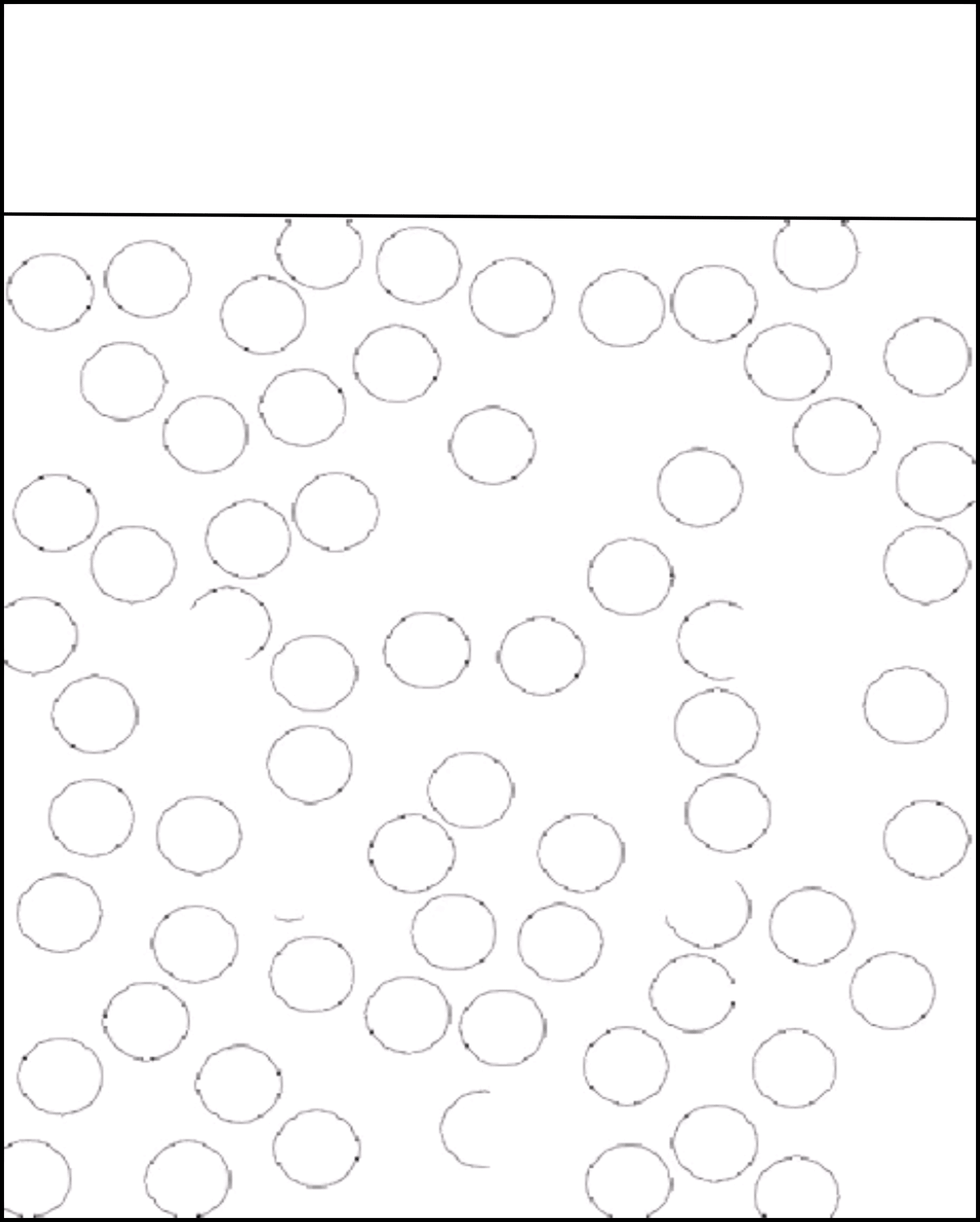}
    ~
    \includegraphics[trim = 0mm 0mm 0mm 0mm, clip, width=0.14\textwidth]{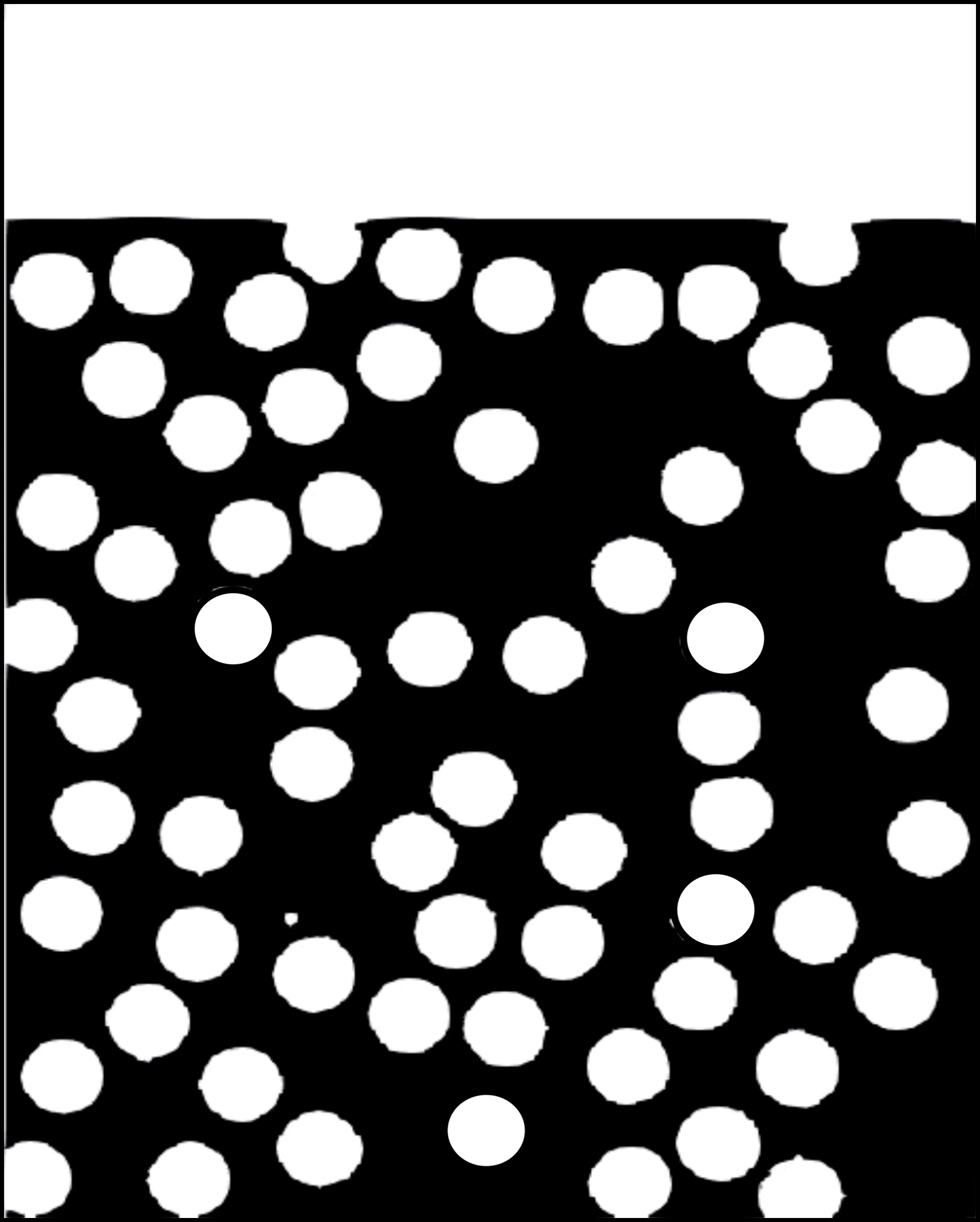} 
    ~
    \includegraphics[trim = 0mm 0mm 0mm 0mm, clip, width=0.14\textwidth]{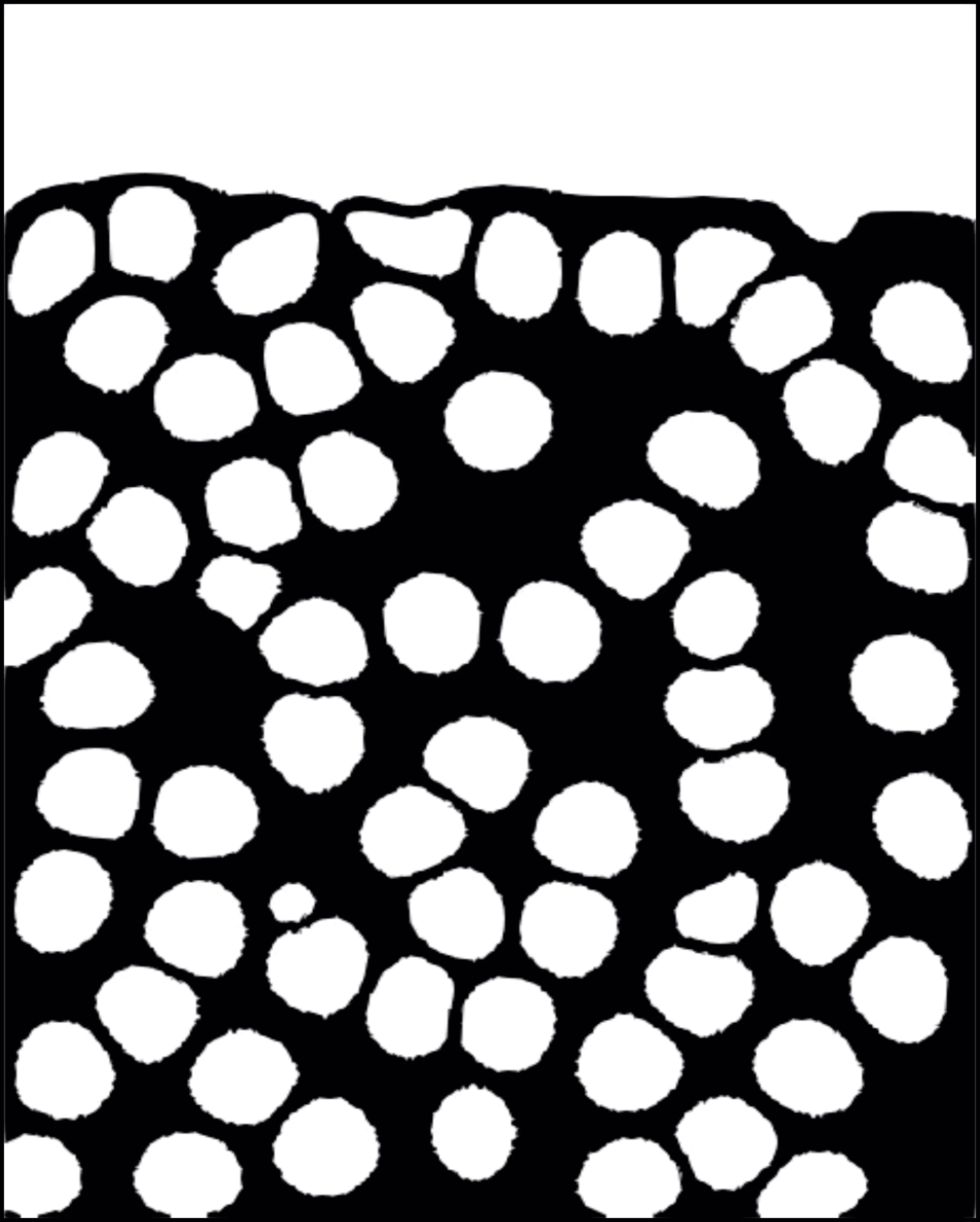} 
    ~
    \includegraphics[trim = 0mm 0mm 0mm 0mm, clip, width=0.14\textwidth]{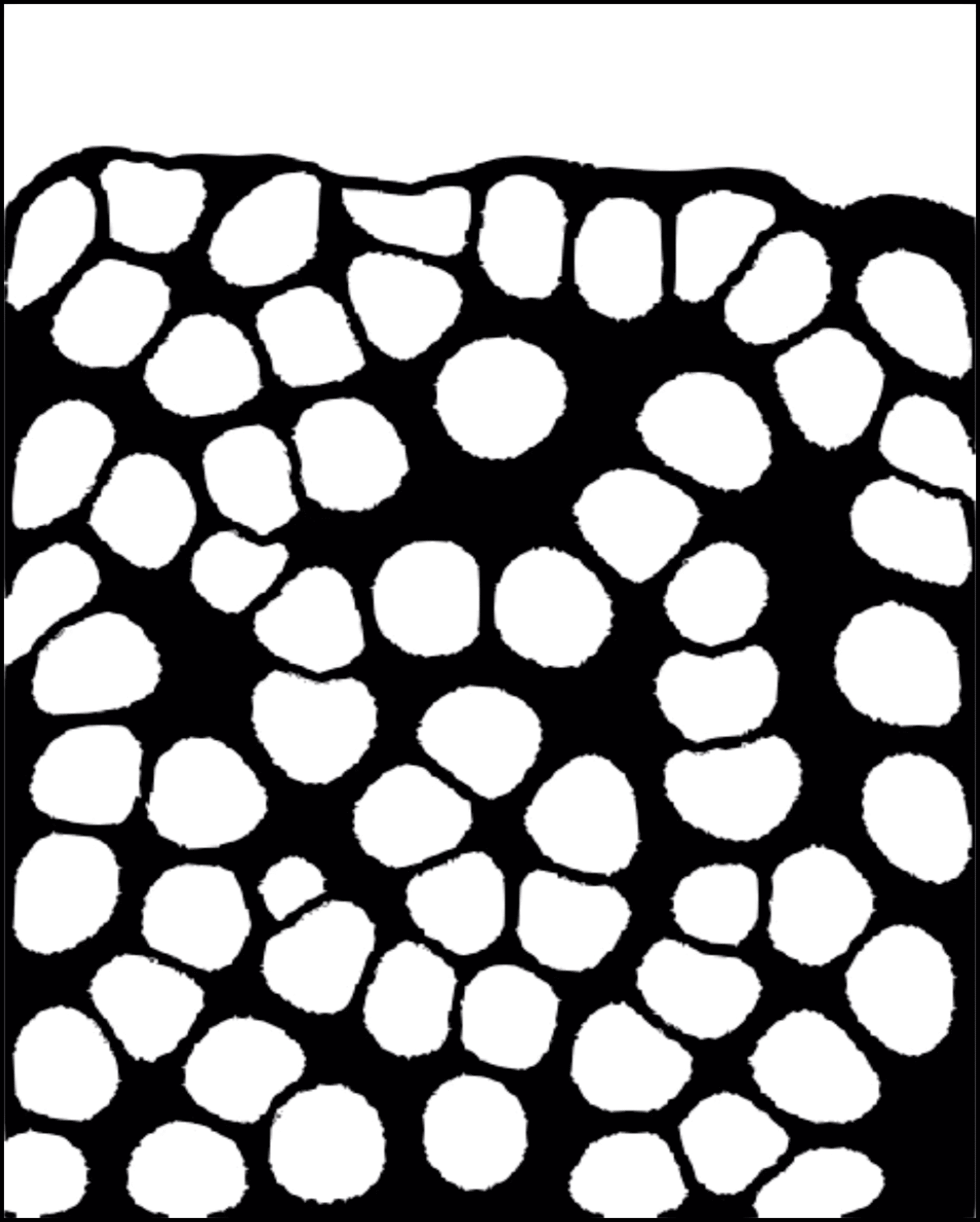} 
    ~
    \includegraphics[trim = 0mm 0mm 0mm 0mm, clip, width=0.14\textwidth]{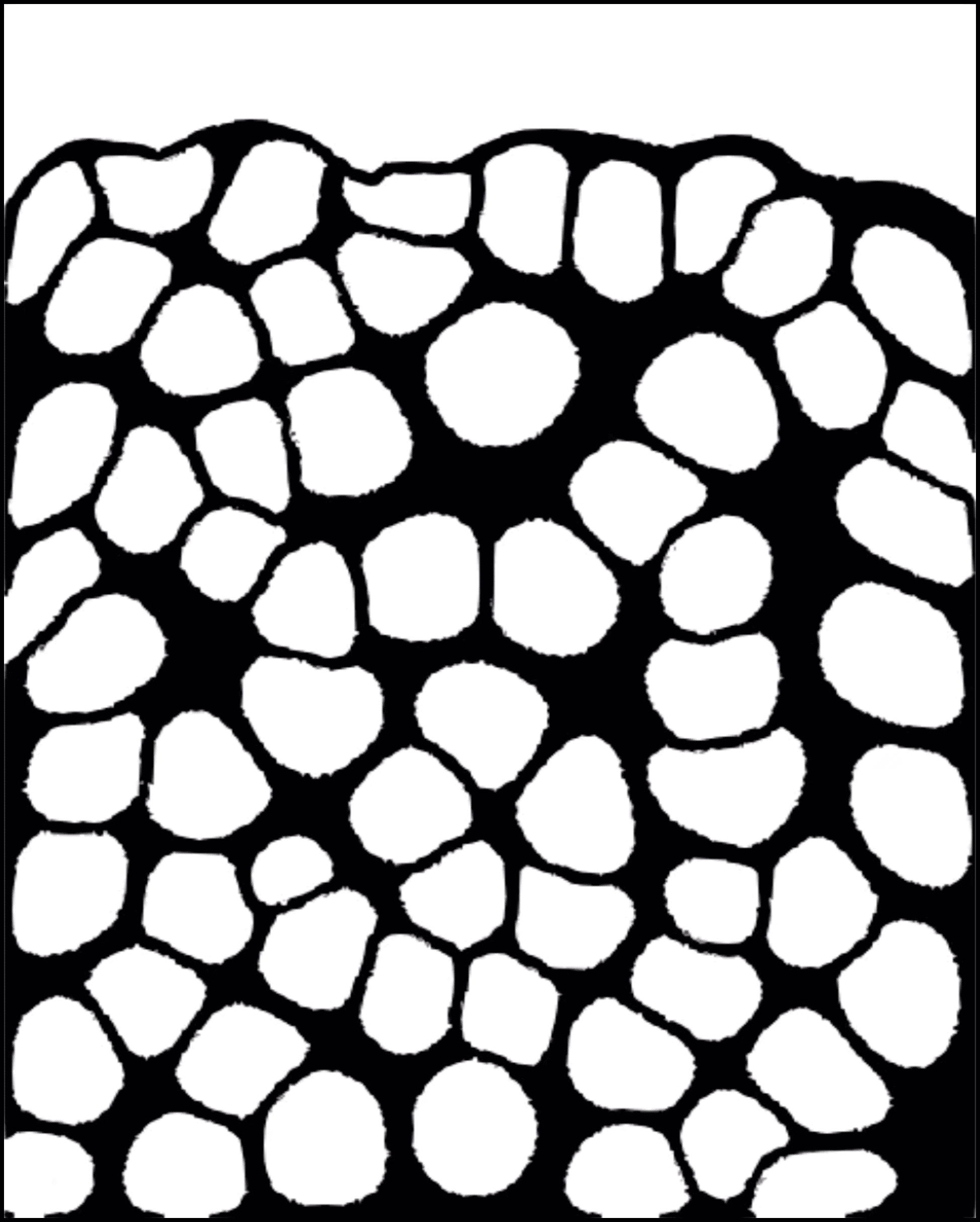} 
    ~
    \includegraphics[trim = 0mm 0mm 0mm 0mm, clip, width=0.14\textwidth]{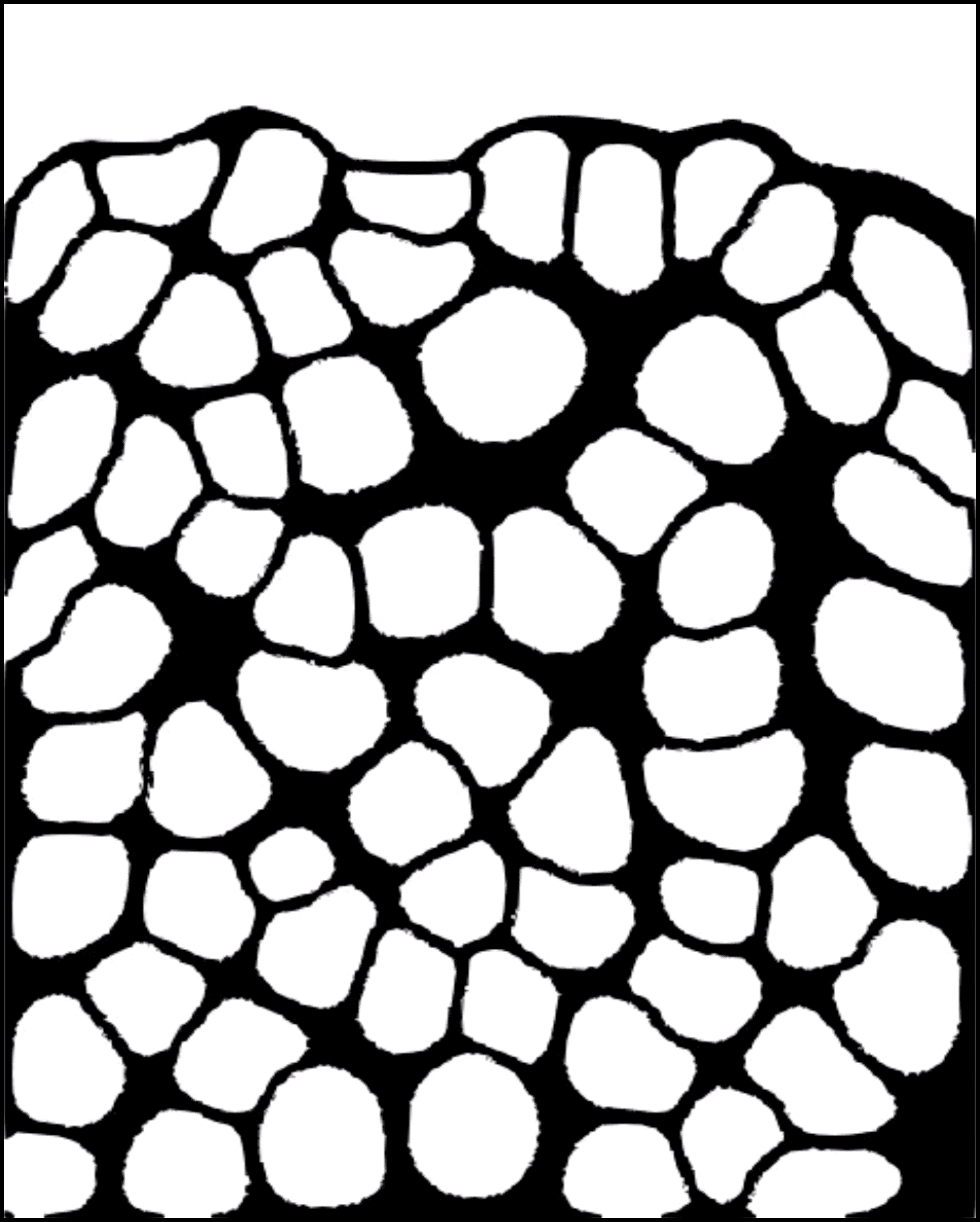}
        \\      (b)     \\
    \includegraphics[trim = 0mm 0mm 0mm 0mm, clip, width=0.14\textwidth]{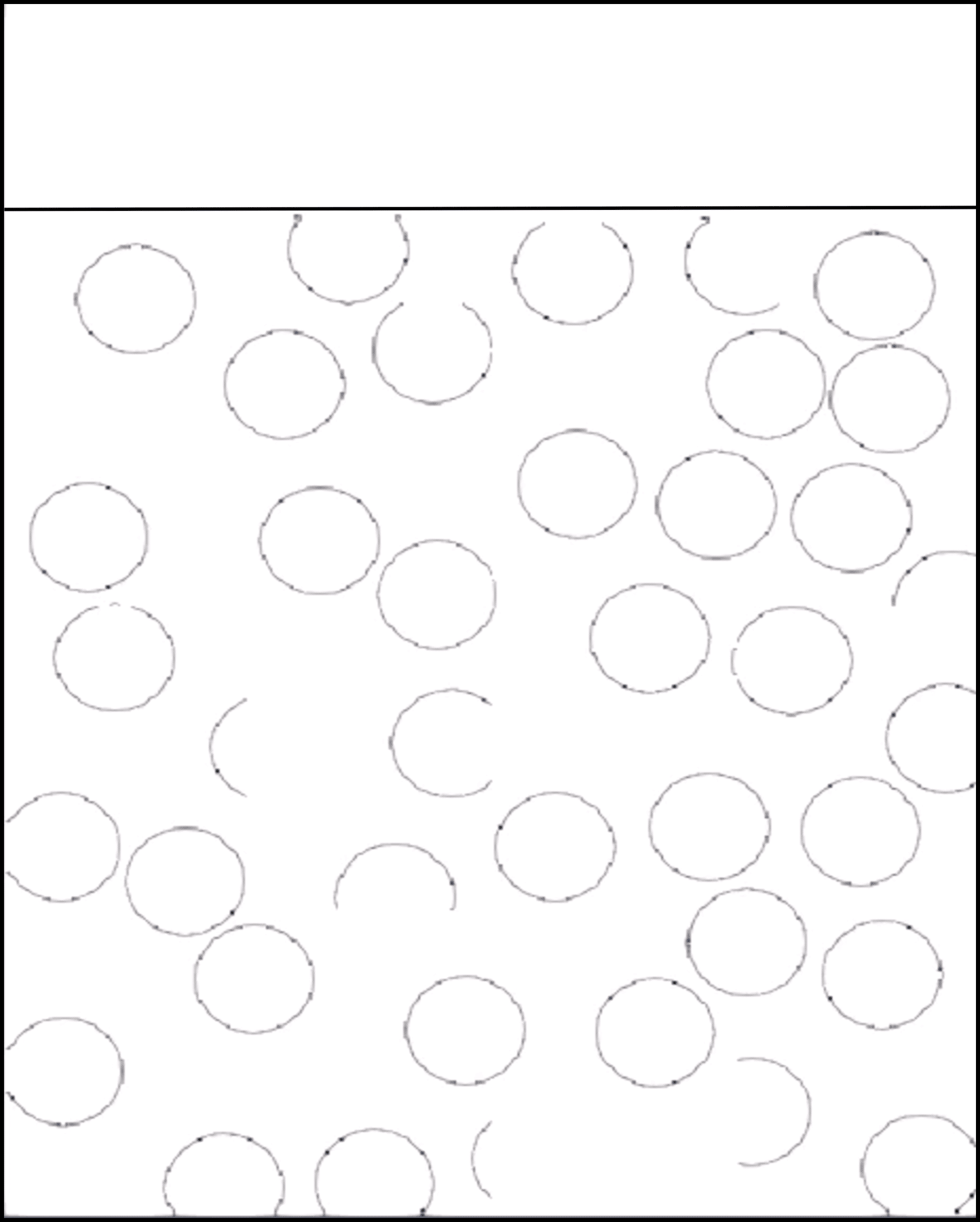}
    ~
    \includegraphics[trim = 0mm 0mm 0mm 0mm, clip, width=0.14\textwidth]{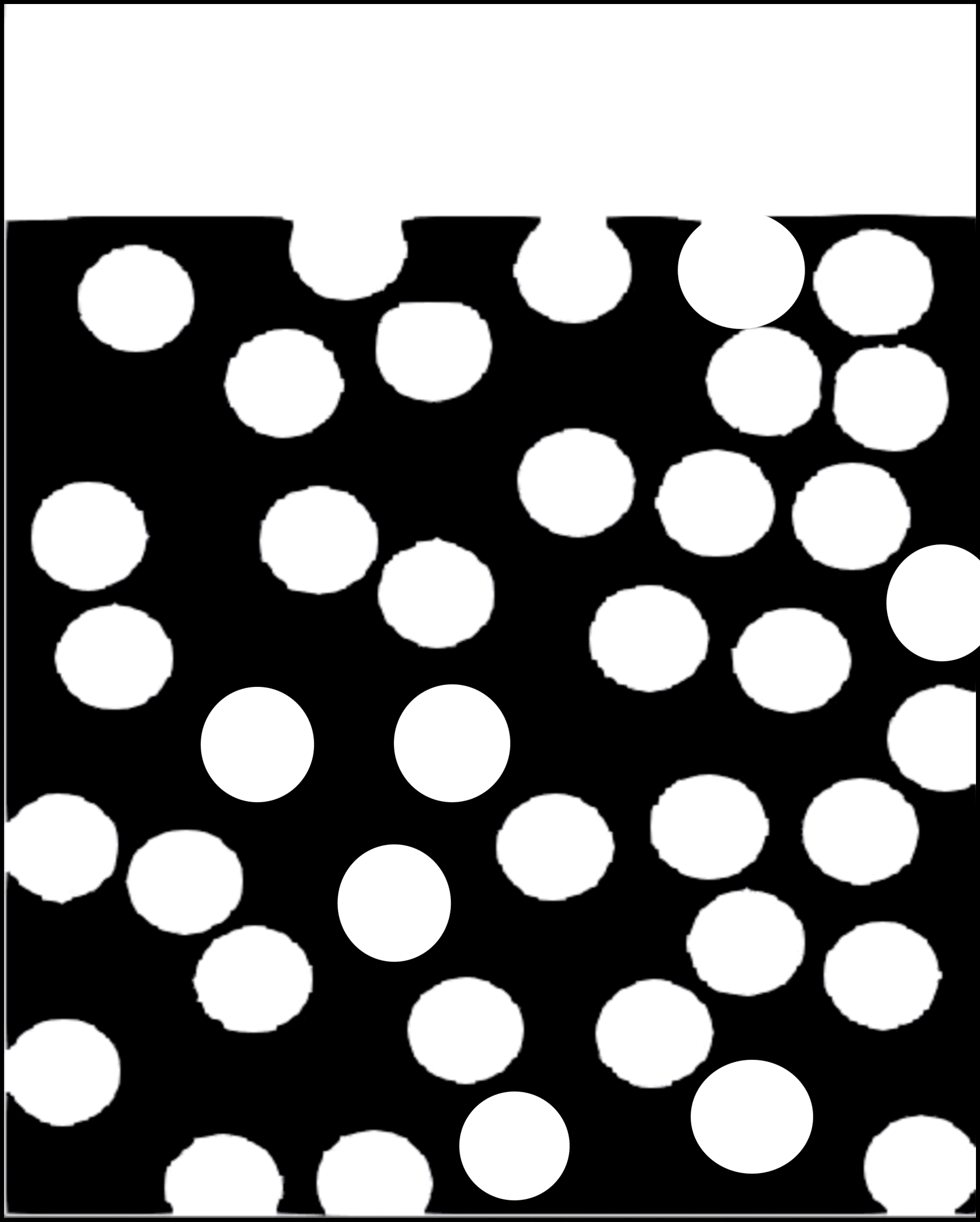} 
    ~
    \includegraphics[trim = 0mm 0mm 0mm 0mm, clip, width=0.14\textwidth]{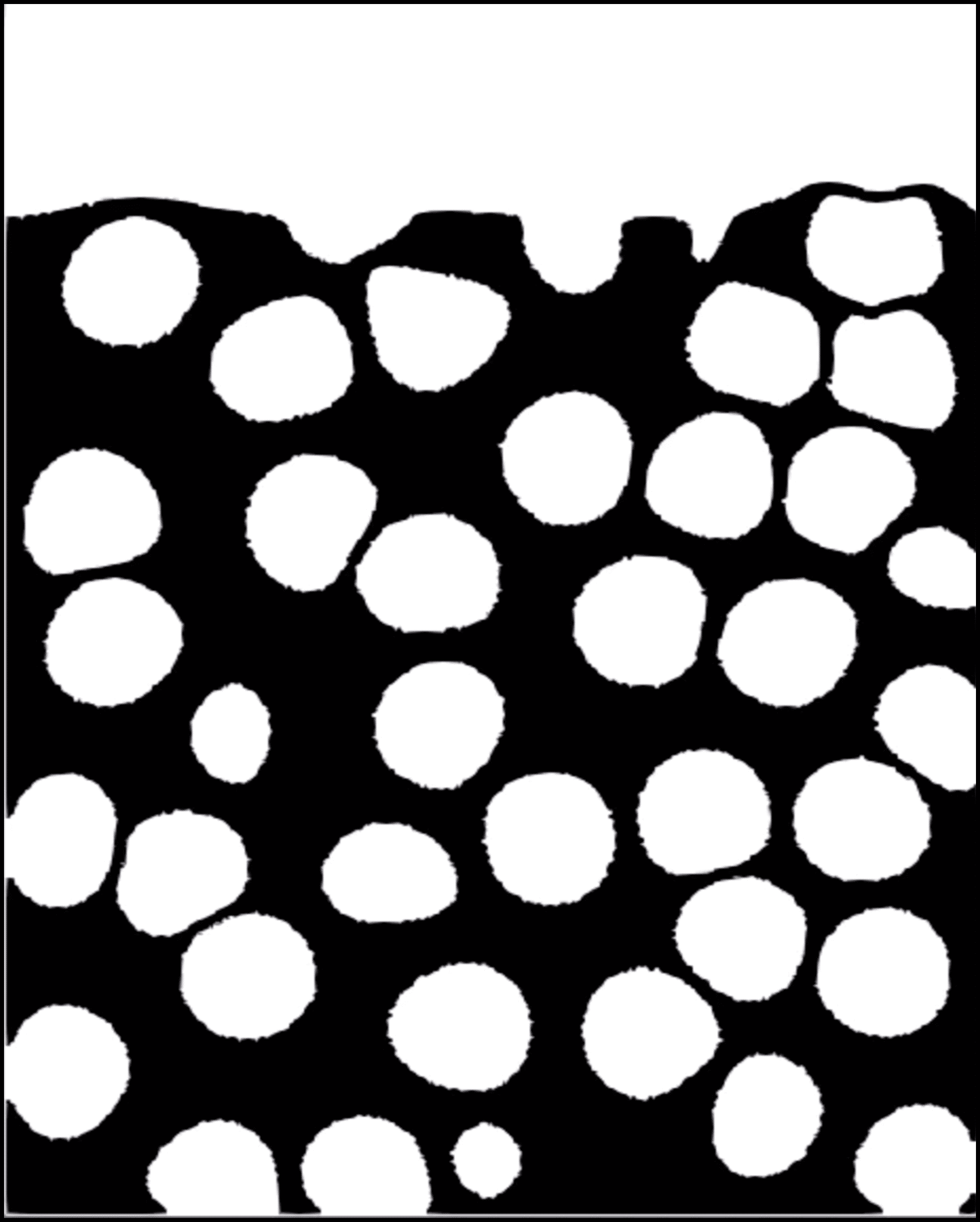} 
    ~
    \includegraphics[trim = 0mm 0mm 0mm 0mm, clip, width=0.14\textwidth]{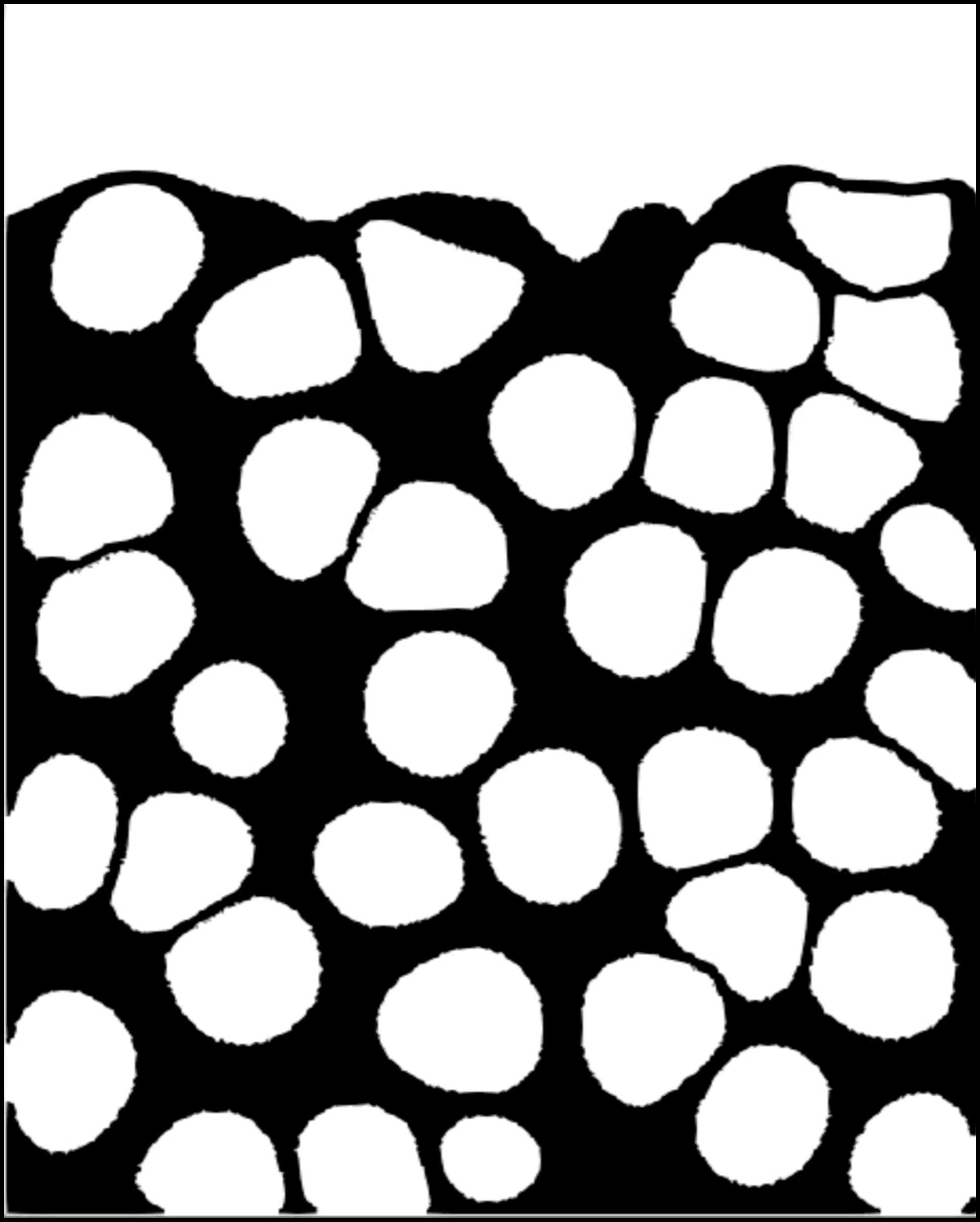} 
    ~
    \includegraphics[trim = 0mm 0mm 0mm 0mm, clip, width=0.14\textwidth]{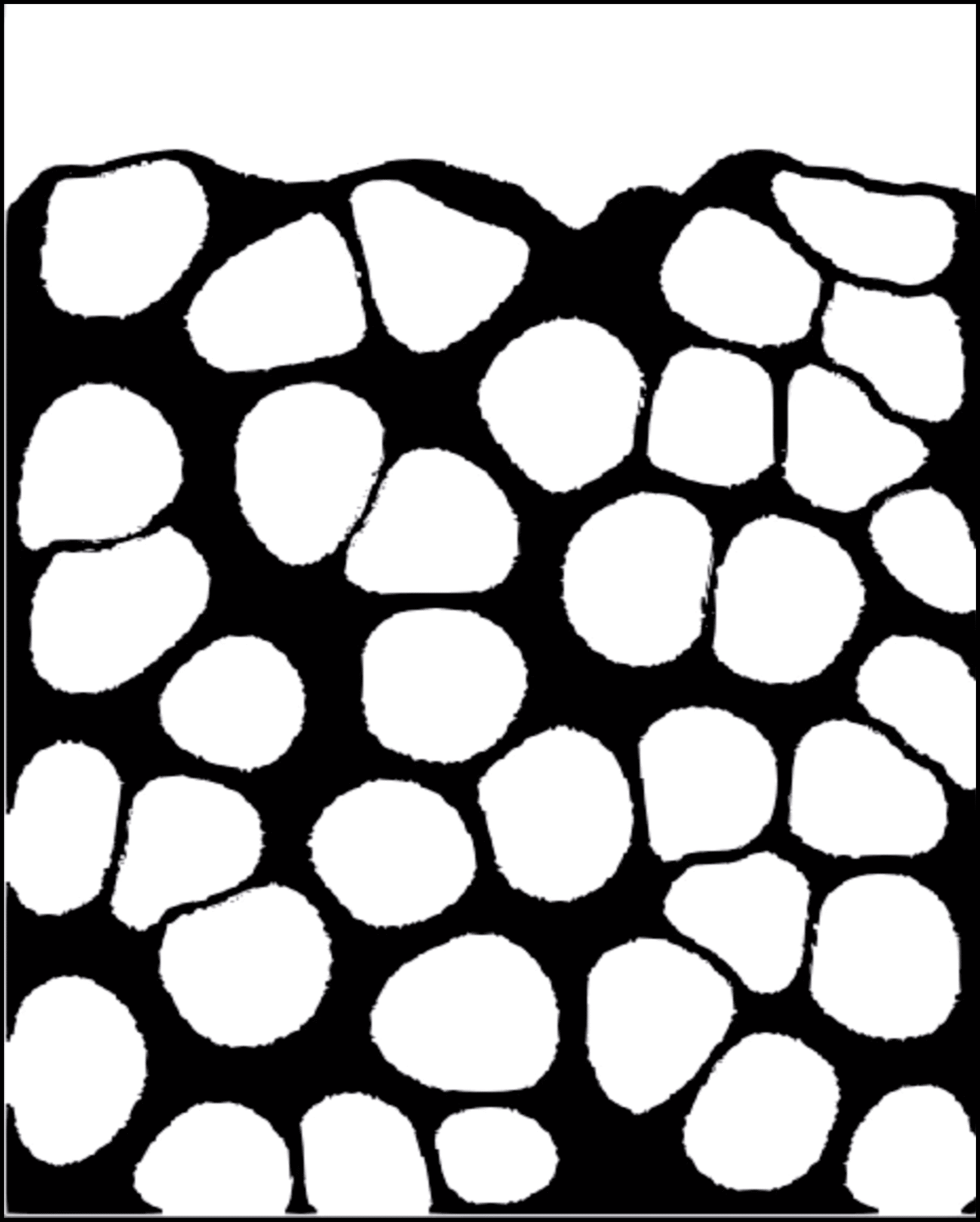} 
    ~
    \includegraphics[trim = 0mm 0mm 0mm 0mm, clip, width=0.14\textwidth]{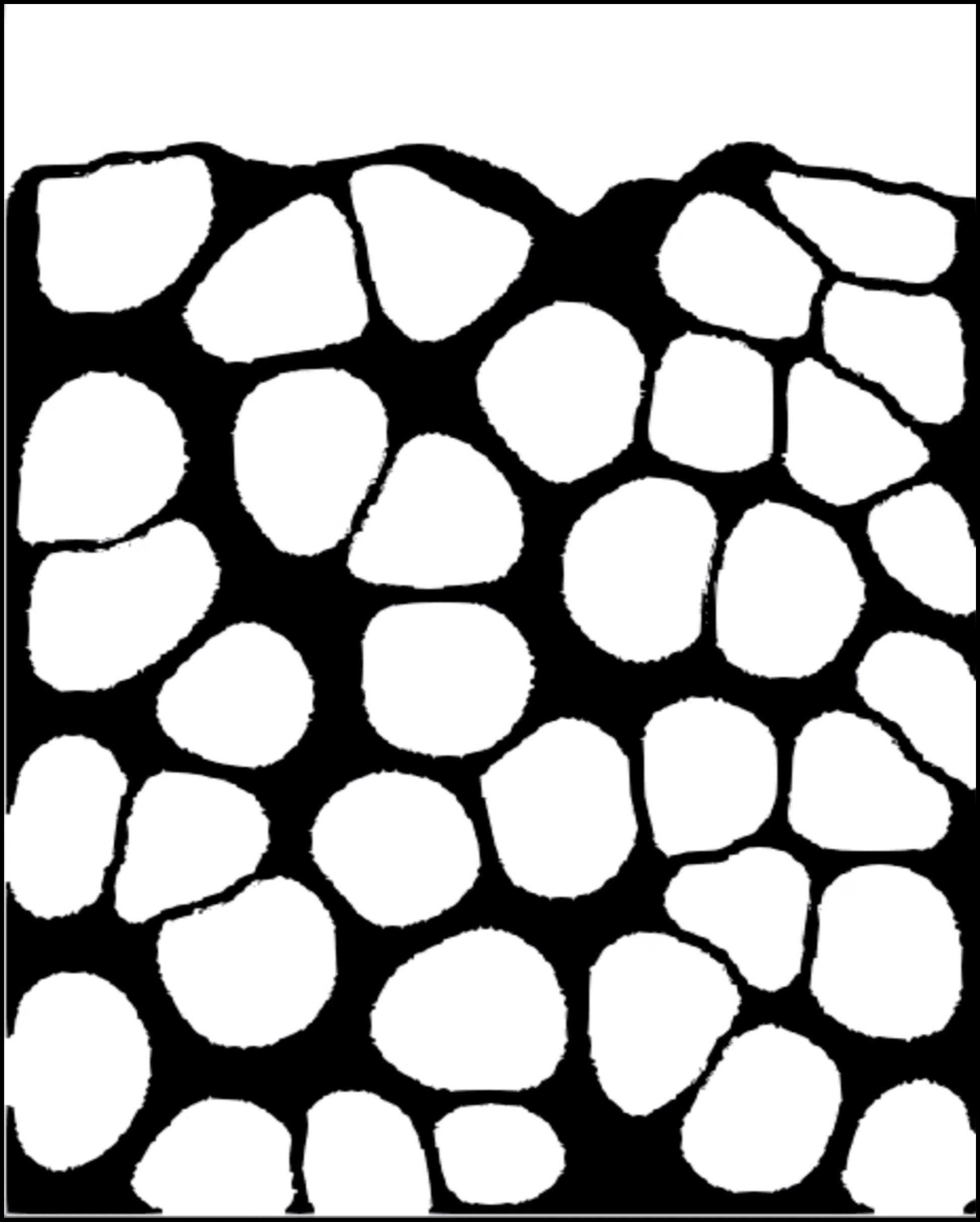}
        \\      (c)     \\    
\caption{FSLBM simulations of the ceramic aerogel foaming process (pores in white) showing the time evolution of bubbles' growth and coalescence in the pre-aerogel to form a porous microstructure. All simulations are conducted over 20 minutes and 5000 timesteps each with $\Delta t=2$s and all the physical parameters are kept unchanged except the initial number of bubbles $n_{bub}$ and initial bubble radius $r_{bub}$ leading to different average aerogel pore size at the end of simulation:
(a) $n_{bub} = 250 $, $r_{bub} = 8 $,  leading to pore radius $8 \mu m$;
(b) $n_{bub} = 50$, $r_{bub} = 18 $,  leading to pore radius $18 \mu m$;
(c) $n_{bub} = 25$, $r_{bub} = 25$,  leading to pore radius $25 \mu m$.
The simulation domain height was 550 $\mu$m and width was 400 $\mu$m, with 400 $\mu$m fluid pool height throughout the simulations.
}
        \label{fig:lbfoam_evol}
        \vspace{-0.2in}
\end{figure}

% Numerical example
Figure \ref{fig:lbfoam_evol} presents snapshots of pre-aerogel foaming simulations conducted using FSLBM. The simulations were initialized with a fluid pool of height 400 $\mu m$, with bubbles of three distinct nuclei sizes (8 $\mu m$, 18 $\mu m$, and 25 $\mu m$) spatially distributed within the simulation domain using a Poisson disk-sampling algorithm \cite{bridson2007}.
The initial number and radii of bubbles serve as proxies for varying concentrations of the foaming agent, one of the key synthesis parameters influencing the resulting ceramic aerogel microstructure. For all simulations, the fluid density was set to 1 $kg/{\mu m}^3$, ambient pressure to 0.33 $N/{\mu m}^2$, surface tension to $\sigma_{LB} = 5 \times 10^{-3}$ $N/\mu m$, and the advection-diffusion source term to $S = 5 \times 10^{-5}$ ${\mu m}^2/s$. We note that, the FSLBM framework accounts for bubble dissolution by coupling the Lattice Boltzmann flow solver with an advection-diffusion equation subject to Henry’s law boundary conditions at bubble interfaces. Bubble rupture is modeled using a disjoining-pressure formulation, where coalescence or collapse occurs if the stabilizing forces between interfaces are insufficient. However, in the ceramic aerogel synthesis process, the combination of CTAB micelles and TEOS-driven silica condensation rapidly forms rigid shells around nascent bubbles, while urea decomposition introduces gas into a gelling silica matrix. This coordinated mechanism effectively stabilizes bubbles against rupture and dissolution, an effect captured by the parameters used in our FSLBM simulations. The variation in the number of initial bubbles and their radii led to distinct classes of ceramic microstructures with different average pore size at the final simulation time.
As demonstrated later, the use of the Poisson disk-sampling algorithm enables FSLBM to generate multiple ensemble realizations for each microstructure class with the same numbers and initial radius of the bubbles to account for stochasticity of the aerogel microstructure.

%++++++++++++++++++++++++++++++++++++++++++++++++++++++++++++++++++++++++
\subsection{Generative surrogate model of aerogel microstructure}\label{sec:gans}
\noindent
FSLBM simulations of aerogel microstructures are computationally expensive, particularly for tasks involving uncertainty propagation through the microstructure–property relationship (see Figure \ref{fig:framework}), which require repeated evaluations of $\mathcal{F}_{LB}(\bs \xi ; \bs \theta_{LB})$ to capture the inherent stochasticity of the microstructure. 
To enable a computationally efficient synthesis-to-microstructure mapping, we introduce a surrogate model for the FSLBM simulations based on the Generative Adversarial Network (GAN) framework originally proposed by \cite{goodfellow2014generative}.
GANs employ two deep convolutional neural networks: the generator $\mathcal{G}_{GAN}(\cdot)$ and the discriminator $\mathcal{D}(\cdot)$. 
These networks are trained an adversarial manner in a two-player minimax game, where the generator, parameterized by $\bs w_{GAN}$, maps a random latent input vector to the data space $p_{G}$, approximating the real data distribution $p_{data}$. Meanwhile, the discriminator, with weight parameters $\boldsymbol{w}_{dis}$, acts as a classifier, outputting the probability of a sample being real or generated. In this unsupervised learning setup, the generator minimizes the divergence between $p_G$ and $p_{data}$, while the discriminator maximizes it, until both networks reach a Nash equilibrium.
Once trained on microstructural imaging data, the generator can efficiently produce ensembles of microstructures that closely replicate the characteristics of the training data. In recent years, the application of GANs to imaging and digital material data has gained significant attention, demonstrating their potential in capturing realistic microstructural features, e.g., \cite{chun2020deep, CHEN2024112976, SCHENK2024113064, FUKATSU2024113143, QIAN2024112942}.

To formulate the generative surrogate model for ceramic aerogel microstructures, we define a \textit{microstructure indicator function} $\varphi(\mathbf{x})$, as illustrated in Figure \ref{fig:framework}. This spatial field assigns values of $\varphi = 0$ to the spatial points $\mathbf{x}$ within the pores and $\varphi = 1$ to those within the solid skeleton of the aerogel. The objective of the surrogate model is to generate a large ensemble of aerogel microstructural patterns, such that
\begin{equation}\label{eq:generative}
	\varphi(\mathbf{x}) = \mathcal{G}_{GAN}(\bs \xi, \bs z ; \bs w_{GAN}),
\end{equation}
where $\mathcal{G}_{GAN}(\bs \xi, \bs z ; \bs w_{GAN})$ is the GAN’s generator, a fully convolutional neural network parameterized by weight vector $\bs w_{GAN}$ that up-convolve the random vector of \textit{latent variable} $\bs z \sim p_{\bs z}(\bs z)$ where $ p_{\bs z}$ is a Gaussian zero mean distribution, scaling its dimension up at each layer, to produce microstructure image. 
A key feature of \eqref{eq:generative} is that the generator is conditioned on the synthesis inputs $\bs \xi$ as an additional input. During training, $\bs z$ varies randomly, while $\bs \xi$ are set to the corresponding synthesis input used to generate the training data, which consist of microstructural images obtained from FSLBM simulations. Once trainined, $\mathcal{G}_{GAN}(\bs \xi, \bs z ; \bs w_{GAN})$ can efficiently generate random microstructure patterns $\varphi(\mathbf{x})$ by sampling $\bs z$  while by varying $\bs \xi$, the model can produce a wide range of diverse and novel morphologies that correspond to different aerogel synthesis conditions, including those interpolated between training data points.

Training vanilla GANs is notoriously challenging, often requiring significant manual intervention to stabilize the training process. To address these challenges, we utilize Wasserstein GANs with Gradient Penalty (WGAN-GP), as proposed by \cite{gulrajani2017improved}. WGAN-GP enhances stability by imposing Lipschitz continuity constraints on the discriminator and introducing a gradient penalty term into the discriminator's loss function. Like vanilla GANs, WGAN-GP is formulated as a two-player zero-sum game, expressed as:
\begin{equation}\label{eq:gans_minmax}
\underset{\boldsymbol{w}_{GAN}}{inf} \; %\underset{\boldsymbol{w}_{dis};\mathcal{D} \; \text{is 1-Lipschitz}}{sup} \;
\underset{\underset{\mathcal{D} \rm{\; is \, 1-Lipschitz}}{\boldsymbol{w}_{dis}}}{sup} \;
\mathbb{E}_{\varphi(\mathbf{x}) \sim p_{data}}
\left[\mathcal{D}(\varphi(\mathbf{x}); \boldsymbol{w}_{dis}) \right ]
-
\mathbb{E}_{\boldsymbol{z}\sim p_{z}(\boldsymbol{z})}
\left[\mathcal{D}(\mathcal{G}(\varphi(\mathbf{x}); \boldsymbol{w}_{GAN}); \boldsymbol{ w}_{dis}) \right ].
\end{equation}
The loss functions for the generator and discriminator in the WGAN-GP are defined as,
\begin{eqnarray}
\mathcal{L}_g & = & - \mathbb{E}_{\boldsymbol{z}\sim p_{z}(\boldsymbol{z})}
\left[\mathcal{D}(\mathcal{G}(\bs \xi, \boldsymbol{z}; \boldsymbol{w}_{GAN}); \boldsymbol{w}_{dis}) \right ]\\
\mathcal{L}_d & = & \mathbb{E}_{\boldsymbol{z}\sim p_{z}(\boldsymbol{z})}
\left[\mathcal{D}(\mathcal{G}(\bs \xi, \boldsymbol{z}; \boldsymbol{w}_{GAN}); \boldsymbol{w}_{dis}) \right ] - \mathbb{E}_{\varphi(\mathbf{x})\sim p_{data}}
\left[\mathcal{D}(\varphi(\mathbf{x}); \boldsymbol{ w}_{dis}) \right ] \nonumber\\
	& + &
c_{\lambda} \mathbb{E}_{\hat{\varphi}(\mathbf{x})\sim p_{\hat{\varphi}(\mathbf{x})}}
\left[ (\| \nabla_{\hat{\varphi}(\mathbf{x})} \mathcal{D}(\hat{\varphi}(\mathbf{x}); \boldsymbol{w}_{dis}) \|_2 - 1)^2 \right],
\end{eqnarray}
where $p_{\hat{\varphi}(\mathbf{x})}$ represents the distribution obtained by uniformly sampling points along straight lines between pairs of samples from $p_{data}$ and $p_{G}$. 
The discriminator's loss consists of three components: the first two terms calculate the Wasserstein-1 distance, measuring the difference between real and generated distributions, while the third term imposes a gradient penalty to enforce Lipschitz continuity, with $c_{\lambda}$ as the regularization weight. The generator's loss minimizes the discriminator's evaluation of the generated samples, encouraging the generator to produce more realistic outputs \cite{yang2020physics}.

%++++++++++++++++++++++++++++++++++++++++++++++++++++++++++++++++++++++++
%++++++++++++++++++++++++++++++++++++++++++++++++++++++++++++++++++++++++
\section{Physical and surrogate models of aerogel mechanical properties}\label{sec:CNN}
This section introduces a physics-based model for analyzing stress and strain distribution within the two-phase aerogel material. Additionally, it describes the development of a microstructure-property surrogate model using a Bayesian convolutional neural network (BayesCNN). Coupled with the microstructural images generated by the WGAN-GP model outlined in Section \ref{sec:gans}, the surrogate models facilitate efficient prediction of the probability distribution of the quantity of interest (QoI), specifically the strain energy.

%++++++++++++++++++++++++++++++++++++++++++++++++++++++++++++++++++++++++
\subsection{Stochastic finite element model of aerogel elastic deformation}
\begin{figure}[h!]
\centering
    \includegraphics[trim = 20mm 0mm 150mm 0mm, clip, width=0.7\textwidth]{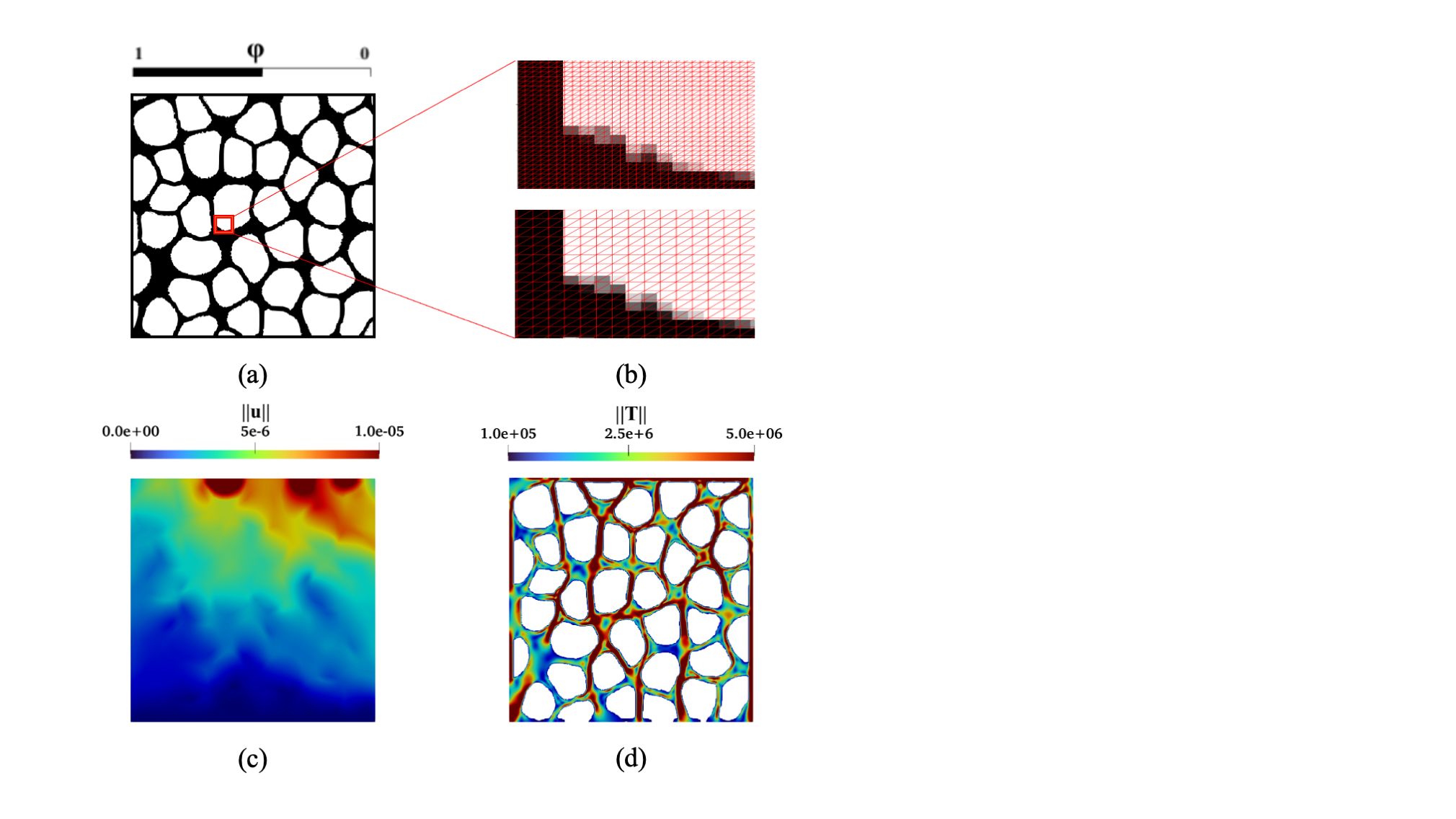}
    \vspace{-0.3in}
    \caption{
    Finite element (FE) simulation results for for a representative microstructure indicator function:
    (a and b) Microstructure image and mesh convergence study comparing two mesh resolutions, 720 $\times$ 720 and 358 $\times$ 358 elements, yielding strain energies of $1382\times10^{-7}$ mJ and $1378\times10^{-7}$ mJ, respectively. 
    The relative error of 0.3\% confirms that the 358 $\times$ 358 mesh provides sufficient resolution for accurate results and is used for all subsequent simulations.
    \blue{(c and d)} Magnitude of the computed displacement (in mm) and Cauchy stress (in mPa=$10^{-3} N/m^2$) fields under uniaxial compression. 
    For all FE simulations, the solid phase is assigned a Young’s modulus $E_s = 100\times10^6$ Pa and Poisson's ratio $\nu_s$ = 0.3. A traction force of $10^5$ N was applied in the negative y-direction.
    }
    \vspace{-0.12in}
        \label{fig:finite_el}
\end{figure}

To accurately simulate the interplay between microstructural features and macroscopic mechanical properties of ceramic aerogels, we employ a finite element (FE) model governed by a stochastic partial differential equation. This model determines the displacement and stress fields based on the microstructural images of the aerogel. 
Let $\Omega$ represent a bounded domain in $\mathbb{R}^d$, where $d$ can be 1, 2, or 3 with Lipschitz boundary, and let the boundary of this domain be denoted by $\partial \Omega$. The goal is to determine the displacement field $\bs u(\mathbf{x})$ at spatial points $\mathbf{x} \in \Omega$. The governing equation is expressed as:
\begin{eqnarray}\label{eq:pde-aerogel}
    \nabla \cdot \bs T(\bs u) & = & \bs f(\mathbf{x}), \quad \mathbf{x} \in \Omega, \nonumber\\
    \bs T(\bs u) \bs n & = & \bs t(\mathbf{x}), \quad \mathbf{x} \in \Gamma_N, \nonumber\\
    \bs u(\mathbf{x}) & = & \bs u^*, \quad \mathbf{x} \in \Gamma_D,
\end{eqnarray}
where,
    $\bs f$ is a prescribed source term, 
    $\bs t$ is traction, 
    $\Gamma_N$ is the subset of the boundary $\partial\Omega$ where the Neumann boundary condition is applied, 
    $\Gamma_D = \partial\Omega \setminus \Gamma_N$ represents the boundary subject to Dirichlet conditions.
The Cauchy stress tensor $\bs T$ is defined as:
\begin{eqnarray}\label{eq:elasticity}
    \bs T(\bs u) & = & 2\mu_s \, \varphi(\bs x) \, \bs E(\bs u) 
    + \lambda_s \varphi(\bs x) \, {\rm tr}(\bs E(\bs u)) \bs I,
\end{eqnarray}
where,
    $\lambda_s$ and $\mu_s$ are the Lamé constants of the solid aerogel phase, equivalent to the material's Young's modulus $E_s$ and Poisson's ratio $\nu_s$,
    $\bs E(\bs u) = \frac{1}{2} \left( \nabla \bs u + (\nabla \bs u)^T \right)$ is the strain tensor,
and $\varphi(\mathbf{x})$ is the stochastic microstructure indicator function generated by the WGAN-GP surrogate model described in the previous section.
Figure \ref{fig:finite_el} illustrates the displacement and stress fields computed for a single realization of aerogel microstructure.
These results are obtained under uniaxial compression conditions, where a traction force of $t_y=10^5$ N is applied in the negative $y$-direction at the top surface. The bottom boundary is fixed $\bs u^*=0$, while roller boundary conditions $u^*_x=0$ are imposed at the left and right boundaries.
The governing equation in \eqref{eq:pde-aerogel} is solved using a uniformly fine mesh to resolve the intricate microstructural patterns represented by $\varphi(\mathbf{x})$. Convergence analysis with respect to mesh resolution indicates that a finite element mesh with 358$\times$358 elements provides sufficient accuracy and numerical stability for this class of microstructures (\blue{see Figure \ref{fig:finite_el}(b)}). 
All finite element simulations are carried out using the DOLFINx library \cite{baratta_2023_10447666} from the FEniCS Project in Python.
The prediction QoI is the scalar strain energy, computed as:
\begin{equation}\label{eq:strain_energy}
    Q = \int_\Omega \bs T(\bs u) : \bs E(\bs u) \; d\mathbf{x}.
\end{equation}
\blue{
We note that in this study, a linear elastic constitutive model is adopted for the solid phase of the ceramic aerogel, to demonstrate the proposed surrogate modeling framework. However, this assumption may not capture nonlinear deformation mechanisms or damage evolution that may arise under loading and ultimately lead to mechanical failure. Ongoing mechanical testing of ceramic aerogels fabricated via in situ bubble-supported synthesis would better inform the development of more representative constitutive models that incorporate these effects (e.g.,  \cite{Ren2019hierarchical,chi2020,ChiShen2022tailoring}).
}

%++++++++++++++++++++++++++++++++++++++++++++++++++++++++++++++++++++++++
\subsection{Bayesian convolutional neural network surrogate model}
\noindent
To develop a surrogate model linking microstructure to mechanical properties, we employ a Bayesian convolutional neural network (CNN) that maps a microstructure image of ceramic aerogel $\varphi(\mathbf{x})$ to the corresponding strain energy output $Q^{CNN} = \mathcal{G}_{CNN}(\varphi_i; \bs w_{CNN})$ (see Figure \ref{fig:framework}). This mapping is parameterized by the weights $\bs{w}_{CNN}$ of the filters of the convolutional layers and the fully-connected layers of the CNN architecture, which are determined through training. The training dataset, denoted as  $\bs{D} = \{(\varphi_i, Q^{FE}_i)\}_{i=1}^{N_D}$ consists of $N_D$ samples furnished by the stochastic finite element model described in previous section, such that $Q^{FE}_i = \mathcal{F}_{FE}(\varphi_i; \bs\theta_{FE})$.
Traditional CNN training rely on maximum likelihood estimation of the weight parameters to minimize the error between the output and the training data. However, these approaches fail to explicitly account for the uncertainty in CNN weight parameters that translate to uncertainty in the output, which becomes critical in scenarios involving small and uncertain datasets. In contrast, a Bayesian framework provides a more robust approach by inferring probability distribution of the weights from data using the Bayes' rule,
\begin{equation}\label{eq:bayes}
\pi_{\text {post }}(\bs w_{CNN} \mid \bs{D}) =
\frac{\pi_{\text {like }}(\bs{D} \mid \bs w_{CNN}) \pi_{\text {prior }}(\bs w_{CNN})}{\pi_{\text {evid }}(\bs{D})},
\end{equation}
where $\pi_{\text{prior }}(\bs{w}_{CNN})$ is prior distribution of the weights,
$\pi_{\text {like }}(\bs{D} \mid \bs{w}_{CNN})$ is the likelihood of observing the data given the weights, and
$\pi_{\text {evid }}(\bs{D})$ is the evidence, a normalization factor ensuring that the posterior integrates to 1. 
The posterior distribution $\pi_{\text {post }}(\bs w_{CNN} \mid \bs{D})$ reflects the updated knowledge of the weights after incorporating the training data. Given this posterior distribution, for any new microstructure $\varphi^{\star}(\mathbf{x})$ the strain energy output can be predicted using the predictive distribution,
\begin{equation}\label{eq:prediction_dis}
\pi\left(Q^{CNN\star} \mid \varphi^{\star}, \bs{D}\right)=\int \pi\left(Q^{CNN\star} \mid \varphi^{\star}, \bs w_{CNN}\right) \pi_{\text {post }}(\bs w_{CNN} \mid \bs{D}) \, d \bs w_{CNN}.
\end{equation}
In practice, a collection of $M_{mc}$ samples are drawn from posterior $\hat{\bs{w}}_{CNN}^{(j)} \sim \pi_{\text {post }}(\bs w_{CNN} \mid \bs{D})$ and the prediction distribution can then be approximated using Monte Carlo estimation, which is expressed as:
\begin{eqnarray}\label{eq:prediction}
\pi\left(Q^{CNN\star} \mid \varphi^{\star}, \bs {D}\right) & = &
\mathbb{E}_{\bs w_{CNN} \mid \bs D} \left[ \pi\left(Q^{CNN\star} \mid \varphi^{\star}, \bs w_{CNN} \right) \right] \nonumber\\
& \approx &
\frac{1}{M_{mc}} \sum_{j=1}^{M_{mc}} \mathcal{G}_{CNN}(\varphi^{\star}; \hat{\bs{w}}_{CNN}^{(j)}).
\end{eqnarray}

\paragraph{Variational Inference}
To achieve computationally efficient Bayesian inference for high-dimensional parameter spaces, we employ variational inference, an optimization-based technique that approximates the true posterior distribution $\pi_{\text{post}}(\bs{w}_{CNN} \mid \bs{D})$ with a tractable variational distribution $q_{\boldsymbol{\eta}}(\bs{w}_{CNN})$, parameterized by $\boldsymbol{\eta}$ and often chosen from a family of distributions such as Gaussians. The optimal variational distribution $q^{\text{opt}}$ is obtained by minimizing the Kullback-Leibler (KL) divergence between the variational distribution and the true posterior,
\begin{equation}\label{eq:q_opt}
q^{\text{opt}}  =  \underset{q}{\arg \min } \; \big\{ \mathrm{KL} \left[q_{\boldsymbol{\eta}}(\bs{w}_{CNN}) \| \pi_{\text{post}}(\bs{w}_{CNN} \mid \bs{D}) \right]  \big\},
\end{equation}
where the KL divergence is defined as:
\begin{eqnarray}\label{eq:KL}
\mathrm{KL} \left[q_{\bs \eta}(\bs w_{CNN}) \| \pi_{\text{post}}(\bs w_{CNN} \mid \bs{D})\right] & = &
\int q_{\bs \eta}(\bs w_{CNN}) \ln \frac{q_{\bs \eta}(\bs w_{CNN})}{\pi_{\text{post}}(\bs w_{CNN} \mid \bs{D})} d \bs w_{CNN} \nonumber\\
& = & 
\mathbb{E}_{q_{\bs \eta}}\left[\ln \frac{q_{\bs \eta}(\bs w_{CNN})}{\pi_{\text{post}}(\bs w_{CNN} \mid \bs{D})}\right].
\end{eqnarray}
Then, the prediction in \eqref{eq:prediction_dis}  is performed with replacing the posterior with the tractable variational approximation $q^{\text{opt}}$.
Substituting the posterior from Bayes’ theorem \eqref{eq:bayes} into the \eqref{eq:KL} gives,
\begin{eqnarray}
\mathrm{KL} \left[q_{\bs \eta}(\bs w_{CNN}) \| \pi_{\text{post}}(\bs w_{CNN} \mid \bs{D})\right] = 
& - & \mathbb{E}_{q_{\bs \eta}}\left[\ln \frac{\pi_{\text{prior}}(\bs w_{CNN} )}{q_{\bs \eta}(\bs w_{CNN})}\right]\nonumber\\
& - & \mathbb{E}_{q_{\bs \eta}}\left[\ln \pi_{\text{like}}(\bs D \mid \bs w_{CNN} )\right] \nonumber\\
& + & \mathbb{E}_{q_{\bs \eta}}\left[\ln \pi_{\text{evid}}(\bs D)\right]. 
\end{eqnarray}
The evidence $\pi_{\text{evid}}(\bs D)$ is independent of the variational distribution $q_{\bs\eta}(\bs w_{CNN})$, so the last term in the above equation $\mathbb{E}_{q_{\bs \eta}} [\ln  \pi_{\text{evid}}(\bs D)] = \ln  \pi_{\text{evid}}(\bs D)$ and does not contribute to the minimization problem in \eqref{eq:q_opt}.
This leads to the definition of the evidence lower bound (ELBO),
\begin{eqnarray}\label{eq:elbo}
\mathrm{L} [q_{\bs\eta}(\bs w_{CNN})] & = & \mathbb{E}_{q_{\bs \eta}}\left[\ln \frac{\pi_{\text{prior}}(\bs w_{CNN} )}{q_{\bs \eta}(\bs w_{CNN})}\right]
+ \mathbb{E}_{q_{\bs \eta}}\left[\ln \pi_{\text{like}}(\bs D \mid \bs w_{CNN} )\right] \nonumber\\
& = &
- \mathrm{KL}\left[q_{\bs \eta}(\bs w_{CNN}) \| \pi_{\text{prior}}(\bs w_{CNN})\right]
+ \mathbb{E}_{q_{\bs \eta}}\left[\ln \pi_{\text{like}}(\bs D \mid \bs w_{CNN} )\right].
\end{eqnarray}
Thus, the KL divergence can be rewritten as:
\begin{equation}
\mathrm{KL} \left[q_{\bs \eta}(\bs w_{CNN}) \| \pi_{\text{post}}(\bs w_{CNN} \mid \bs{D})\right] = 
-\mathrm{L}[q_{\bs\eta}(\bs w_{CNN})] + \ln  \pi_{\text{evid}}(\bs D),
\end{equation}
where minimizing the KL divergence is equivalent to maximizing $ L[q_{\bs\eta}(\bs w_{CNN})] $ over $q_{\bs\eta}(\bs w_{CNN}) $ and ELBO provides a lower bound on $\ln  \pi_{\text{evid}}(\bs D)$. 
Considering an additive noise model, assuming Gaussian-distributed total error, the log-likelihood function is defined as:
\begin{eqnarray}\label{eq:likelihood}
\ln \pi_{\text{like}}(\bs{D} \mid \bs w_{CNN})
&=&-\frac{N_D}{2} \ln(2\pi\sigma_{noise}) - \frac{1}{2\sigma_{noise}^2} \sum_{i=1}^{N_D} \left(Q^{CNN}_i - Q^{FE}_i\right)^2.
\end{eqnarray}
where 
$R(\bs w_{CNN}) = \sum_{i=1}^{N_D} \left(Q^{CNN}_i - Q^{FE}_i\right)^2$ is called the data misfit and
$\sigma_{noise}^2$ is a noise variance hyperparameter that regulates the relative contribution of the prior (regularization) and the likelihood (data fit) in the posterior inference.
The likelihood formulation in \eqref{eq:likelihood}, equipped with a tunable noise parameter, allows the BayesCNN surrogate to account for both aleatoric uncertainty stemming from microstructural variability in the stochastic finite element simulations, and epistemic uncertainty due to limited training data availability.

The expectation of the log-likelihood is evaluated using Monte Carlo sampling, 
\begin{equation}\label{eq:expected_likelihood}
\mathbb{E}_{q_{\bs \eta}(\bs w_{CNN})}[\ln \pi_{\text{like}}(\bs{D} \mid \bs w_{CNN})]
=
\frac{1}{M_{mc}} \sum_{j=1}^{M_{mc}} \ln \pi_{\text{like}}\left(\bs{D} \mid \hat{\bs w}_{CNN}^{(j)}\right),
\end{equation}
where $\hat{\bs w}_{CNN}^{(j)} \sim q_{\bs \eta}(\bs w_{CNN})$ are samples drawn from the variational distribution, and $M_{mc}$ is the number of samples.
%
% %%%==========
To achieve computational efficiency in the high-dimensional parameter space of CNNs, where full-rank posterior approximations are intractable, the mean-field approximation is employed. It expresses the variational distribution as a product of independent Gaussian distributions,
\begin{equation}\label{eq:meanfield}
q_{\bs\eta}(\bs w_{CNN}) = \prod_{i=1}^{s} \mathcal{N}(w_{CNN_i}; \mu_i, \sigma_i),
\end{equation}
where the variational parameters $\boldsymbol{\eta}$ consist of the means and standard deviations of each Gaussian component, $\boldsymbol{\eta} = \{\mu_i, \sigma_i\}_{i=1}^d$.
Bayesian inference is then framed as optimizing the variational parameters $\boldsymbol{\eta}$ to ensure that the variational density closely approximates the true posterior distribution. Using the ELBO from \eqref{eq:newcost}, this can be written as,
\begin{equation}\label{eq:eta_opt}
\bs \eta^{\text{opt }}= \underset{\bs \eta}{\arg \min } \;\; 
\left\{
\mathrm{KL}\left[q_{\bs \eta}(\bs w_{CNN}) \| \pi_{\text{prior}}(\bs w_{CNN})\right]
- \mathbb{E}_{q_{\bs \eta}}\left[\ln \pi_{\text{like}}(\bs D \mid \bs w_{CNN} )\right]
\right\}.
\end{equation}
Considering Gaussian prior, the first term in \eqref{eq:eta_opt} has a closed-form solution, while the second term can be evaluated using \eqref{eq:expected_likelihood}.
Beyond computational convenience, the use of a Gaussian prior in BayesCNNs is motivated by the maximum entropy principle \cite{jaynes2003}, which states that among all distributions with specified mean and variance, the Gaussian has the highest entropy and thus represents the least informative prior.
In this work, we adopt a zero-mean Gaussian prior with variance 0.01, reflecting an unbiased assumption that positive and negative weights are equally likely before observing data. The chosen variance expresses moderate prior confidence, providing regularization to mitigate overfitting without overly constraining the capacity to learn from data \cite{tran2022all, nalisnick2018priors}.

For numerical optimization, the ELBO from \eqref{eq:elbo} is reformulated as, 
\begin{equation}\label{eq:newcost}
\mathrm{L} [q_{\bs\eta}(\bs w_{CNN})] \propto
\beta\mathrm{KL}\left[q_{\bs \eta}(\bs w_{CNN}) \| \pi_{\text{prior}}(\bs w_{CNN})\right]
+ \frac{1}{M_{mc}}\sum_{j=1}^{M_{mc}} R(\hat{\bs w}_{CNN}^{(j)}),
\end{equation}
where $\beta = 2\sigma_\text{noise}^2$ and the right-hand side of the \eqref{eq:newcost}
is employed in the optimization process to determine the variational parameters.
To enhance the efficiency of gradient-based optimization, we applied the local reparameterization trick to the convolutional layers. This approach preserves computational performance while enabling efficient updates of variational parameters. The optimization technique, commonly referred to as ``Bayes by Backprop" \cite{blundell2015weight}, facilitates the training of BayesCNN by leveraging this reparameterization framework.
While the mean-field approximation used in this work offers a computationally tractable solution for high-dimensional BayesCNNs, its assumption of independence among weights, as defined in \eqref{eq:meanfield}, limits its expressiveness by neglecting posterior correlations. This simplification is particularly restrictive in convolutional architectures, where weights often exhibit structured dependencies across spatial and channel dimensions. A promising alternative is the use of matrix variate Gaussian variational posteriors \cite{louizos2016structured}, which explicitly model covariance structures across input and output dimensions of each layer, capturing inter-parameter correlations more accurate uncertainty estimation. The applicability of this approach, along with a comparative evaluation of its predictive accuracy and computational cost relative to the mean-field approximation, will be investigated in future work within the context of microstructure–property surrogate modeling.

%++++++++++++++++++++++++++++++++++++++++++++++++++++++++++++++++++++++++
%++++++++++++++++++++++++++++++++++++++++++++++++++++++++++++++++++++++++
\section{Numerical results}\label{sec:results}
This section presents computational results from the integrated modeling framework (Figure \ref{fig:framework}), focusing on the development and evaluation of WGAN-GP and BayesCNN surrogate models. Their predictive accuracy is assessed against high-fidelity methods, including the Free-surface Lattice Boltzmann Method (FSLBM) and stochastic Finite Element (FE) simulations. The surrogate models are then used to propagate uncertainty through the microstructure-property relationships, analyzing the impact of microstructural randomness and surrogate model errors on the reliability of the predicted quantities of interest (QoI).
The dataset generation was performed on a single CPU, with each FSLBM simulation requiring approximately 10 minutes and each FE simulation taking around 35 minutes. Model training was conducted on GPU-enabled systems, with WGAN-GP and BayesCNN training times of approximately 8 hours and 6 hours, respectively.

%++++++++++++++++++++++++++++++++++++++++++++++++++++++++++++++++++++++++
\subsection{Generative surrogate model of microstructure}\label{sec:gans_results}
\noindent
We detail the construction of the WGAN-GP surrogate model and quantitatively evaluate its performance by comparing microstructure images and morphological features generated by the WGAN-GP to the ones from FSLBM simulations.

\begin{figure}[h!]
    \centering     
    \includegraphics[scale=0.4]{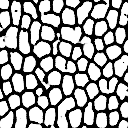}
    \includegraphics[scale=0.4]{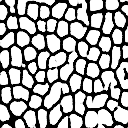}
    \includegraphics[scale=0.4]{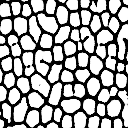}
    \includegraphics[scale=0.4]{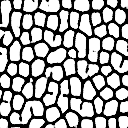}
    \includegraphics[scale=0.4]{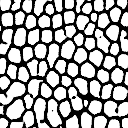}
    \\\vspace{0.05in}    
    \includegraphics[scale=0.4]{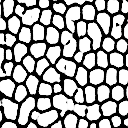}
    \includegraphics[scale=0.4]{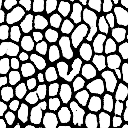}
    \includegraphics[scale=0.4]{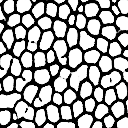}
    \includegraphics[scale=0.4]{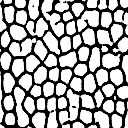}
    \includegraphics[scale=0.4]{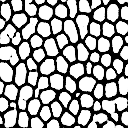}
    \\ (a) \\
    \includegraphics[scale=0.4]{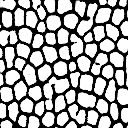}
    \includegraphics[scale=0.4]{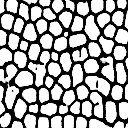}
    \includegraphics[scale=0.4]{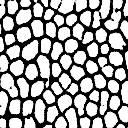}
    \includegraphics[scale=0.4]{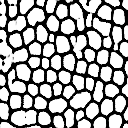}
    \includegraphics[scale=0.4]{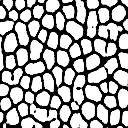}
    \\\vspace{0.05in}
    \includegraphics[scale=0.4]{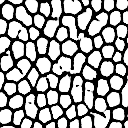}
    \includegraphics[scale=0.4]{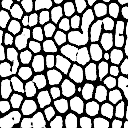}
    \includegraphics[scale=0.4]{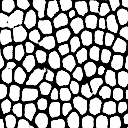}
    \includegraphics[scale=0.4]{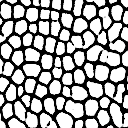}
    \includegraphics[scale=0.4]{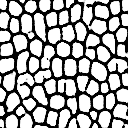}
    \\ (b) \\
    \caption{
    Representative microstructures of aerogels with an average pore radius of 6$\mu m$, demonstrating that the trained WGAN-GP model can successfully generate the morphological characteristics of microstructures obtained from FSLBM simulations.
    (a) Training images produced by the high-fidelity FSLBM simulations of the foaming process, denoted by $\mathcal{F}_{LB}(\bs \xi; \bs \theta_{LB})$.
    (b) Microstructure images generated by the WGAN-GP generator network following training $\mathcal{G}_{GAN}(\bs \xi, \bs z ; \bs w_{GAN})$.
    }
    \label{fig:gans_training}
    \vspace{-0.2in}
\end{figure}
%------
\paragraph{Networks architecture and training}
The WGAN-GP generator network consists of four layers, each with a Convolutional Transpose, Batch Normalization, and ReLU activation, followed by a final layer with a Convolutional Transpose and Tanh activation to output the generated microstructure image. A latent vector $\bs z$ of size 4 $\times$ 1 is progressively upscaled by a factor of 2 in each subsequent layer to produce the 128 $\times$ 128 $\times$ 1 microstructure. The discriminator network mirrors this structure, using Convolutional, Batch Normalization, and Leaky ReLU layers, with a final Convolutional layer producing a scalar output representing the Wasserstein distance between synthetic and generated image distributions. Both networks employ a 4 $\times$ 4 kernel, 2 $\times$ 2 stride, and 1 $\times$ 1 padding, achieving a scaling factor of $2^5$ (=32) over five layers. The architectures, summarized in Table \ref{tab:gan_arch} (see Appendix) , are adapted from previous GAN-based material microstructure models \cite{chun2020deep, CHEN2024112976}, with modifications tailored to this study.
The training dataset comprises 128 $\times$ 128 pixel microstructure images created using FSLBM simulations, $\mathcal{F}_{LB}(\bs \xi; \bs \theta_{BL})$. 
The WGAN-GP was trained on 2,500 synthetic images from FSLBM simulations (Figure \ref{fig:lbfoam_evol}(a)) with an average pore size of $6 \mu m$. Training employed the Adam optimizer with a learning rate of 0.0002, a batch size of 10, and a discriminator penalty coefficient $c_{\lambda} = 10$, over 2000 epochs.

\begin{figure}[h!]
    \centering
    \includegraphics[trim = 0mm 0mm 0mm 0mm, clip, width=0.48\textwidth]{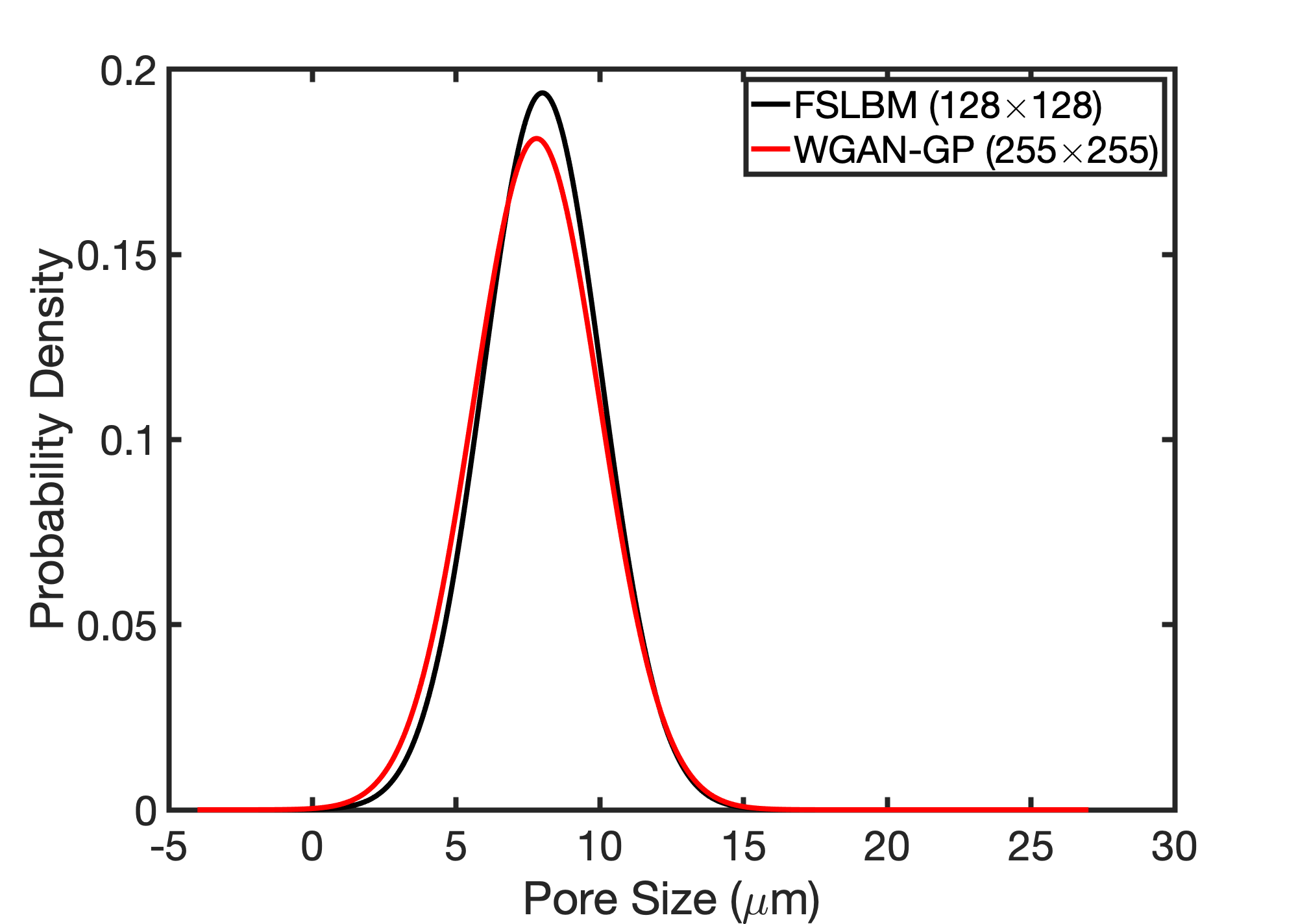}
    ~
    \includegraphics[trim = 0mm 0mm 0mm 0mm, clip,width=0.48\textwidth]{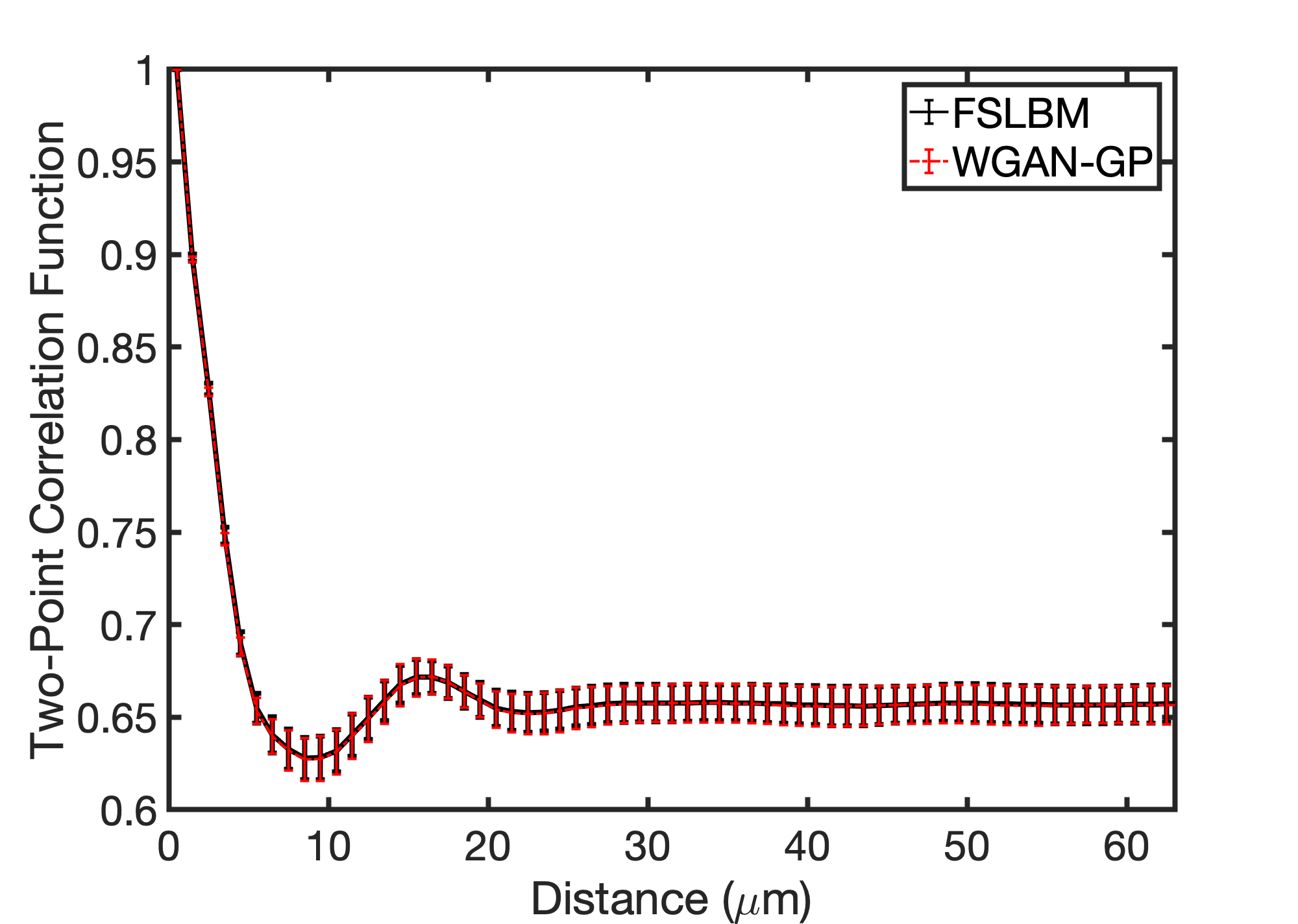}
    \\\hspace{0.1in}(a) \hspace{2.5in} (b)
    \caption{
    Quantitative evaluation of the WGAN-GP generated microstructures against microstructures obtained from FSLBM simulations.
    Morphological features are compared across 30 microstructure samples from each model:
    (a) Probability distributions of pore diameters, and
    (b) one-dimensional two-point correlation functions.
    }
    \label{fig:kde_corr}
\end{figure}

%----------
\paragraph{Accuracy of the generated microstructure images}
Figure \ref{fig:gans_training} compares microstructure images of aerogels generated by the WGAN-GP surrogate model with those obtained from high-fidelity FSLBM simulations.
To quantitatively assess the accuracy of the WGAN-GP and ensure it does not overfit or generate spurious artifacts, morphological metrics including pore size distributions and two-point correlation functions are computed and compared between the generated and reference microstructures. Figure \ref{fig:kde_corr}(a) shows the Kernel Density Estimate (KDE) of pore diameters for 30 samples, revealing close agreement between data and generated microstructures. Pore diameters were calculated using an image segmentation method that labels individual pores, determines their areas (based on pixel counts), and computes the equivalent diameter of a circle with the same area. Further details on this method are provided in \cite{bhattacharjee2023integrating}.
The two-point correlation function, $S_2(\boldsymbol{r})$, is a statistical measure characterizing the morphology of two-phase media \cite{bostanabad2018computational}, representing the probability that two points separated by $\lvert \boldsymbol{r} \rvert$ belong to the same phase. Here, we focus on the pore phase (pixels with a value of 255) to quantify pore distribution within a given distance $\boldsymbol{r}$. Figure \ref{fig:kde_corr}(b) shows 1D plots of $S_2(\boldsymbol{r})$ for 30 synthetic and generated samples, with error bars capturing sample variability. The close agreement demonstrates the WGAN-GP's ability to reproduce microstructural features consistent with FSLBM simulations.

%----
\paragraph{Scalable microstructure generator}
Despite the widespread application of GANs in modeling material microstructures, their utility is fundamentally constrained by the fixed size of training images. However, material properties of practical interest are often required at the component level, involving domains several orders of magnitude larger than those captured by microscale imaging or modeling techniques.
To overcome this limitation, we propose a novel approach that enables the generation of arbitrary-sized microstructure images using WGAN-GP trained on smaller image datasets. This method utilizes a latent vector as input to the generator, trained in a low-dimensional latent space, allowing efficient generation of microstructures at various scales while maintaining control over the output dimensions.

\begin{figure}[h]
    \centering
    \includegraphics[trim = 170mm 10mm 170mm 10mm, clip, scale=0.12]{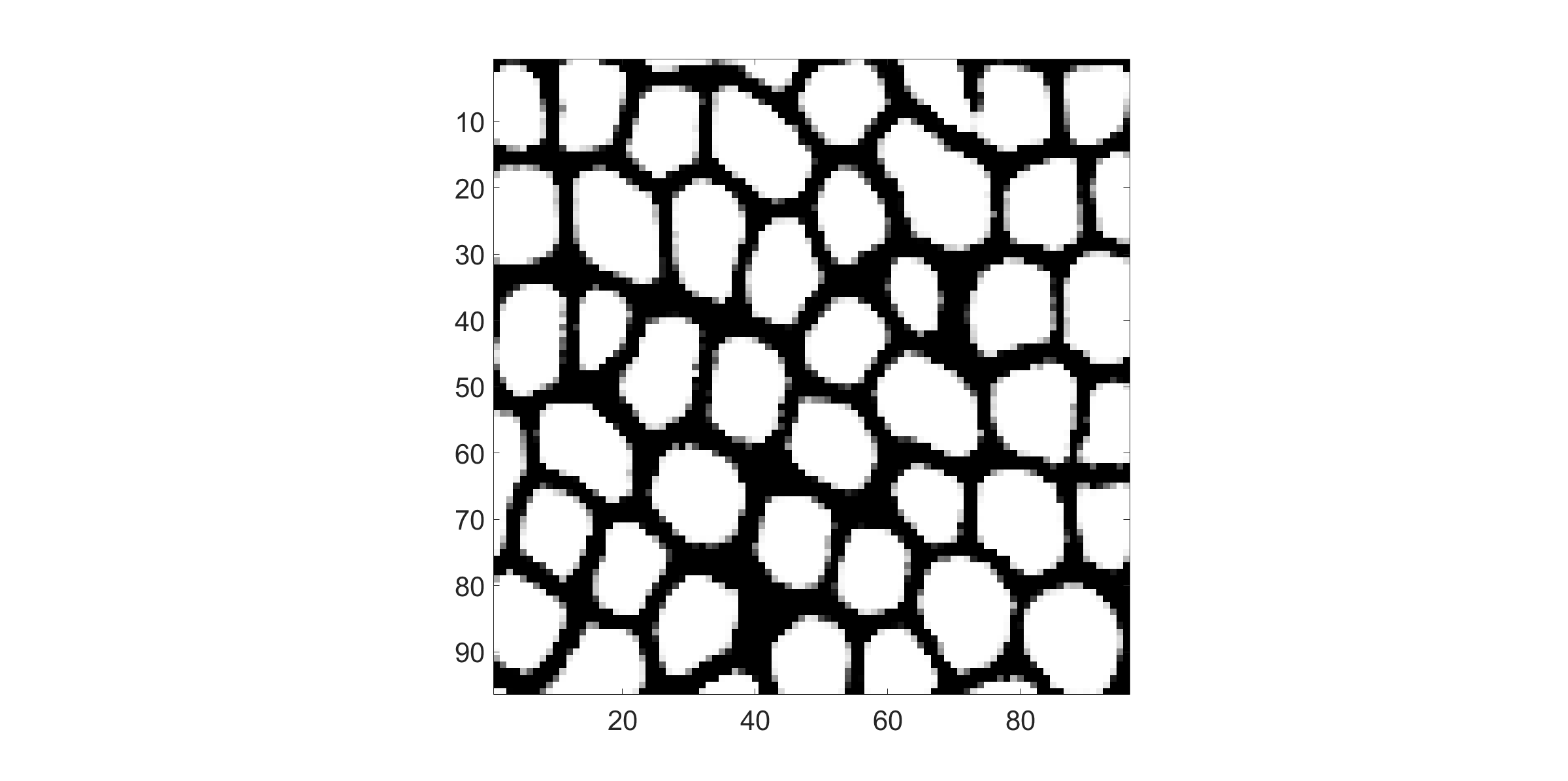}
    \includegraphics[trim = 170mm 10mm 170mm 10mm, clip, scale=0.12]{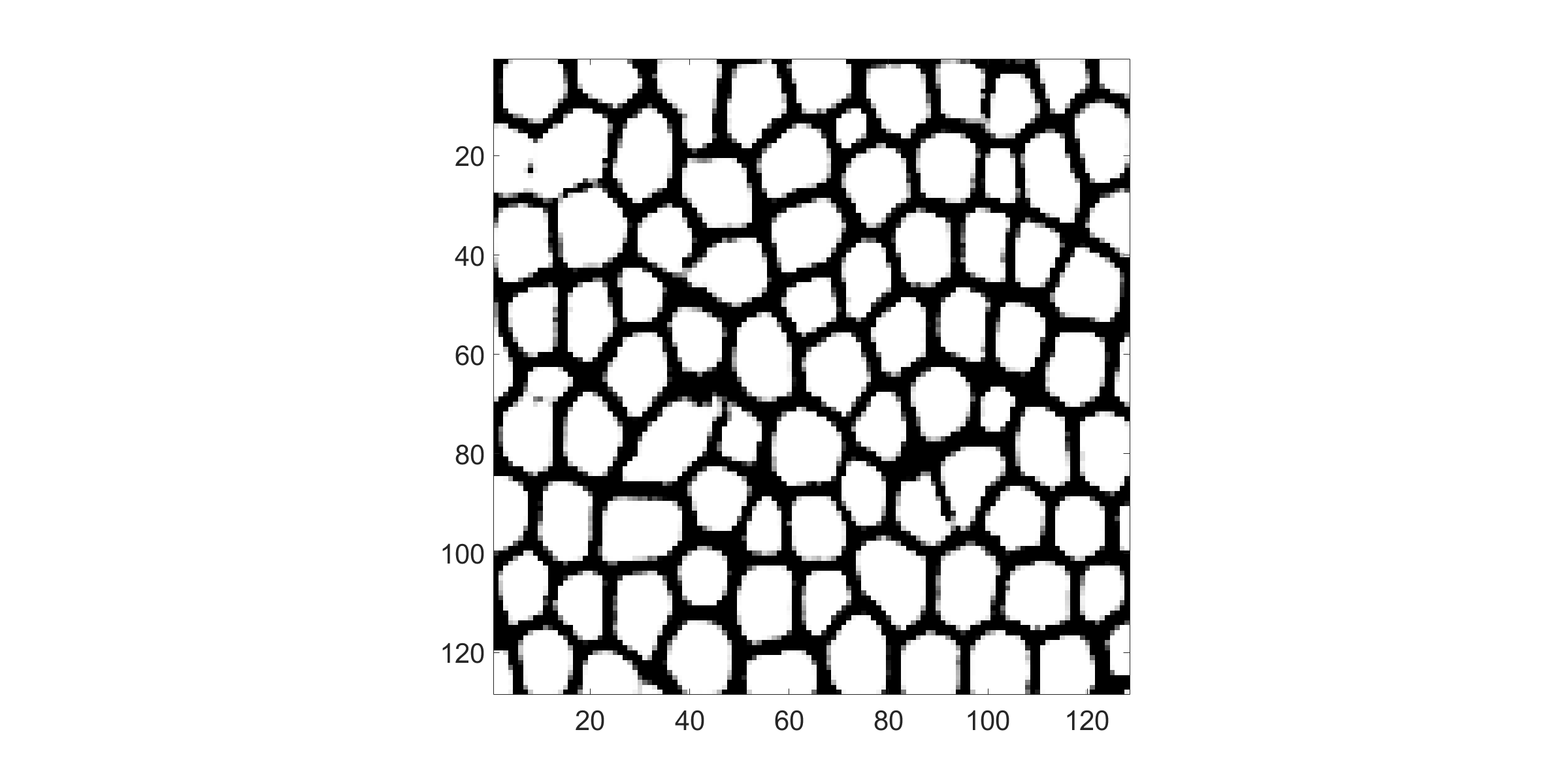}
    \includegraphics[trim = 170mm 10mm 170mm 10mm, clip, scale=0.12]{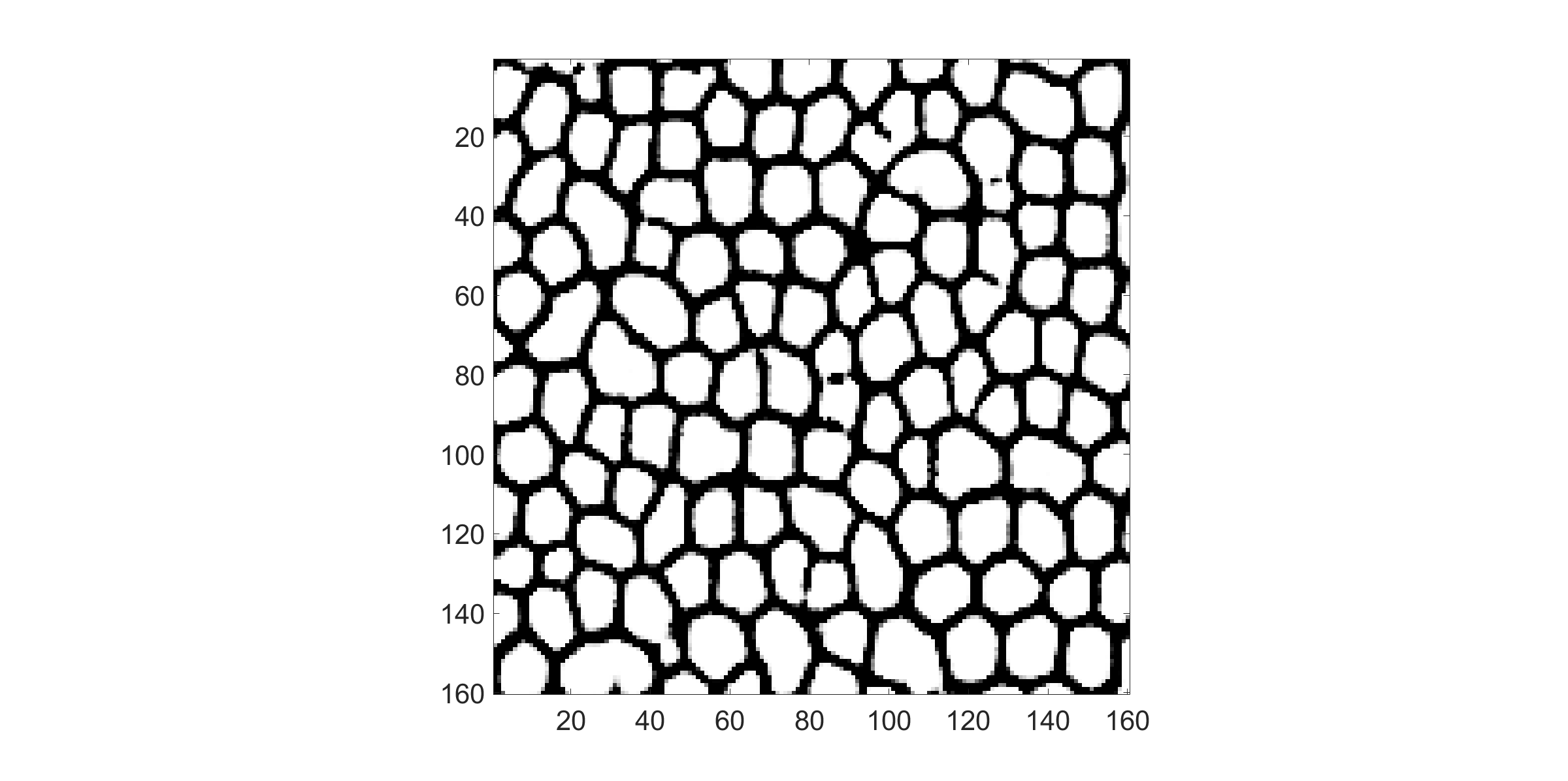}
    \\\hspace{0in} (a) 96$\times$96 \hspace{0.5in} (b) 128$\times$128 \hspace{0.5in} (c) 160$\times$160\\
    \includegraphics[trim = 170mm 10mm 170mm 10mm, clip, scale=0.12]{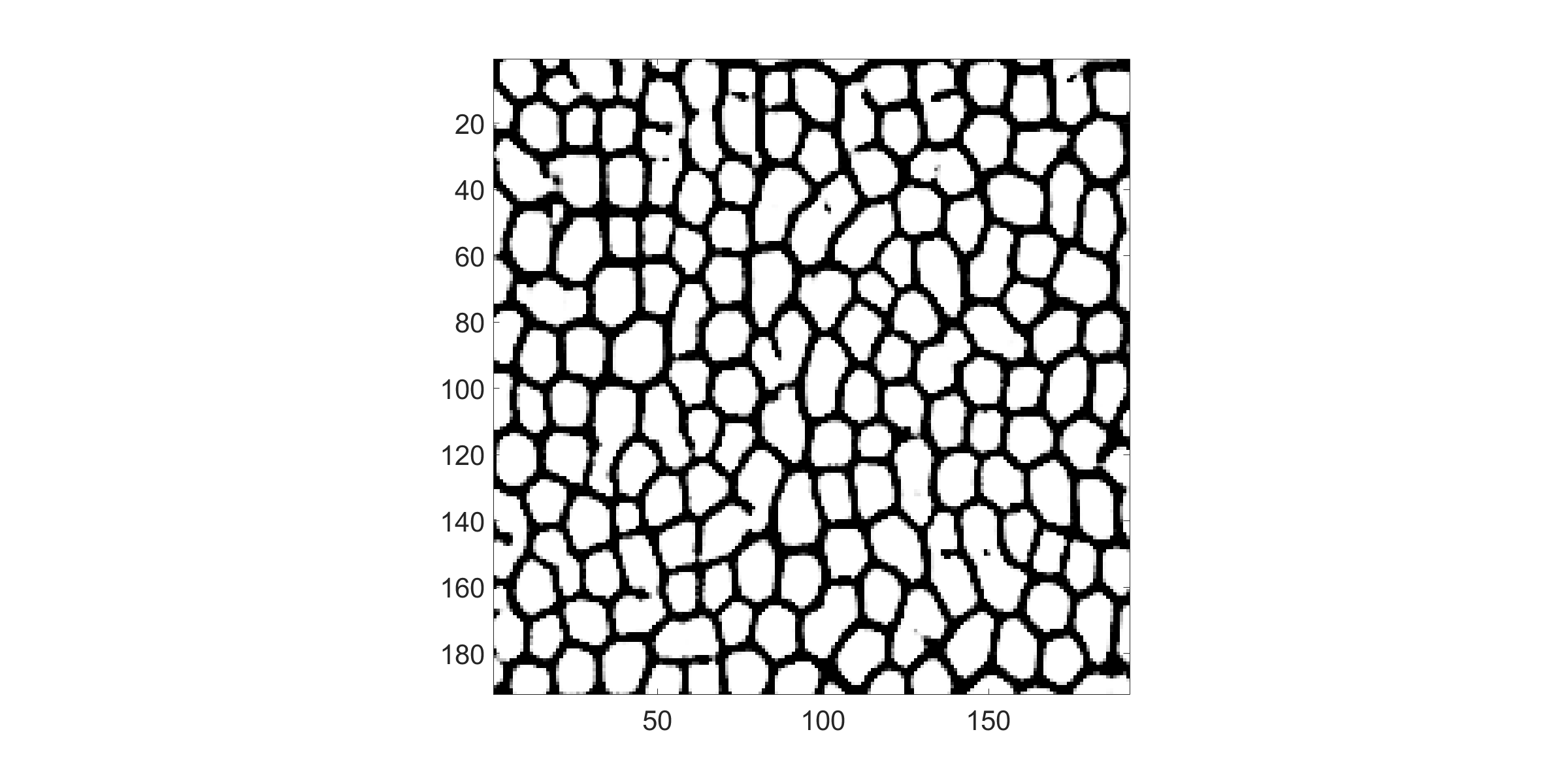}
    \includegraphics[trim = 170mm 10mm 170mm 10mm, clip, scale=0.12]{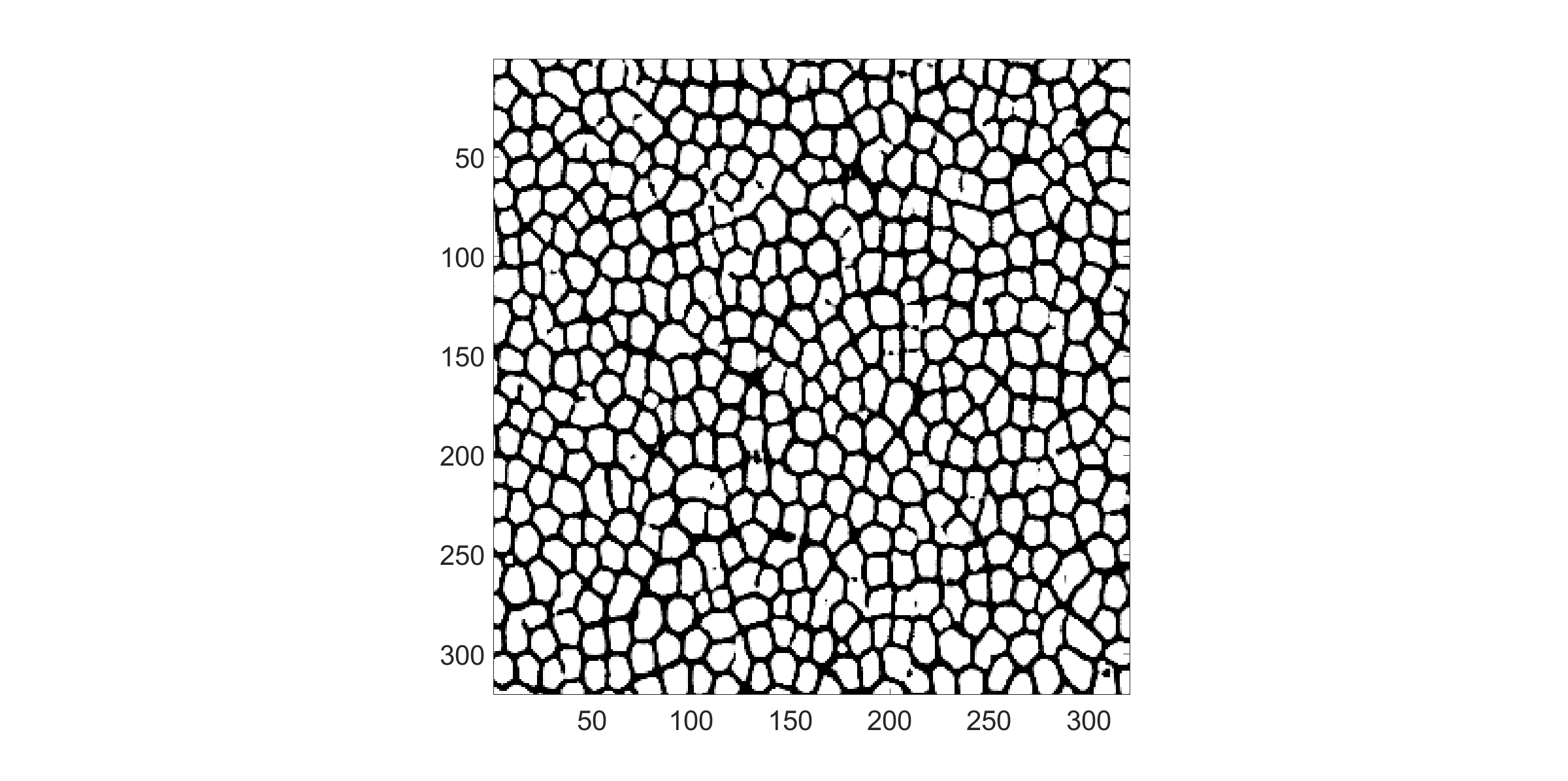}
    \includegraphics[trim = 170mm 10mm 170mm 10mm, clip, scale=0.12]{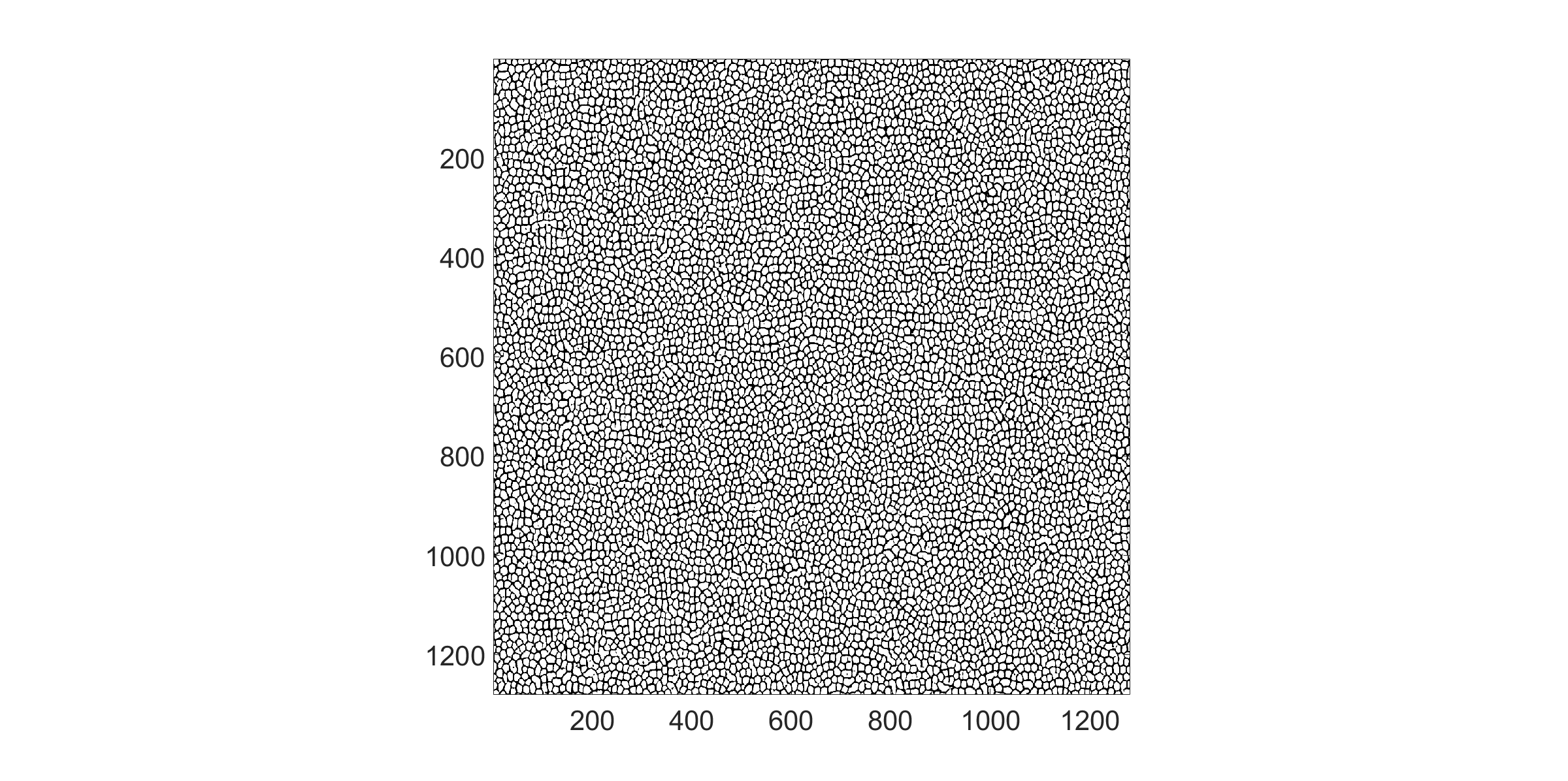}
     \\\hspace{0in} (d) 192$\times$192 \hspace{0.5in} (e) 384$\times$384 \hspace{0.5in} (f) 1280$\times$1280\\
    \caption{
    Demonstration of the scalability of the WGAN-GP generator network, trained on microstructure images with a domain size of 70 $\mu m$, in synthesizing realistic microstructures across a range of domain sizes. The generator produces images with varying pixel dimensions corresponding to ceramic aerogel domains of:
    (a) 52.5 $\mu m$,
    (b) 70 $\mu m$,
    (c) 87.5 $\mu m$,
    (d) 105 $\mu m$,
    (e) 210 $\mu m$, and
    (f) 700 $\mu m$.
    }
    \label{fig:scaled}
\end{figure}
\begin{figure}[h!]
    \centering
   \includegraphics[trim = 0mm 0mm 0mm 0mm, clip, width=0.48\textwidth]{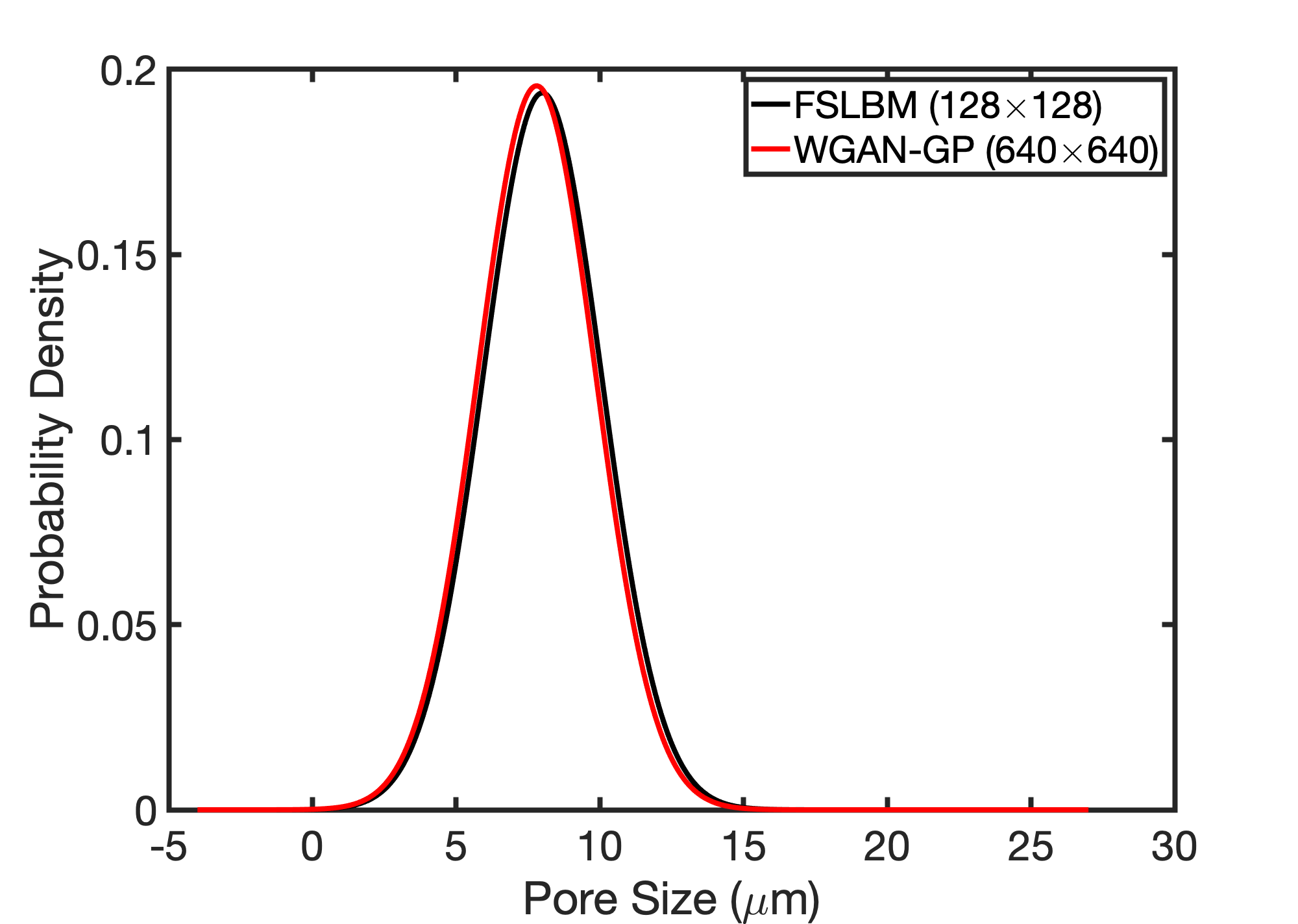}
    \hspace{0in}
    \includegraphics[trim = 0mm 0mm 0mm 0mm, clip, width=0.48\textwidth]{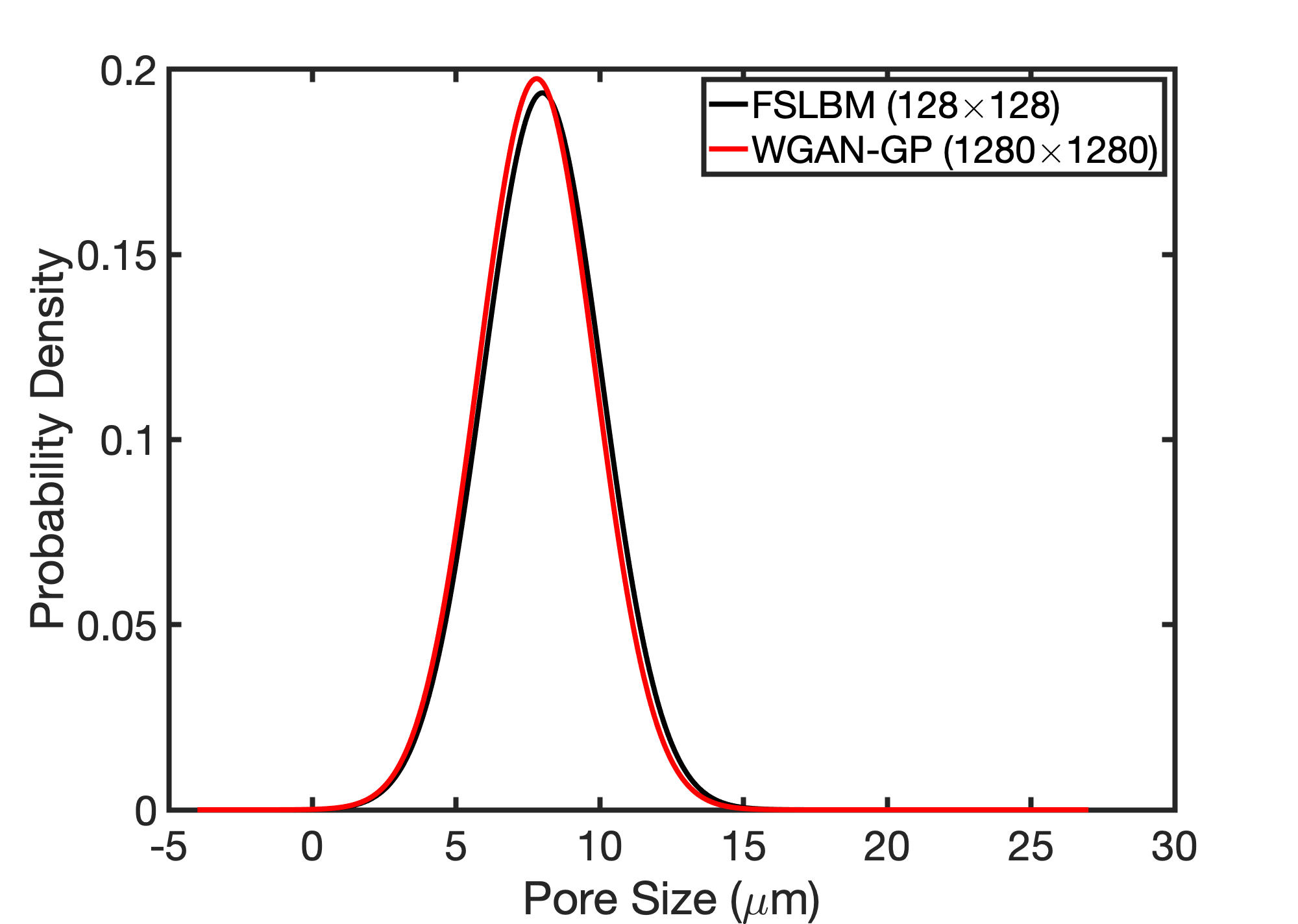}
    \\\hspace{0.1in}(a) \hspace{2.5in} (b)
    \caption{
    \blue{Comparison of pore diameter probability distributions between the 128$\times$128 microstructure training data obtained from FSLBM simulations and microstructures generated by the scalable WGAN-GP generator network (refer to Figure \ref{fig:scaled}) at larger domain sizes: (a) 640$\times$640 pixels (corresponding to 350$\mu m$), (b) 1280$\times$1280 pixels (corresponding to 700$\mu m$).
    }
    }
    \label{fig:kde_scale}
    \vspace{-0.2in}
\end{figure}

As described, the generator architecture, trained with a latent vector $\bs z$ of size 4 $\times$ 1, provides an upscaling factor of 2 per layer. However, increasing the image dimensions requires adding more convolutional layers, which also necessitates corresponding layers in the discriminator. This increases network complexity, training time, and computational cost.
To address these issues, we reformulate the generator architecture by separating it from the fully connected layers and using linear transformations to reshape the latent vector $\bs z$. Specifically, $\bs z$ of dimension 4 $\times$ 1 is mapped to a latent variable of dimension (-1, hidden\textunderscore dim, 4, 4) where hidden\textunderscore dim is the hidden dimension size, and -1 denotes the arbitrary batch size. This approach provides more flexibility over the size of the latent vector we input to the generator, and this also results in the reduction of the total number of layers. 
Importantly, this method preserves the microstructural features while allowing for the efficient generation of images at arbitrary dimensions.
Figure \ref{fig:scaled} shows microstructures of various sizes generated by the scalable WGAN-GP, trained on 128$\times$128 pixel synthetic microstructures (corresponding to a 70 $\mu m$ aerogel domain). The output size is controlled by the dimensions of the latent variable, providing flexibility. Figure \ref{fig:kde_scale} compares the KDEs of pore sizes from 30 synthetic microstructures and larger generated images. The agreement highlights the scalable generator's ability to produce larger microstructure images while preserving key statistical features across different sizes.

%++++++++++++++++++++++++++++++++++++++++++++++++++++++++++++++++++++++++
\subsection{BayesCNN surrogate model of mechanical properties}\label{sec:bcnn_results}
\noindent
This section describes the construction of the BayesCNN surrogate model, using high-fidelity simulations from the stochastic FE model, to predict the mapping between microstructure images and the strain energy of ceramic aerogels (QoI). The surrogate model’s performance with quantified uncertainty, is validated for both \textit{in-distribution} and \textit{out-of-distribution} predictions. \textit{In-distribution} refers to microstructures with pore sizes and morphology similar to the training data, while \textit{out-of-distribution} includes microstructures with characteristics that interpolate between the training data.

\begin{figure}[h!]
    % First column
        \centering
        \includegraphics[trim = 0mm 0mm 0mm 0mm, clip, width=0.48\textwidth]{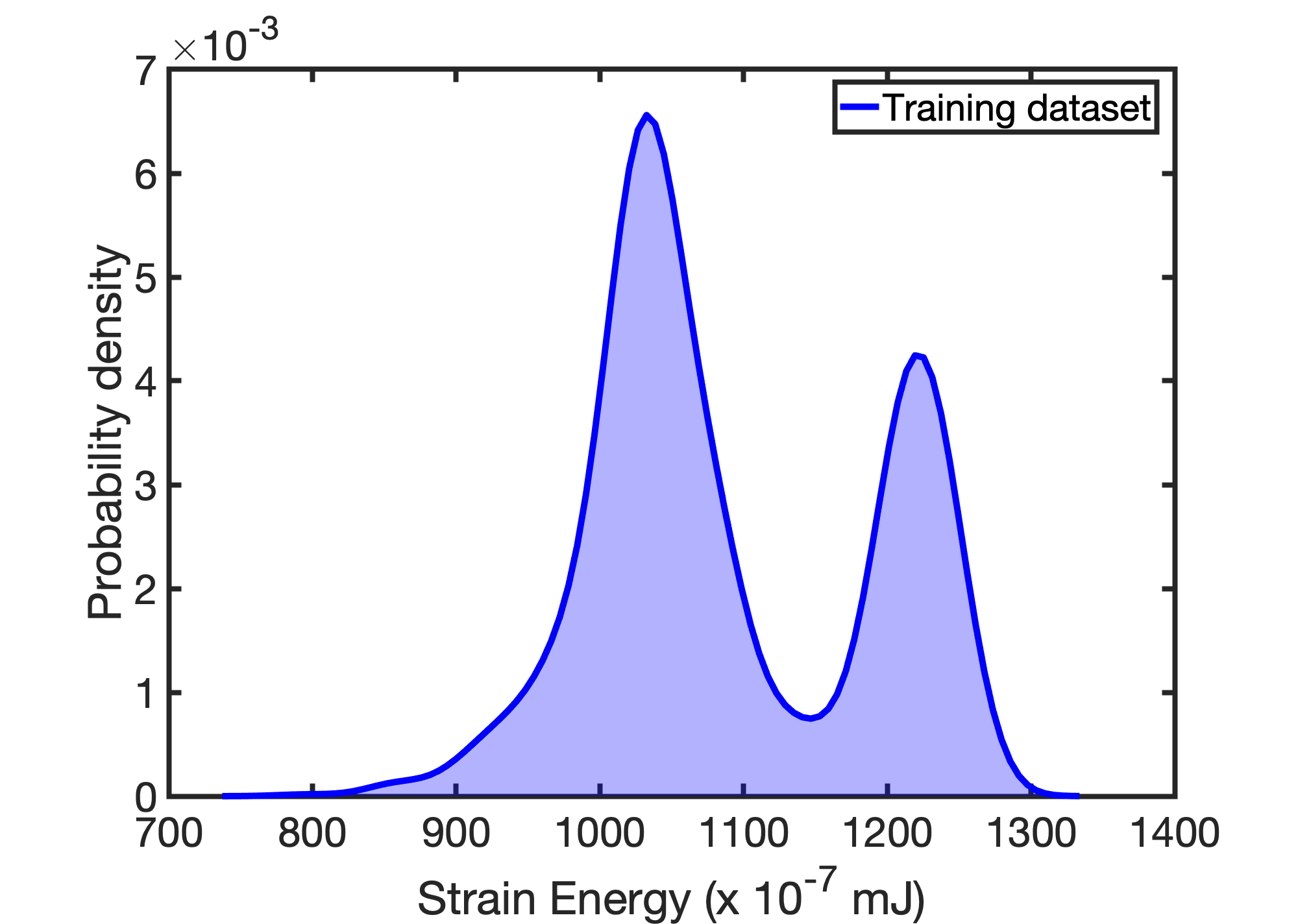}\\
        (a)\\
    \hfill
    \begin{minipage}[t]{0.3\textwidth}
        \centering
        \includegraphics[scale=0.15]{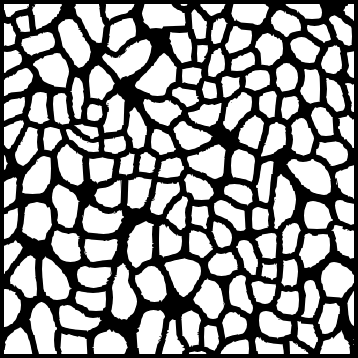}
        \includegraphics[scale=0.15]{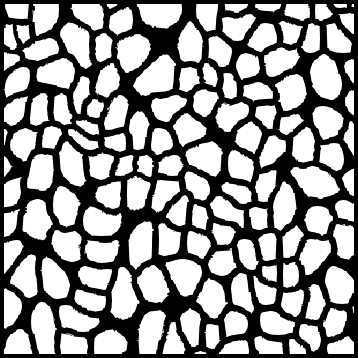}
        \includegraphics[scale=0.15]{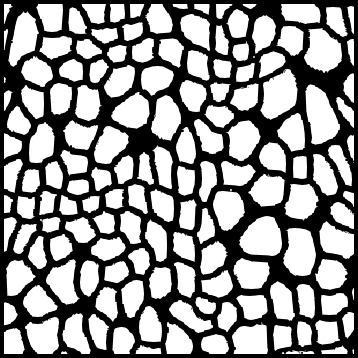}
        \includegraphics[scale=0.15]{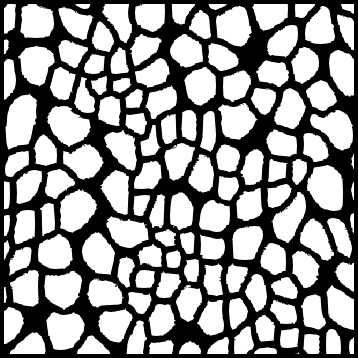} \\
        (b)
    \end{minipage}
    \hfill
    % Second column
    \begin{minipage}[t]{0.3\textwidth}
        \centering
        \includegraphics[scale=0.15]{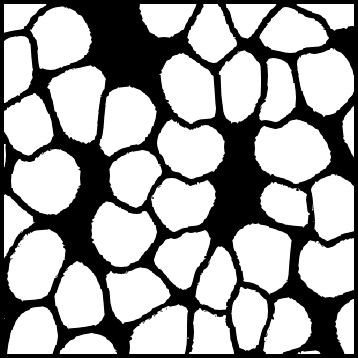}
        \includegraphics[scale=0.15]{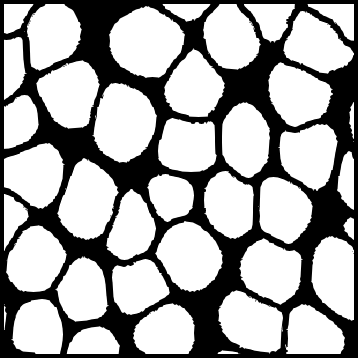}
        \includegraphics[scale=0.15]{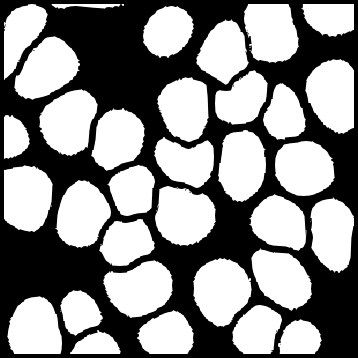}
        \includegraphics[scale=0.15]{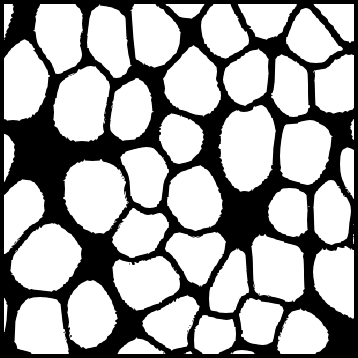} \\
       (c)
    \end{minipage}
    \hfill
    \begin{minipage}[t]{0.3\textwidth}
        \centering
        \includegraphics[scale=0.15]{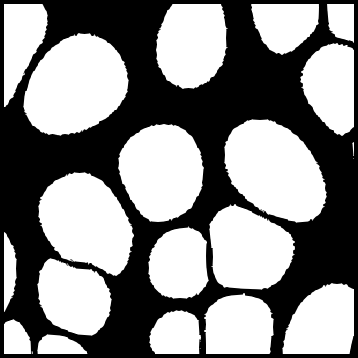}
        \includegraphics[scale=0.15]{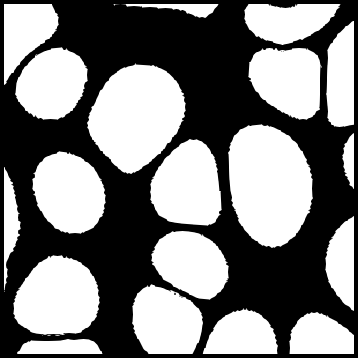}
        \includegraphics[scale=0.15]{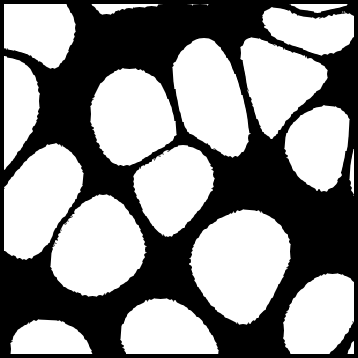}
        \includegraphics[scale=0.15]{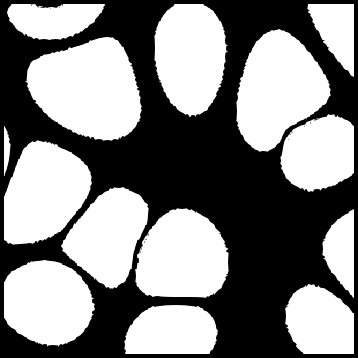} \\
        (d)
    \end{minipage}
    \hfill
    \vspace{-0.1in}
    \caption{
    Training data of microstructure-strain energy relationship to construct the BayesCNN surrogate model:
    (a) Probability distribution of strain energy values $Q^{FE}$ computed using the stochastic FE model $\mathcal{F}_{FE}(\varphi, \bs\theta_{FE})$.
    Representative microstructure samples from three microstructural classes used for training:
    (b) \textit{small pores} with average pore sizes of $6 \mu m$;
    (c) \textit{medium pores} with average pore sizes of $18 \mu m$;
    (d) \textit{large pores} with average pore sizes of $25 \mu m$.
    }
    \label{fig:training_microstructure}
    \vspace{-0.2in}
\end{figure}

%-------------
\paragraph{High-fidelity data}
Three categories of ceramic aerogel microstructures: \textit{small pores}, \textit{medium pores}, and \textit{large pores}, with average pore sizes of $6 \mu m$, $18 \mu m$, and $25 \mu m$, respectively, is taken into account (Figure \ref{fig:training_microstructure}). A total of 1,188 samples (396 per category) were simulated using the stochastic FE model
$\mathcal{F}_{FE}(\varphi, \bs\theta_{FE})$ to compute the strain energy $Q^{FE}$. 
For all samples, uniaxial loading is applied by imposing a downward traction of ${t}_y = -10^6 N$ on the top surface, while the displacements at the bottom boundary are fully constrained (see Figure \ref{fig:finite_el}).
The resulting microstructure-strain energy data was randomly divided into 90\% training dataset and 10\% in-distribution dataset.
\begin{figure}[h!]
\centering
    \includegraphics[trim = 0mm 0mm 0mm 0mm, clip, width=0.19\textwidth]{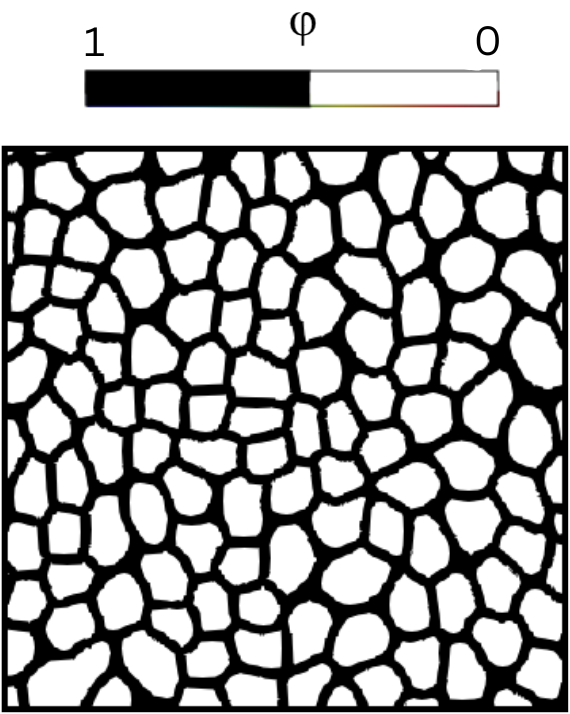}
    \includegraphics[trim = 0mm 0mm 0mm 0mm, clip, width=0.19\textwidth]{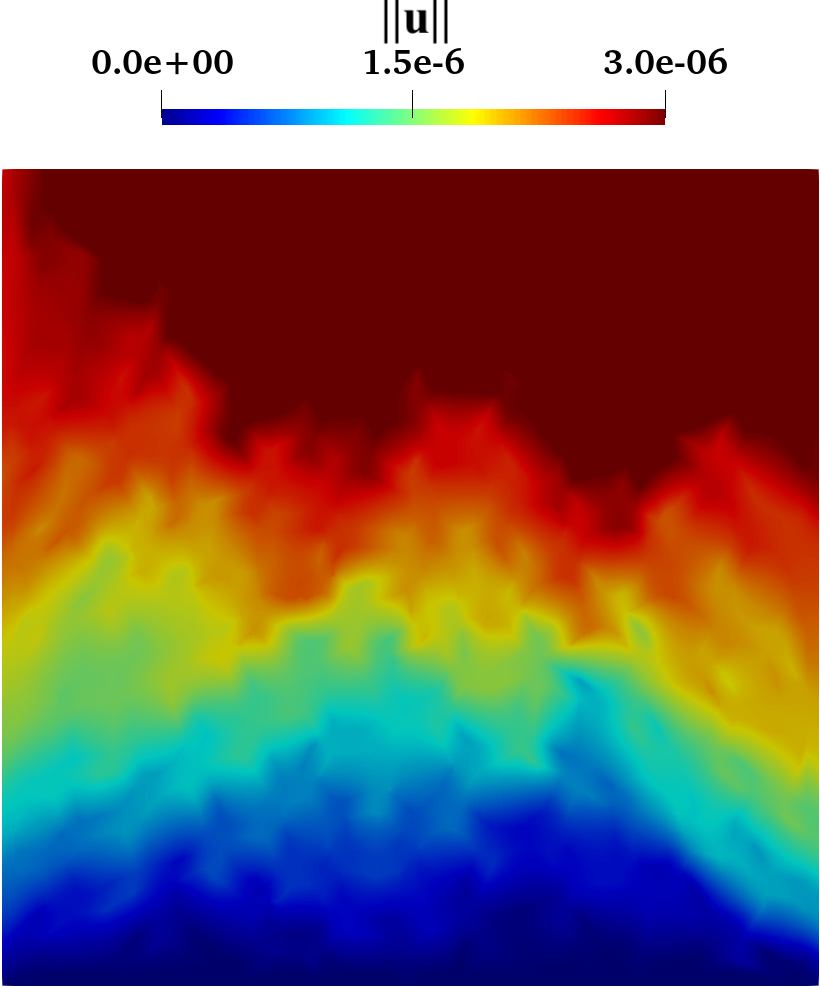} 
    \includegraphics[trim = 0mm 0mm 0mm 0mm, clip, width=0.19\textwidth]{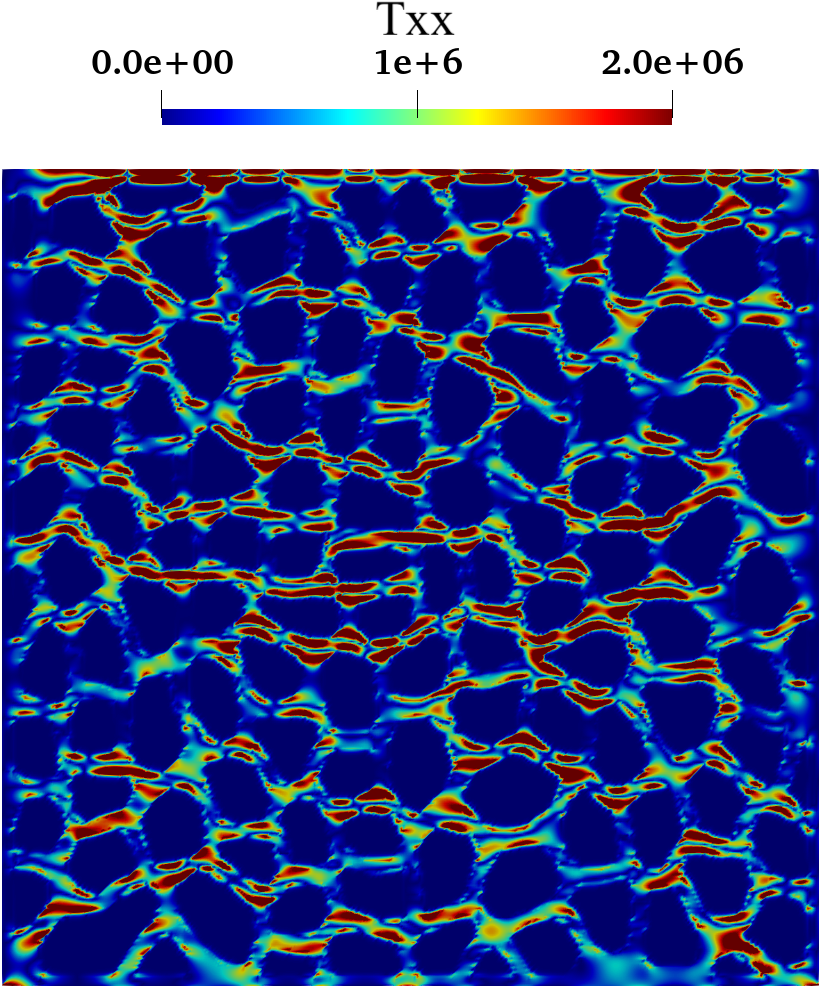} 
    \includegraphics[trim = 0mm 0mm 0mm 0mm, clip, width=0.19\textwidth]{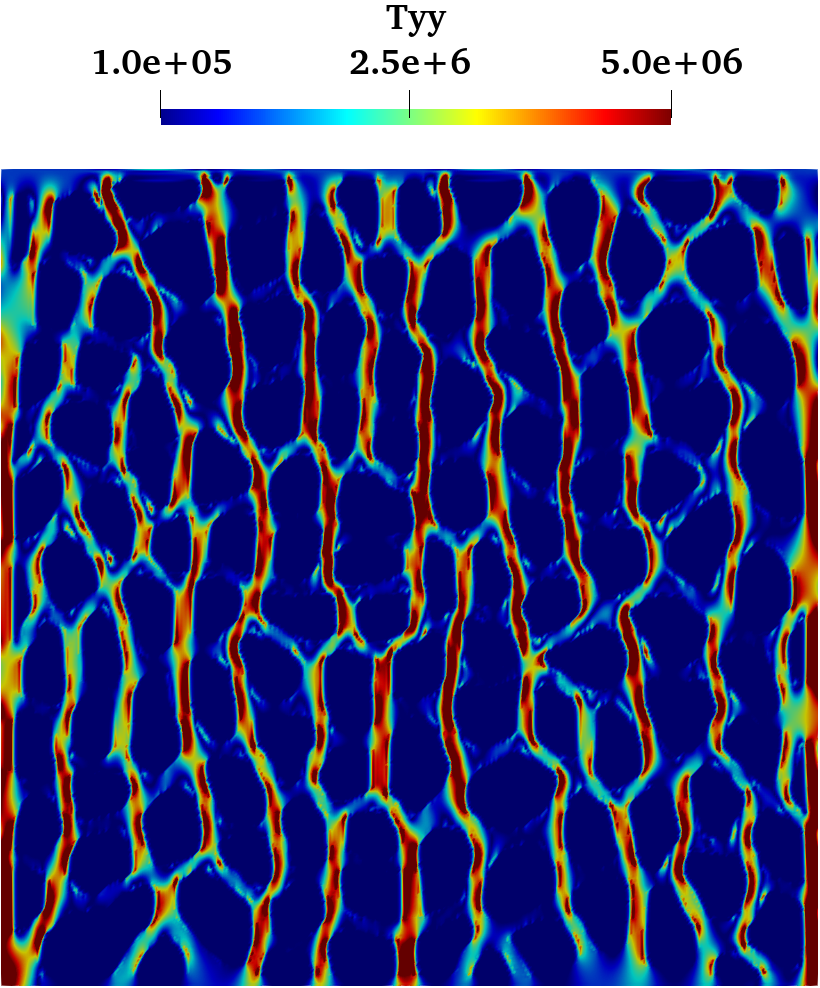} 
    \includegraphics[trim = 0mm 0mm 0mm 0mm, clip, width=0.19\textwidth]{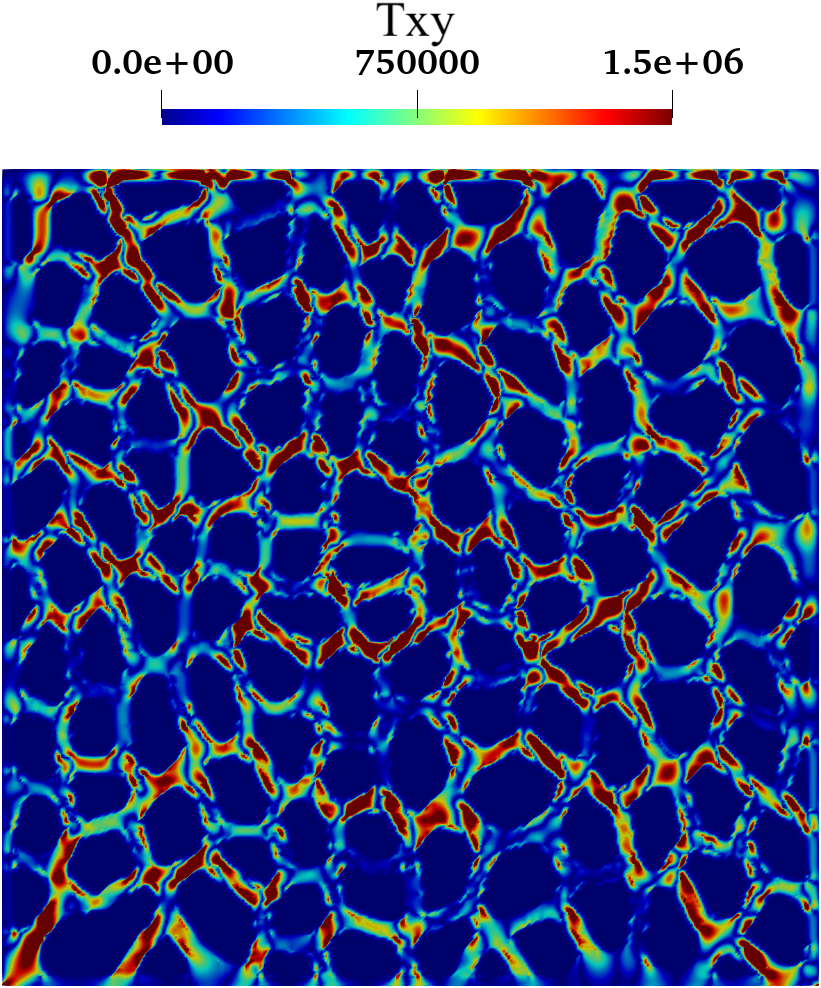}
        \\      (a)     \\
    \includegraphics[trim = 0mm 0mm 0mm 0mm, clip, width=0.19\textwidth]{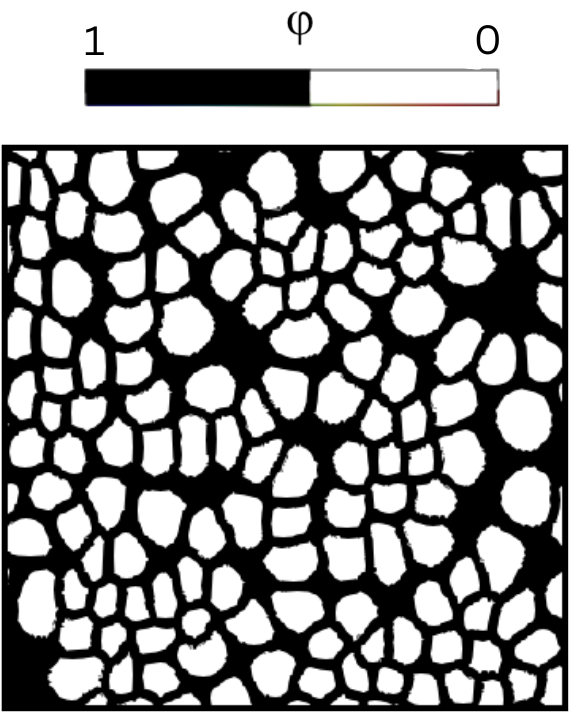}
    \includegraphics[trim = 0mm 0mm 0mm 0mm, clip, width=0.19\textwidth]{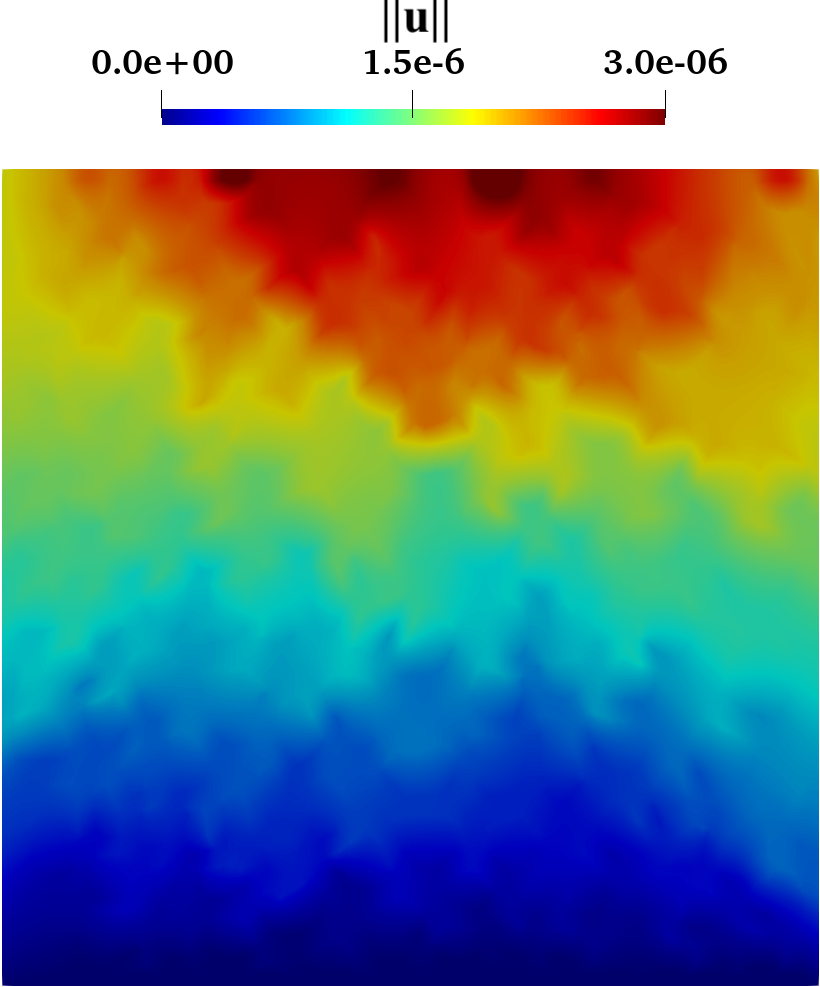} 
    \includegraphics[trim = 0mm 0mm 0mm 0mm, clip, width=0.19\textwidth]{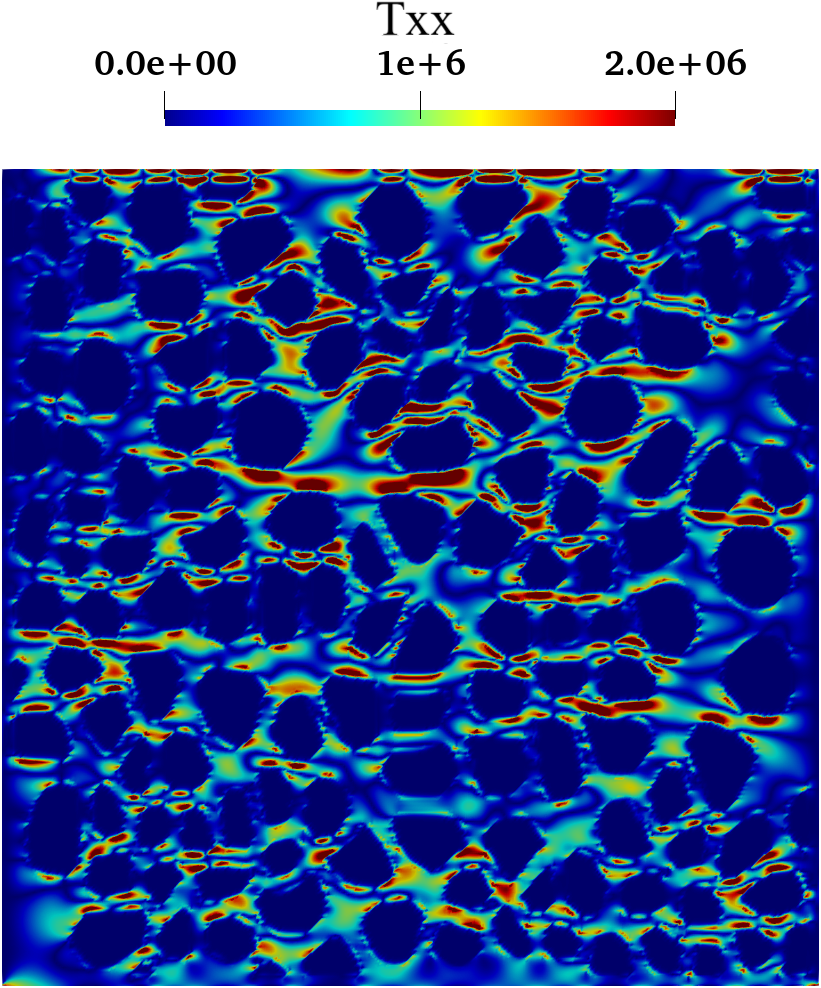} 
    \includegraphics[trim = 0mm 0mm 0mm 0mm, clip, width=0.19\textwidth]{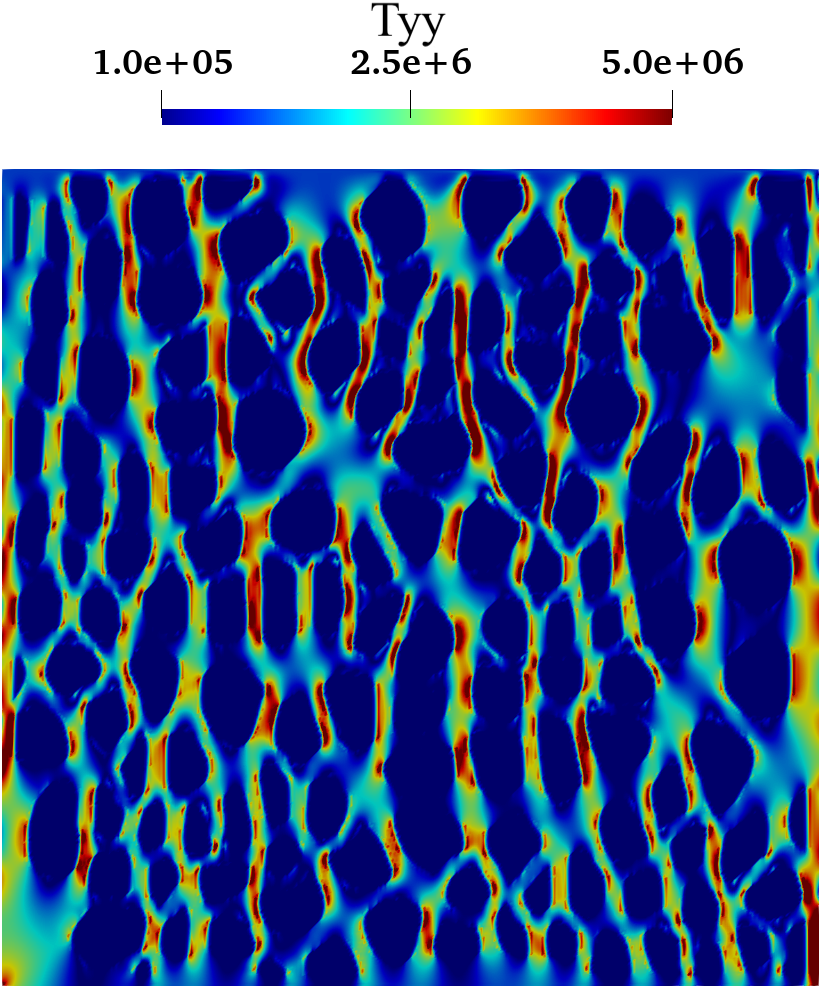} 
    \includegraphics[trim = 0mm 0mm 0mm 0mm, clip, width=0.19\textwidth]{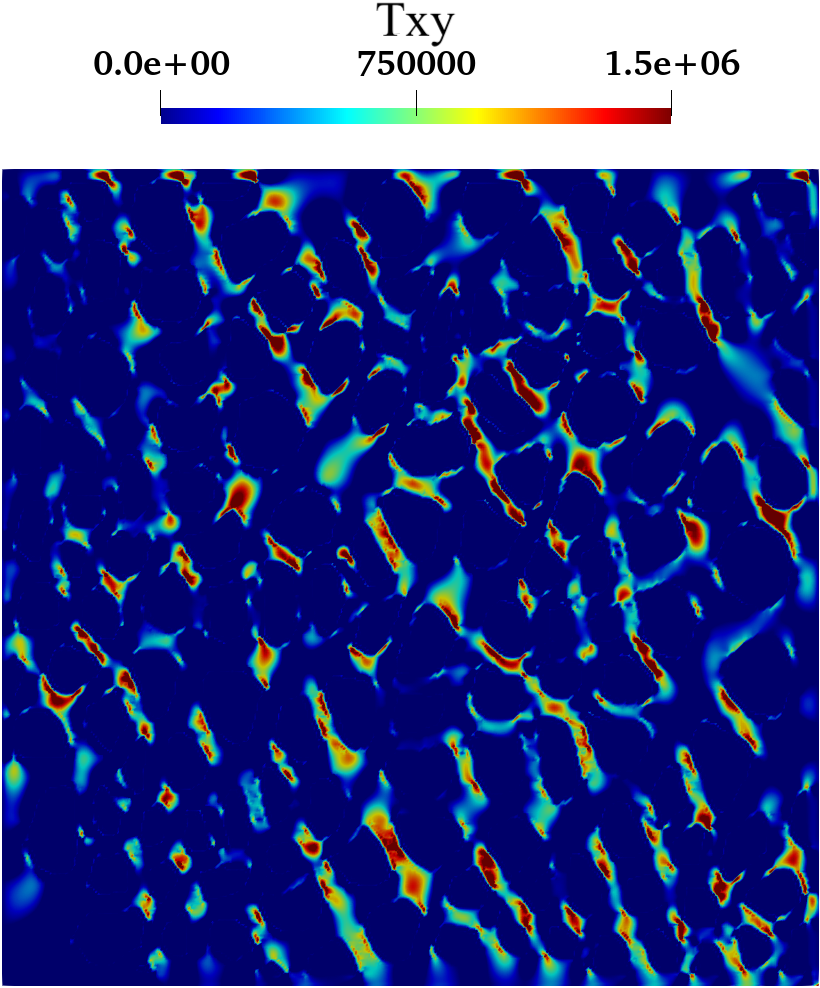}
        \\      (b)     \\
        \vspace{-0.1in}
\caption{
Finite element simulation results for \textit{small pores} microstructures, showing the microstructure indicator function $\phi(\mathbf{x})$, the magnitude of displacement $\|\mathbf{u}\|$ (in mm), and components of Cauchy stress $\mathbf{T}$ (in mPa):
Two examples of microstructures
(a) A representative microstructure (typical of 98\% of samples) with a strain energy of $1098 \times 10^{-7} \, \text{mJ}$; 
(b) An anomalous sample exhibiting a lower strain energy of $919 \times 10^{-7} \text{mJ}$, attributed to isolated solid-phase regions that disrupt continuous load-bearing pathways between the top and bottom boundaries.
}
        \label{fig:fe_small}
        \vspace{-0.1in}
\end{figure}
\begin{figure}[h!]
\centering
    \includegraphics[trim = 0mm 0mm 0mm 0mm, clip, width=0.19\textwidth]{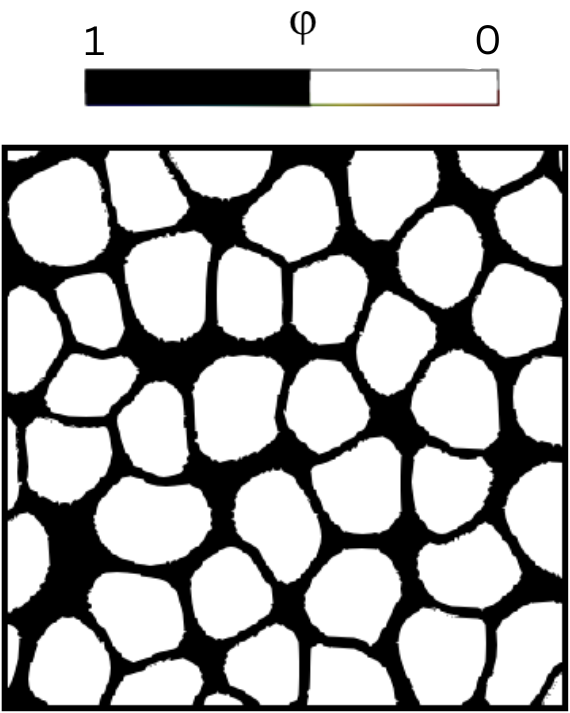}
    \includegraphics[trim = 0mm 0mm 0mm 0mm, clip, width=0.19\textwidth]{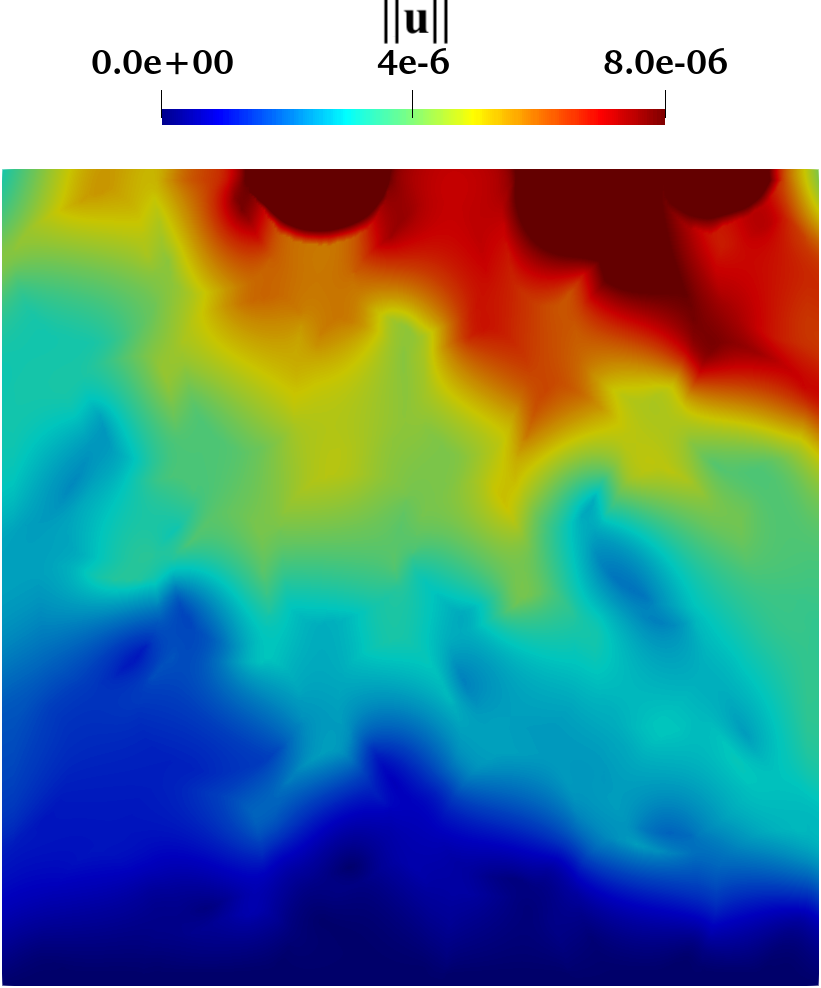} 
    \includegraphics[trim = 0mm 0mm 0mm 0mm, clip, width=0.19\textwidth]{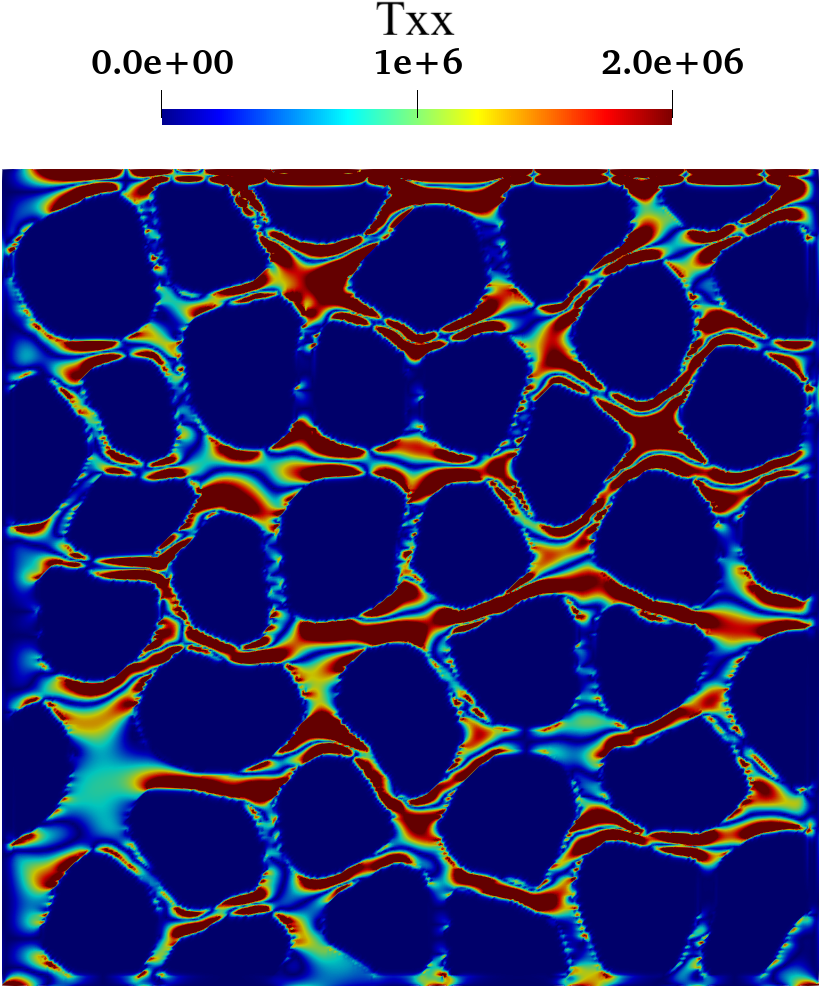} 
    \includegraphics[trim = 0mm 0mm 0mm 0mm, clip, width=0.19\textwidth]{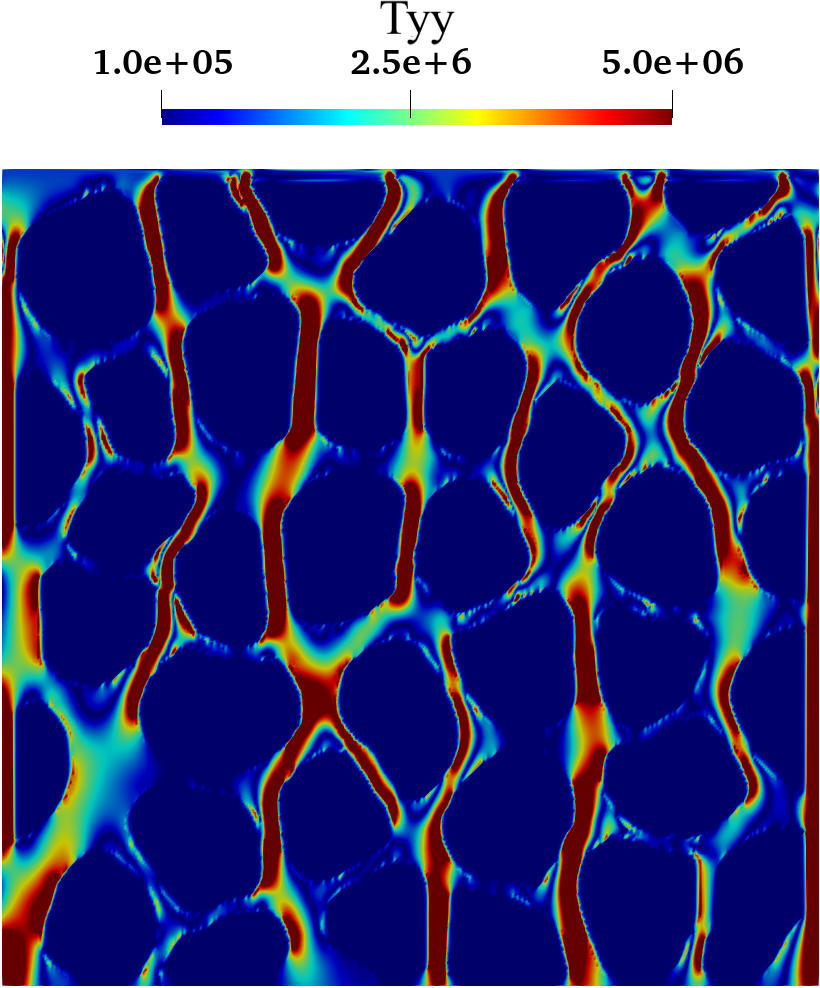} 
    \includegraphics[trim = 0mm 0mm 0mm 0mm, clip, width=0.19\textwidth]{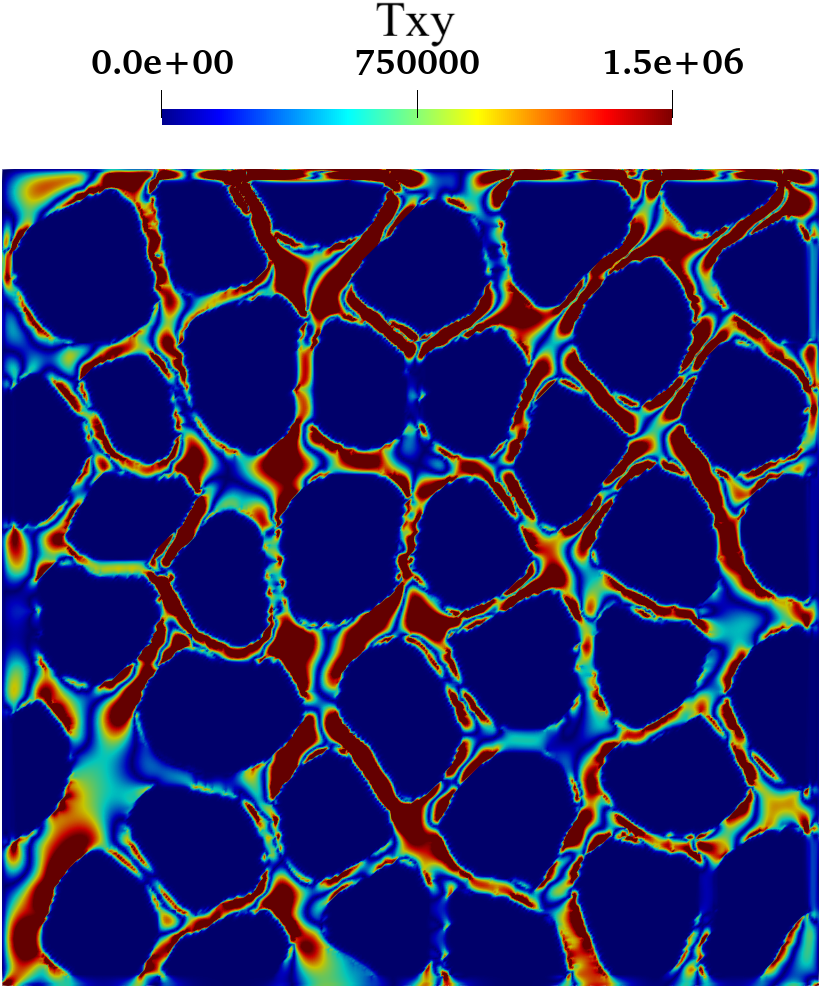}
        \\      (a)     \\
    \includegraphics[trim = 0mm 0mm 0mm 0mm, clip, width=0.19\textwidth]{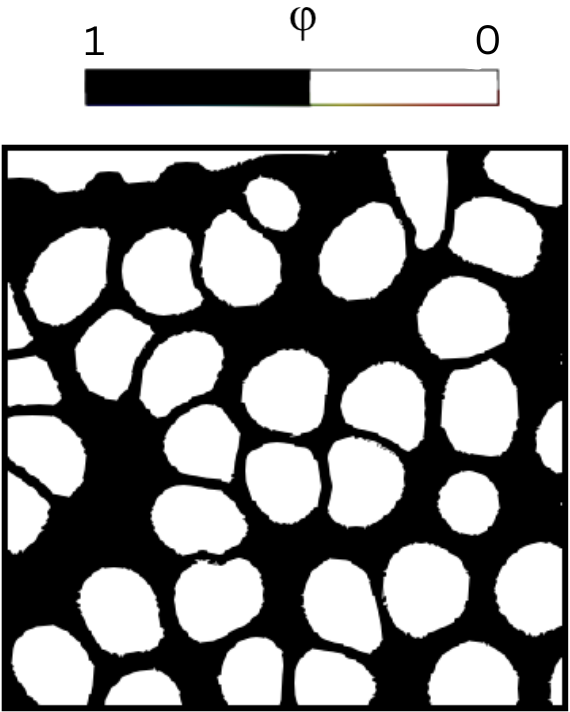}
    \includegraphics[trim = 0mm 0mm 0mm 0mm, clip, width=0.19\textwidth]{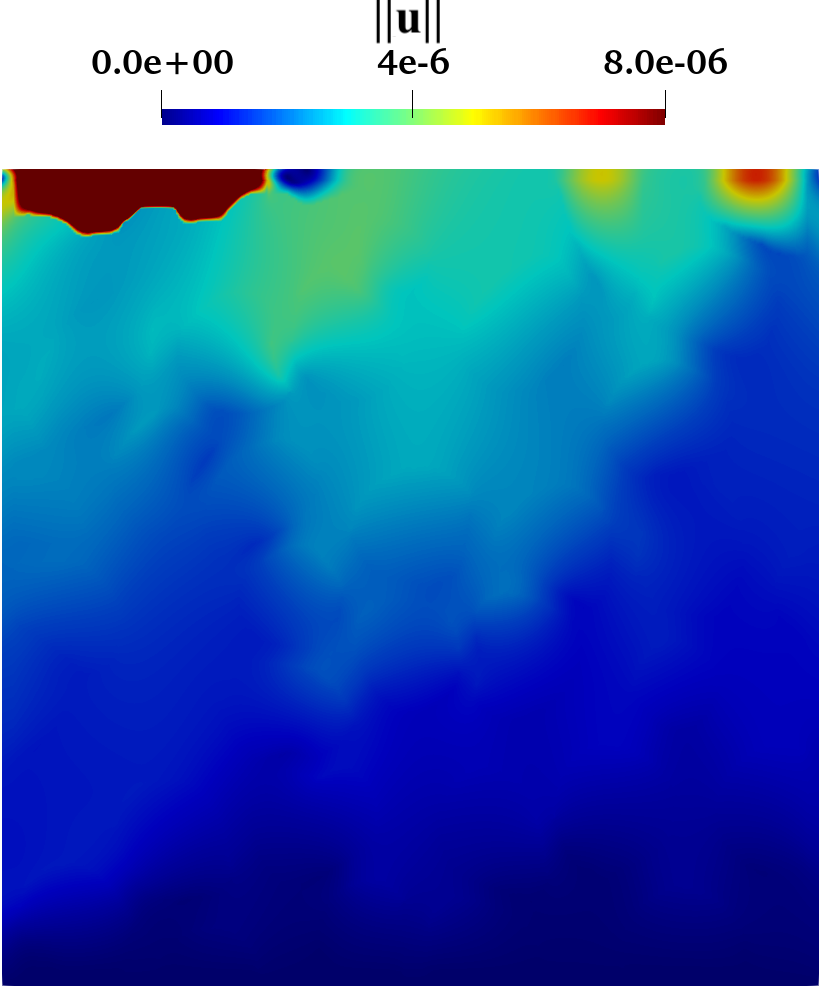} 
    \includegraphics[trim = 0mm 0mm 0mm 0mm, clip, width=0.19\textwidth]{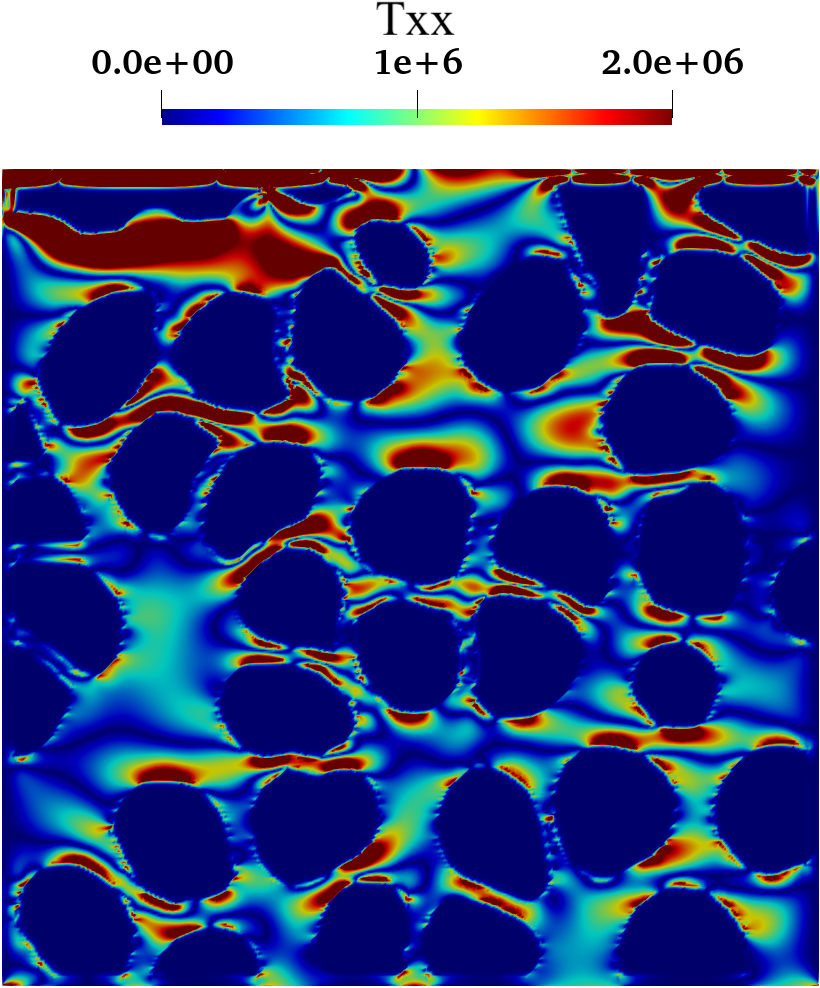} 
    \includegraphics[trim = 0mm 0mm 0mm 0mm, clip, width=0.19\textwidth]{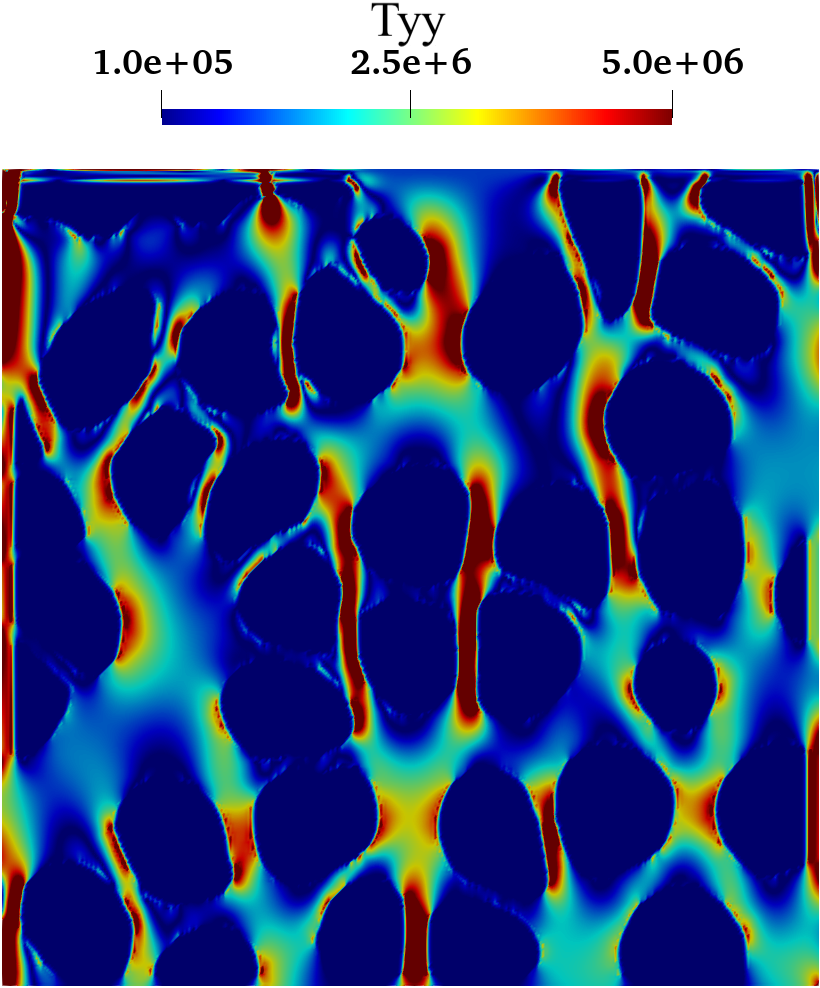} 
    \includegraphics[trim = 0mm 0mm 0mm 0mm, clip, width=0.19\textwidth]{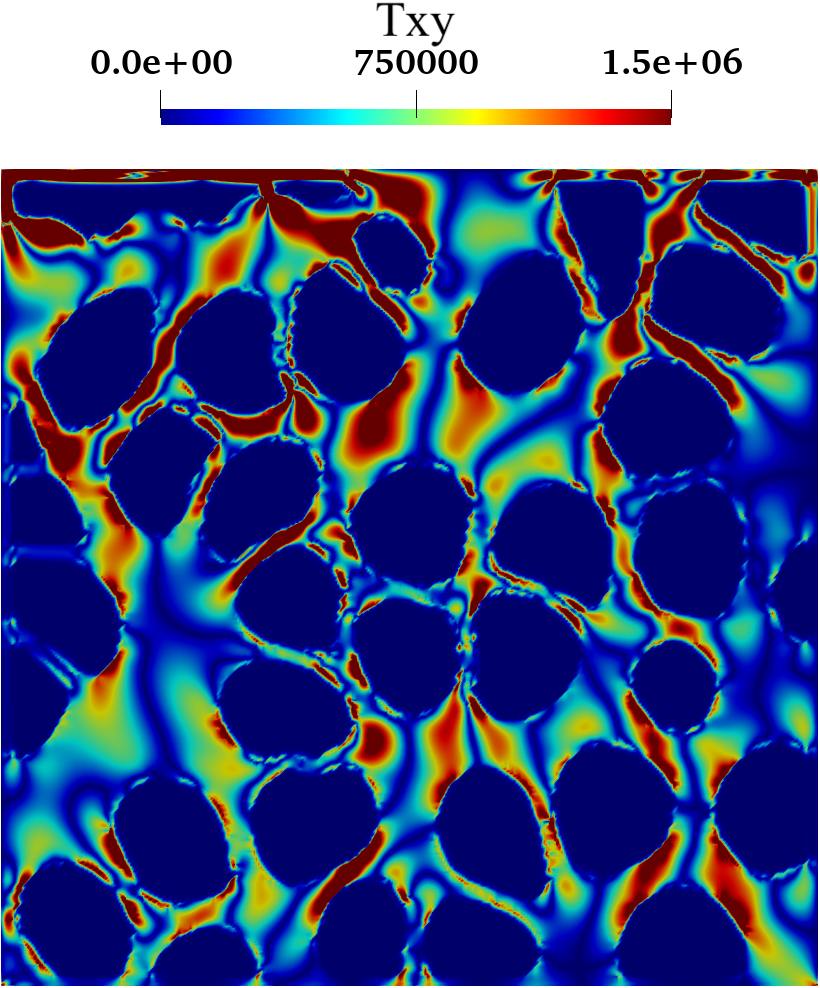}
        \\      (b)     \\
        \vspace{-0.15in}
\caption{
Finite element simulation results for \textit{medium pores} microstructures, showing the microstructure indicator function $\phi(\mathbf{x})$, the magnitude of displacement $\|\mathbf{u}\|$ (in mm), and components of Cauchy stress $\mathbf{T}$ (in mPa):
Two examples of microstructures
(a) A representative microstructure (typical of 97\% of samples) with strain energy value of $1265 \times 10^{-7} \text{mJ}$; 
(b) An anomalous sample with lower strain energy ($994 \times 10^{-7} \text{mJ}$). 
}
        \label{fig:fe_medium}
        \vspace{-0.15in}
\end{figure}
\begin{figure}[h!]
\centering
    \includegraphics[trim = 0mm 0mm 0mm 0mm, clip, width=0.19\textwidth]{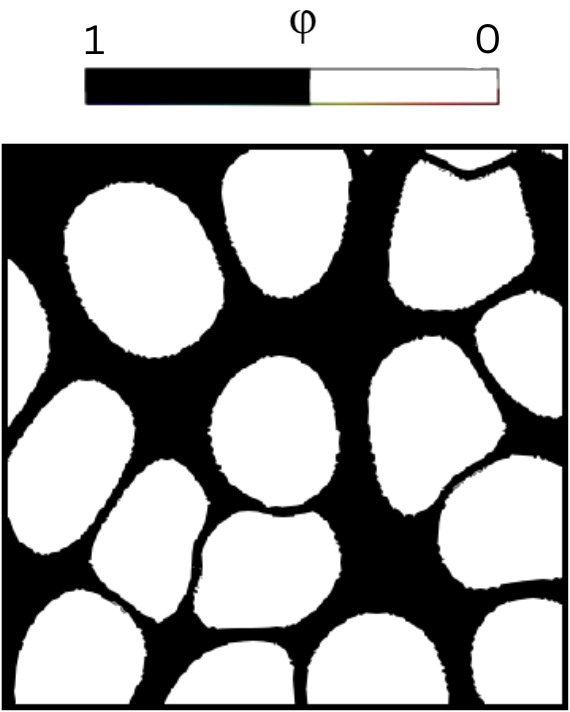}
    \includegraphics[trim = 0mm 0mm 0mm 0mm, clip, width=0.19\textwidth]{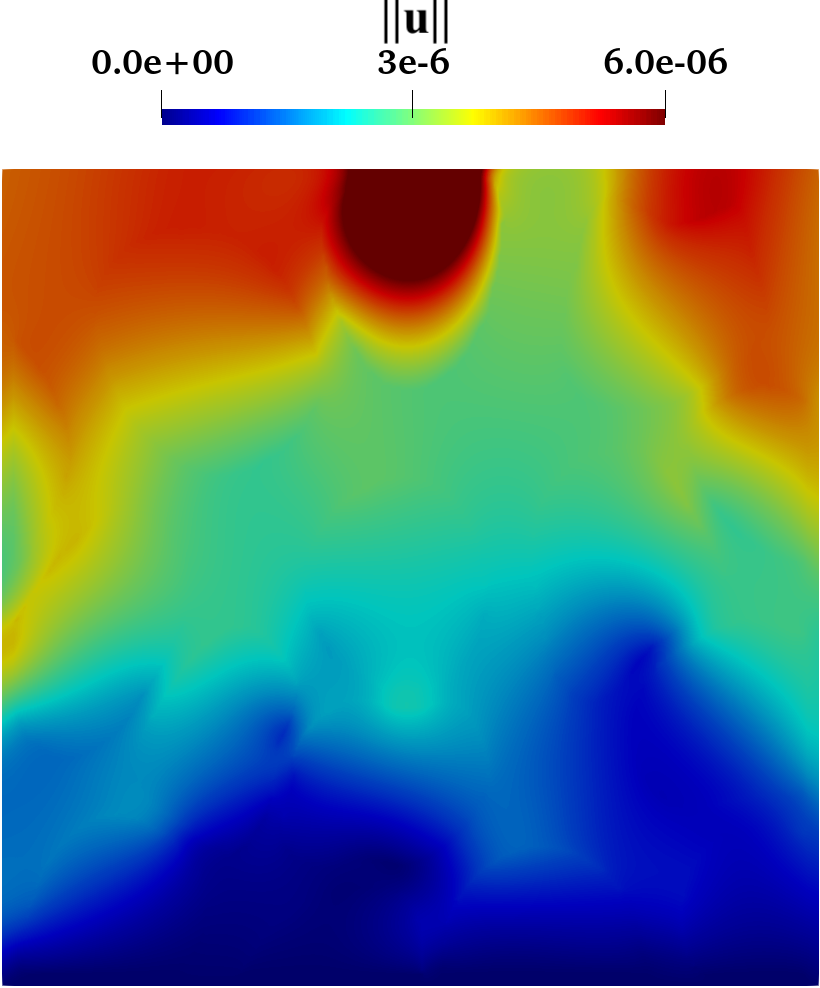} 
    \includegraphics[trim = 0mm 0mm 0mm 0mm, clip, width=0.19\textwidth]{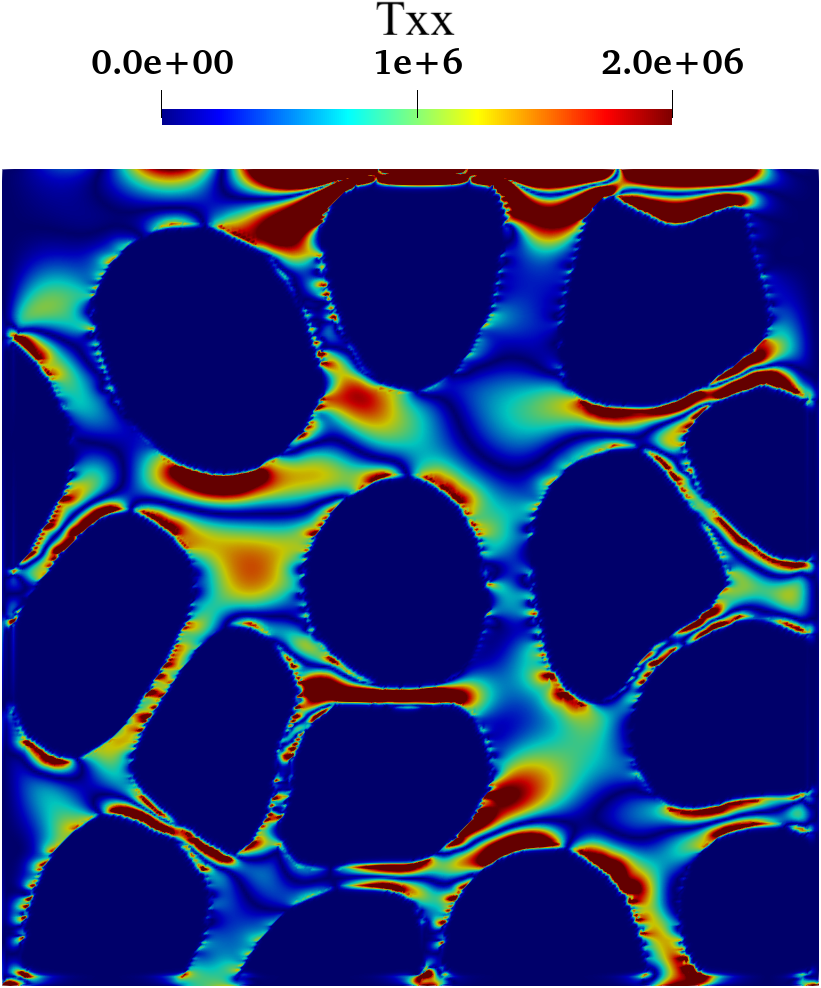} 
    \includegraphics[trim = 0mm 0mm 0mm 0mm, clip, width=0.19\textwidth]{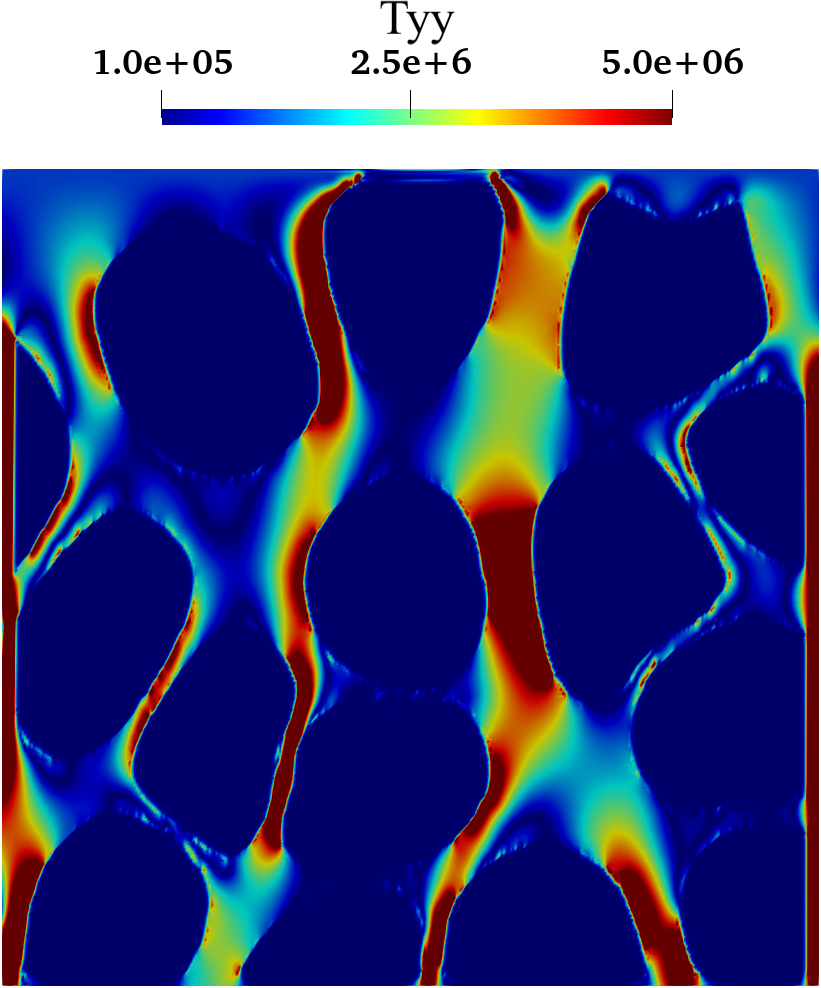} 
    \includegraphics[trim = 0mm 0mm 0mm 0mm, clip, width=0.19\textwidth]{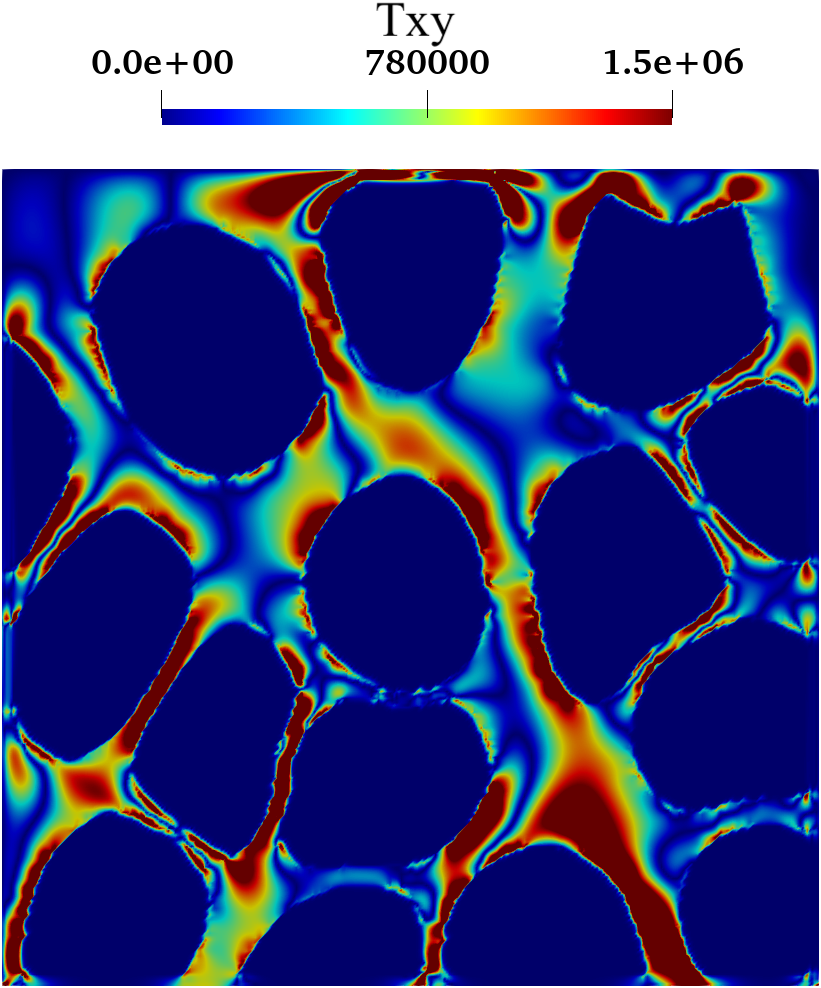}
        \\      (a)     \\
    \includegraphics[trim = 0mm 0mm 0mm 0mm, clip, width=0.19\textwidth]{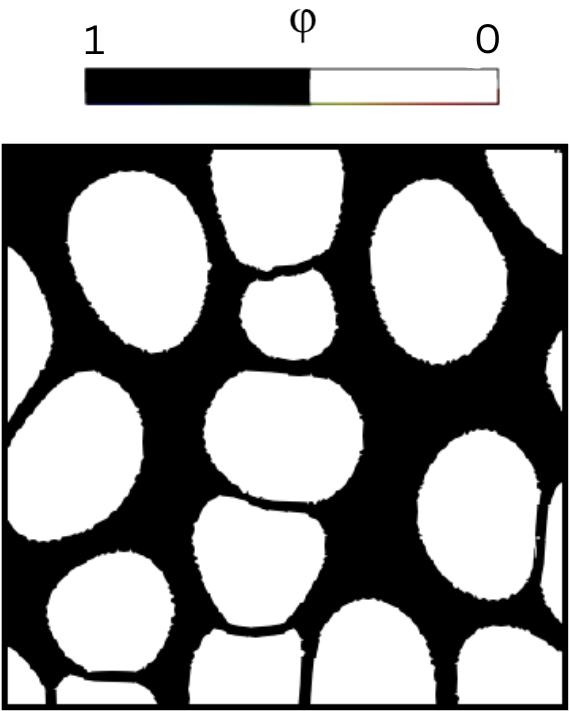}
    \includegraphics[trim = 0mm 0mm 0mm 0mm, clip, width=0.19\textwidth]{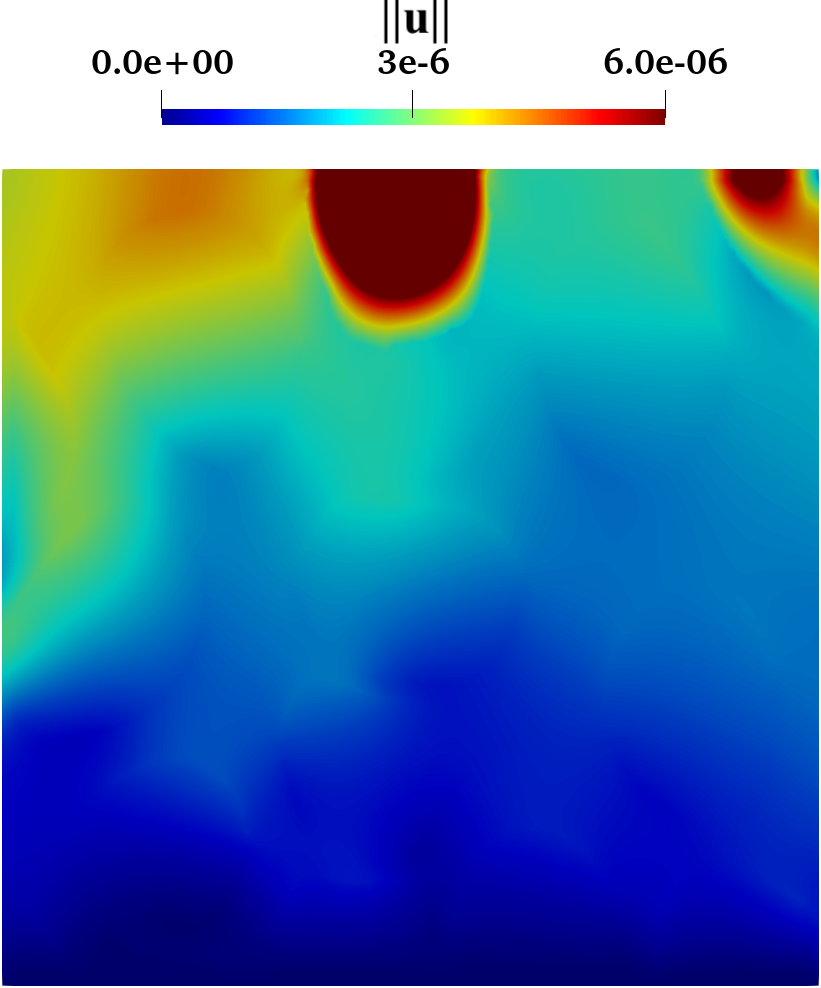}  
    \includegraphics[trim = 0mm 0mm 0mm 0mm, clip, width=0.19\textwidth]{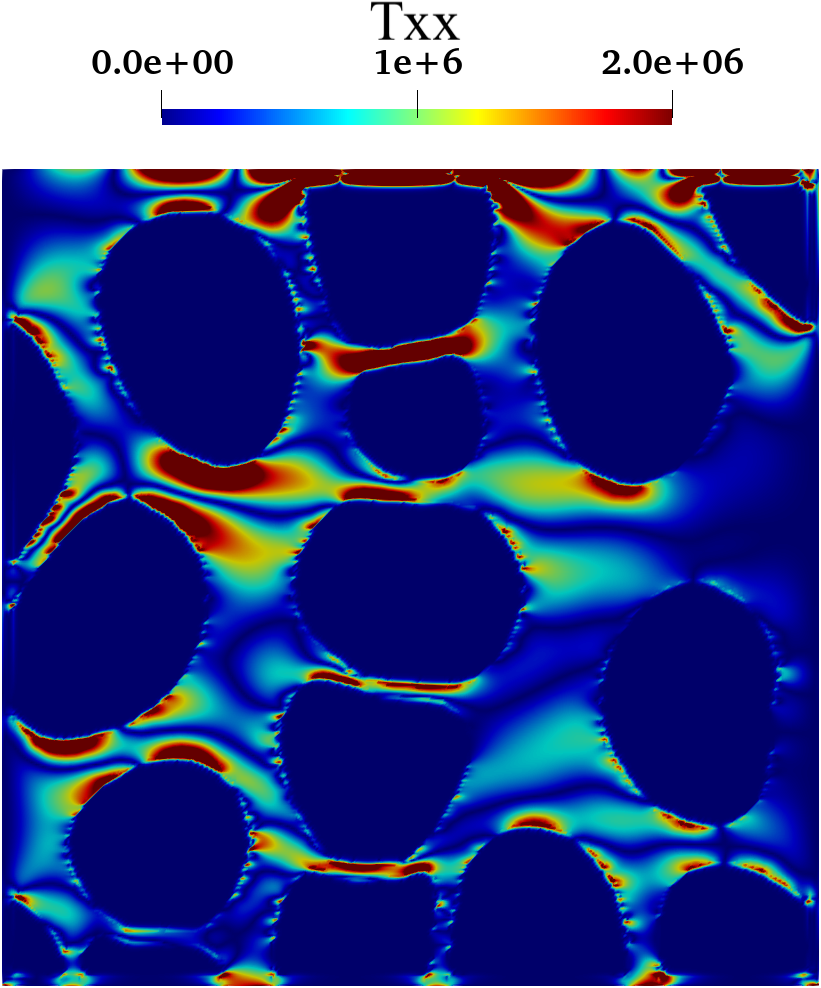} 
    \includegraphics[trim = 0mm 0mm 0mm 0mm, clip, width=0.19\textwidth]{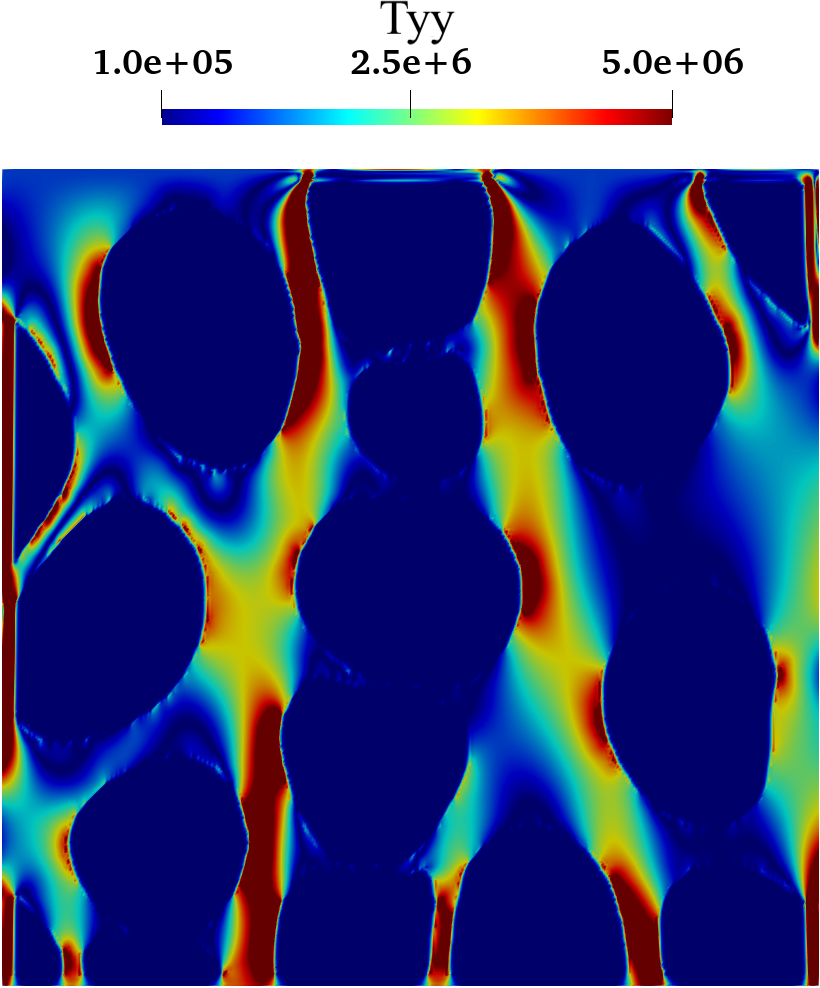} 
    \includegraphics[trim = 0mm 0mm 0mm 0mm, clip, width=0.19\textwidth]{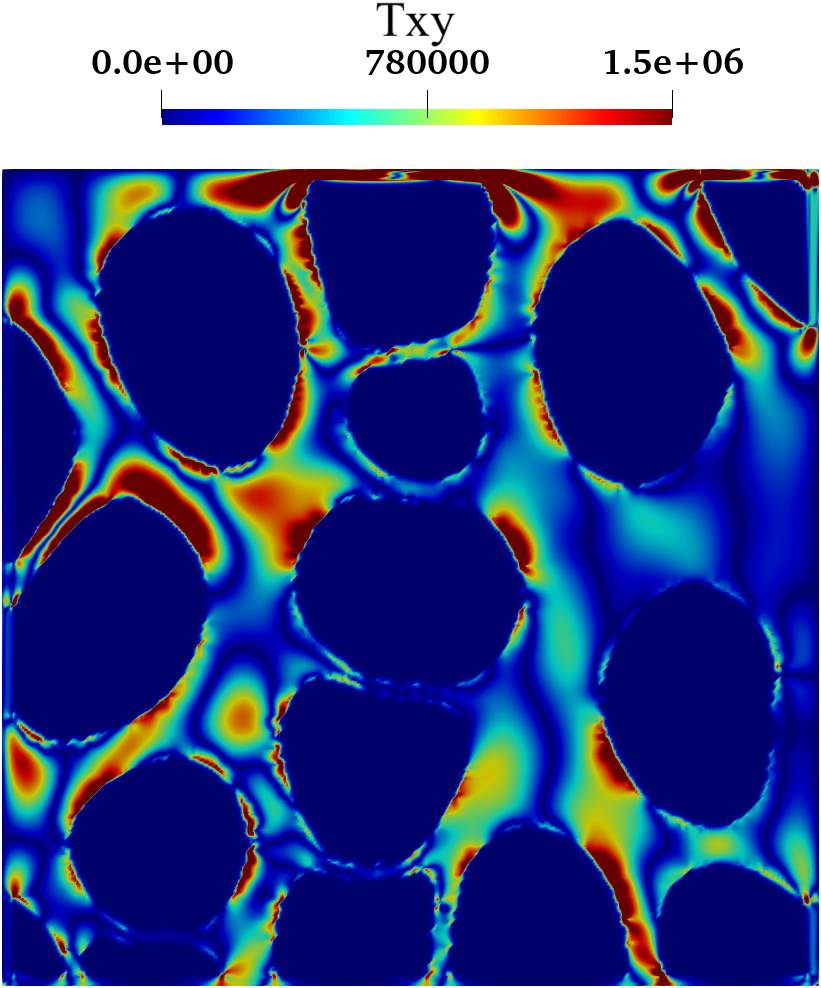}
        \\      (b)     \\
    \includegraphics[trim = 0mm 0mm 0mm 0mm, clip, width=0.19\textwidth]{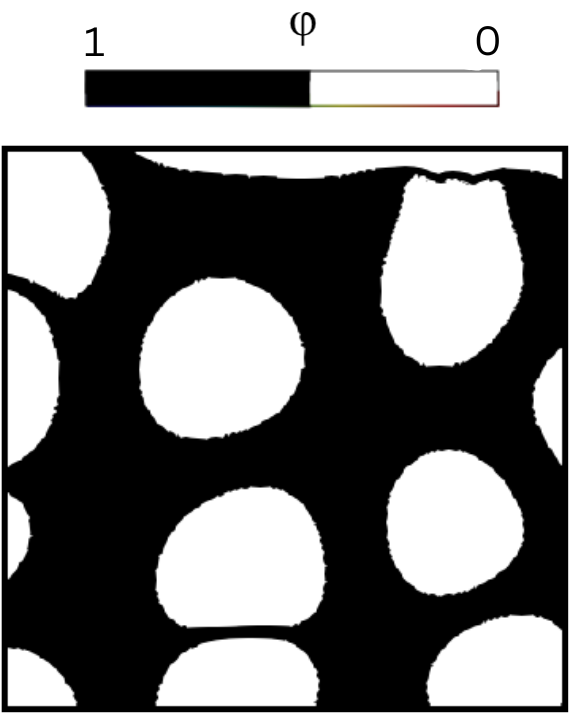}
    \includegraphics[trim = 0mm 0mm 0mm 0mm, clip, width=0.19\textwidth]{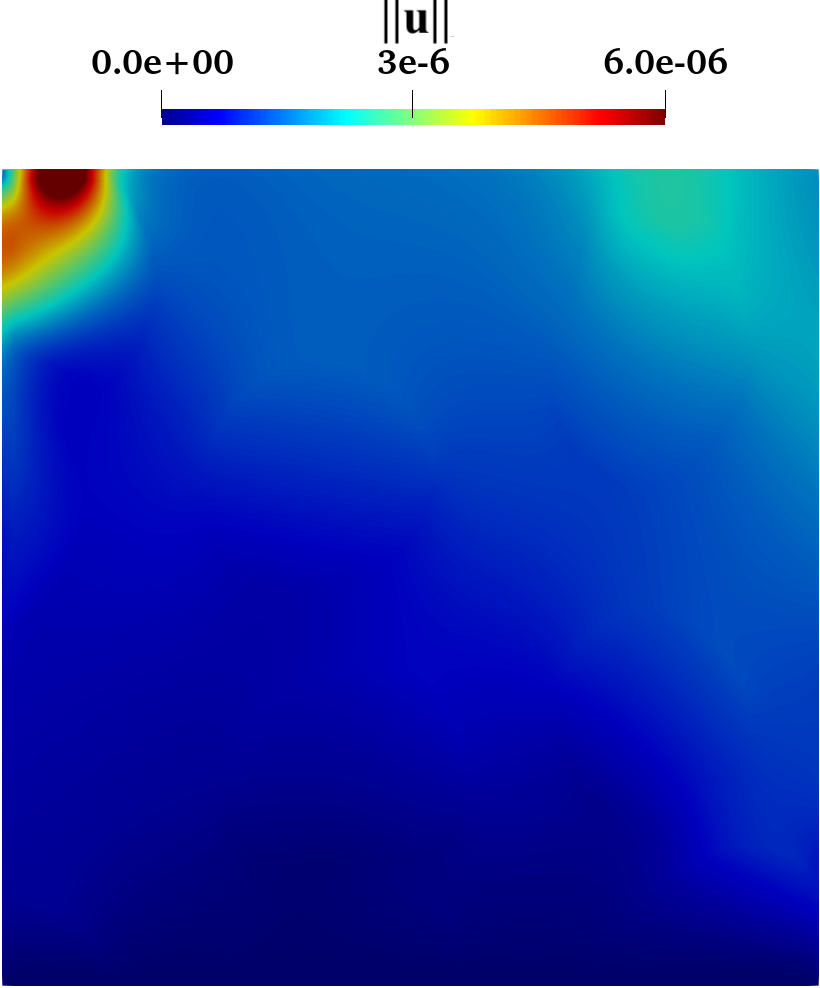}  
    \includegraphics[trim = 0mm 0mm 0mm 0mm, clip, width=0.19\textwidth]{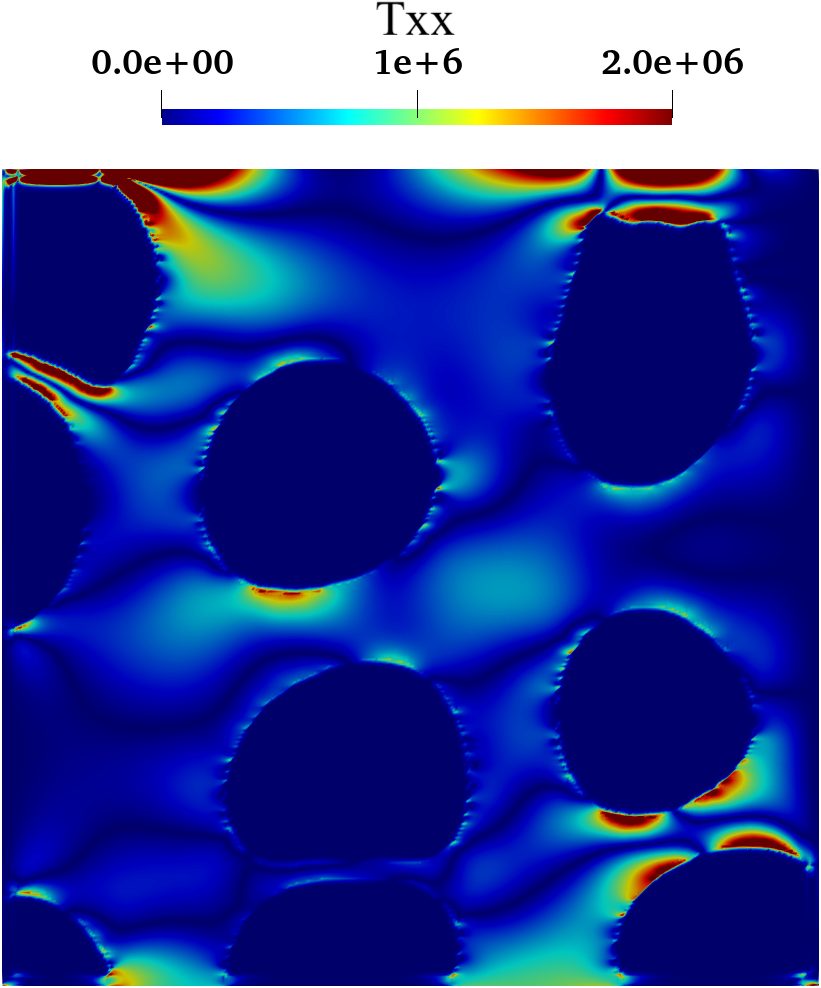} 
    \includegraphics[trim = 0mm 0mm 0mm 0mm, clip, width=0.19\textwidth]{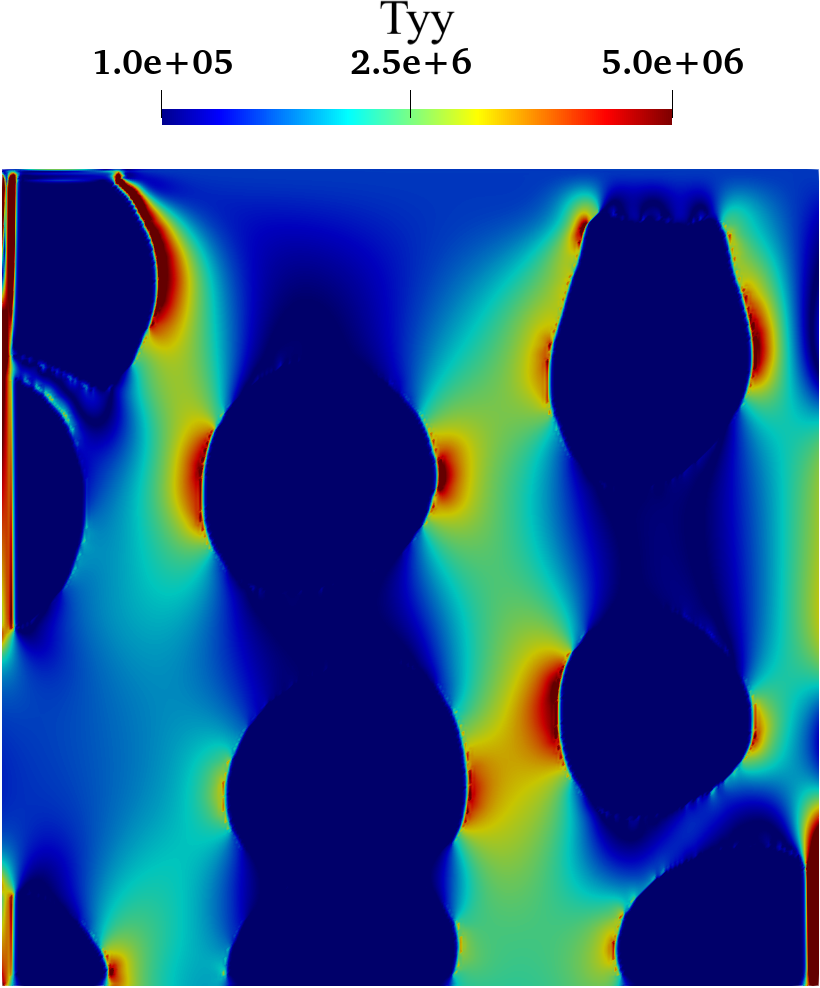} 
    \includegraphics[trim = 0mm 0mm 0mm 0mm, clip, width=0.19\textwidth]{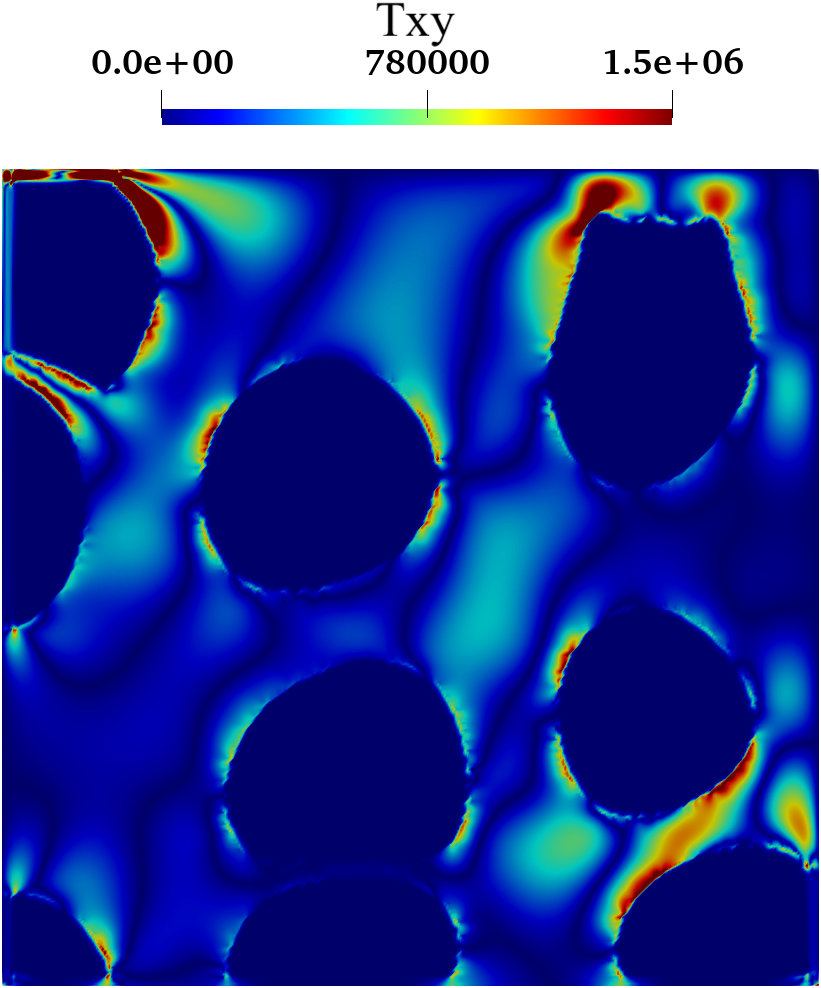}
        \\      (c)     \\
        \vspace{-0.15in}
\caption{
Finite element simulation results for \textit{large pores} microstructures, showing the microstructure indicator function $\phi(\mathbf{x})$, the magnitude of displacement $\|\mathbf{u}\|$ (in mm), and components of Cauchy stress $\mathbf{T}$ (in mPa):
Three examples of microstructures
(a) A low strain-energy sample;
(b) A representative typical sample with a strain energy of $1049 \times 10^{-7} \text{mJ}$ (characteristic of approximately 50\% of the dataset);
(c) A high strain-energy sample.
}
        \label{fig:fe_large}
        \vspace{-0.15in}
\end{figure}
Additionally, to assess the BayesCNN's generlizability, an out-of-distribution dataset was generated using FSLBM simulations. This dataset includes 10 microstructures with an average pore size of 12 $\mu m$ (intermediate between \textit{small pores} and \textit{medium pores}) and 10 with an average pore size of 22 $\mu m$ (between \textit{medium pores} and \textit{large pores}). The FSLBM parameters matched those used for training data generation, except for the average pore size and initial bubble count: 75 bubbles for the 12 $\mu m$ microstructures and 35 for the 22 $\mu m$ cases.

The bimodal distribution of strain energy probabilities, illustrated in Figure \ref{fig:training_microstructure}(a), reflects the distinct mechanical behaviors exhibited within and across the three categories of ceramic aerogel microstructures. Representative FE solutions for these categories are presented in Figures \ref{fig:fe_small}--\ref{fig:fe_large}.
For the \textit{small pores} category, the majority of microstructural samples (approximately 98\%) exhibit strain energies ranging from $1000\times 10^{-7}$ to $1100 \times 10^{-7} \, \text{mJ}$ with occasional anomalies at lower strain energy levels. As shown in Figure \ref{fig:fe_small}, most microstructures demonstrate mechanical deformation dominated by high-strain pathways connecting the top and bottom boundaries. In some cases, due to unexpected microstructural patterns, these pathways are interrupted by isolated solid-phase regions, leading to deviations in mechanical response.
A similar trend is observed for the \textit{medium pores} category, where strain energies predominantly range from $1150 \times 10^{-7} $ to $1300 \times 10^{-7} \, \text{mJ}$, with sporadic anomalies indicating increased mechanical strength (Figure \ref{fig:fe_medium}).
In contrast, the \textit{large pores} category exhibits a distinct mechanical behavior characterized by a near-normal strain energy distribution ranging from $800 \times 10^{-7} $ to $1300 \times 10^{-7} \, \text{mJ}$ (Figure \ref{fig:fe_large}). These variations highlight the strong influence of pore size and distribution on the mechanical properties of ceramic aerogels.

\begin{figure}[h!]
    \centering
    \includegraphics[trim = 0mm 0mm 0mm 0mm, clip, width=0.48\textwidth]{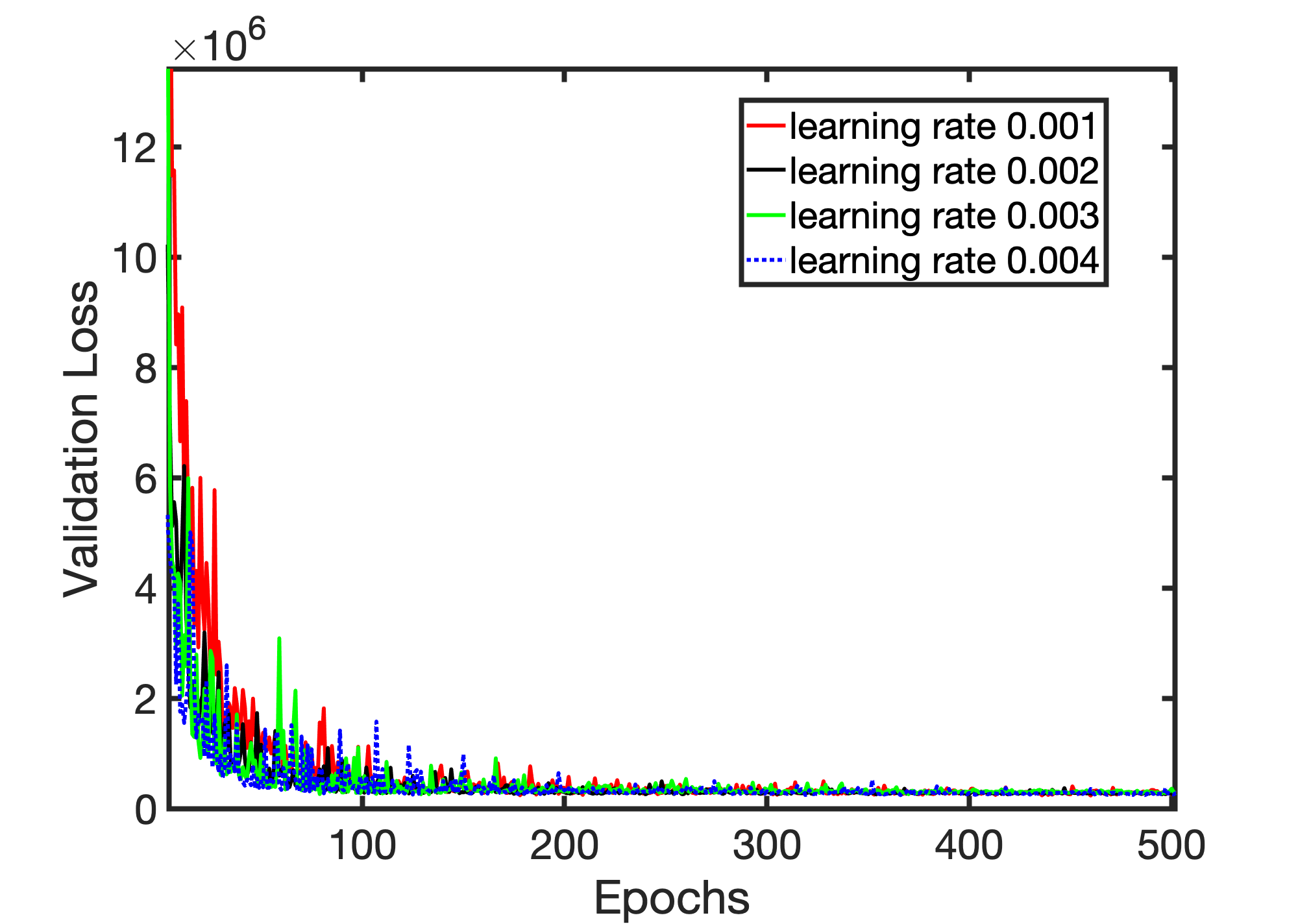}
    ~
    \includegraphics[trim = 0mm 0mm 0mm 0mm, clip, width=0.48\textwidth]{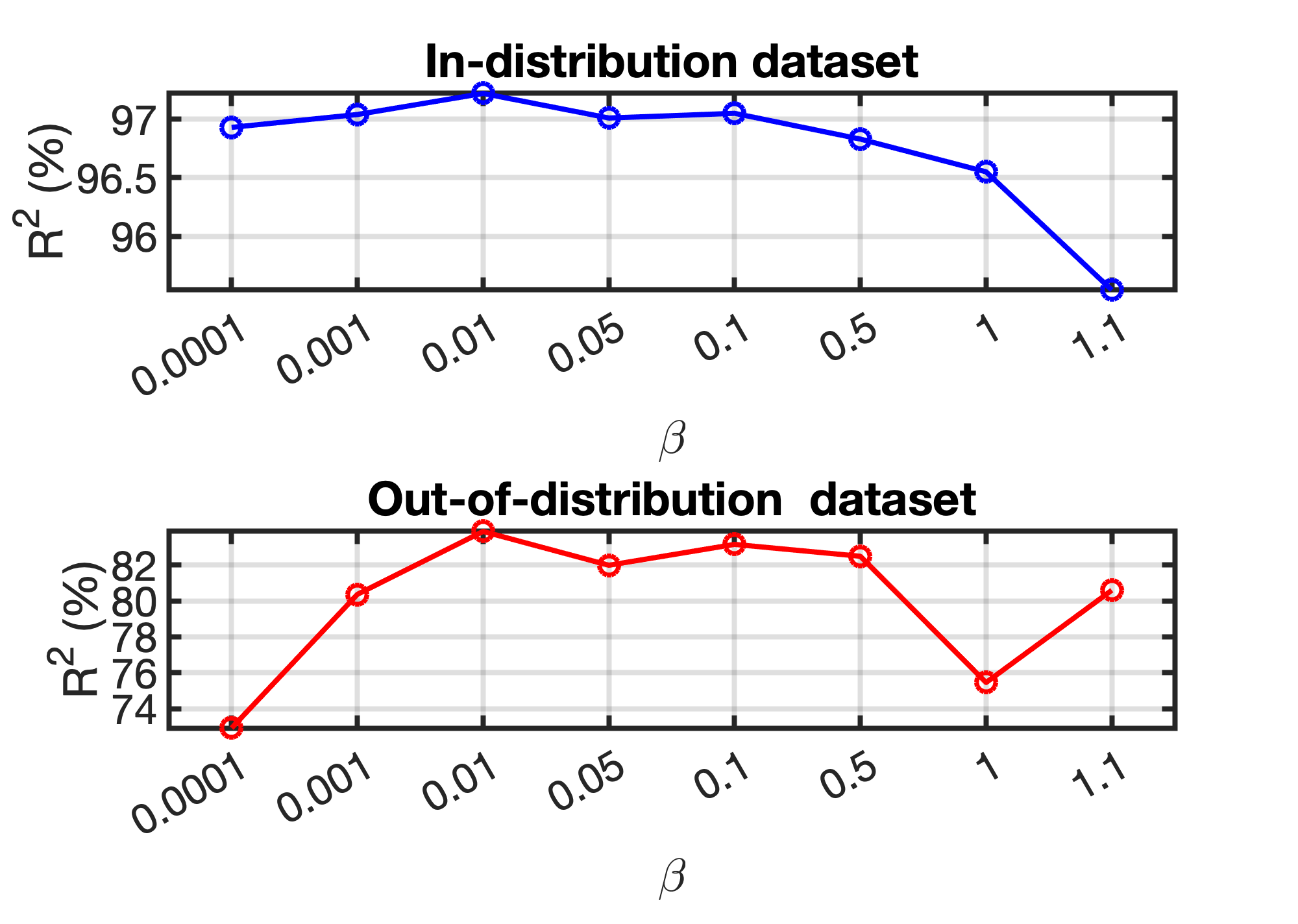}
    \\ (a) \hspace{2.5in} (b) \\
    \caption{Selection of hyperparameters for the BayesCNN model:
(a) Validation loss as a function of training epochs for various learning rates used in the optimizer.
(b) Comparison of $R^2$ values across different $\beta$ values in both in-distribution and out-of-distribution datasets.
}
    \label{fig:hyperparam}
    \vspace{-0.0in}
\end{figure}

%--------------------
\paragraph{Network architecture and training}
The BayesCNN's architecture and hyperparameters were determined through a systematic validation process using the in-distribution dataset (10\% of high-fidelity data generated using stochastic FE model). Table \ref{tab:cnn_arch} (see Appendix) details the architecture, including input and output shapes and the total number of parameters. Key hyperparameters for the variational inference include the learning rate employed by the Adam optimizer and $\beta$ in \eqref{eq:newcost}, which balances data misfit and prior regularization in the Bayesian framework.
Figure \ref{fig:hyperparam} illustrates the process of selecting these hyperparameters. The validation loss, defined as the mean squared error between validation data and the corresponding BayesCNN mean predictions, is plotted over training epochs for four different learning rates in Figure \ref{fig:hyperparam}(a). The four learning rates yield similar validation losses over the chosen course of epochs. A learning rate of 0.002, yielding the most stable validation loss decrease (fewest number of spikes), was chosen for Bayesian training. Additionally, the validation loss stopped decreasing around 400 epochs, corresponding to the minimum loss.
The accuracy of Bayesian inference depends critically on the proper selection of $\beta$. Overestimating $\beta$ diminishes the influence of valuable data, while underestimating it leads to over-fitting, biased parameter estimates. To avoid such biases, $\beta$ was tuned based on its impact on $R^2$ values computed as,
\begin{equation}
R^2 
\,=\, 1 \;-\; 
\frac{ \sum_{j=1}^{N_{id}} \bigl(Q^{FE}_j - \mathbb{E}(Q^{CNN}) \bigr)^2 }
      { \sum_{j=1}^{N_{id}} \bigl(Q^{FE}_j -  \mathbb{E}(Q^{FE}) \bigr)^2 }\,,    
\end{equation}
where $N_{id}$ is the number of in-distribution datapoints, $Q^{FE}$ is the deterministic strain energy computed from FE model for each microstructure sample, and $Q^{CNN}$ the BayesCNN prediction computed using 40 samples of the weight parameters for each microstructure. 
As shown in Figure \ref{fig:hyperparam}(b), $\beta = 0.01$ consistently achieves the highest $R^2$ for both in-distribution and out-of-distribution data, and thus selected for the Bayesian inference.
This result also aligns with the cold posterior effect \cite{wenzel2020good}, demonstrating improved predictive performance for $\beta$ below 1.
%

%--------------------
%
\begin{figure}[h]
    \centering
    \includegraphics[trim = 0mm 0mm 0mm 0mm, clip, width=0.48\textwidth]{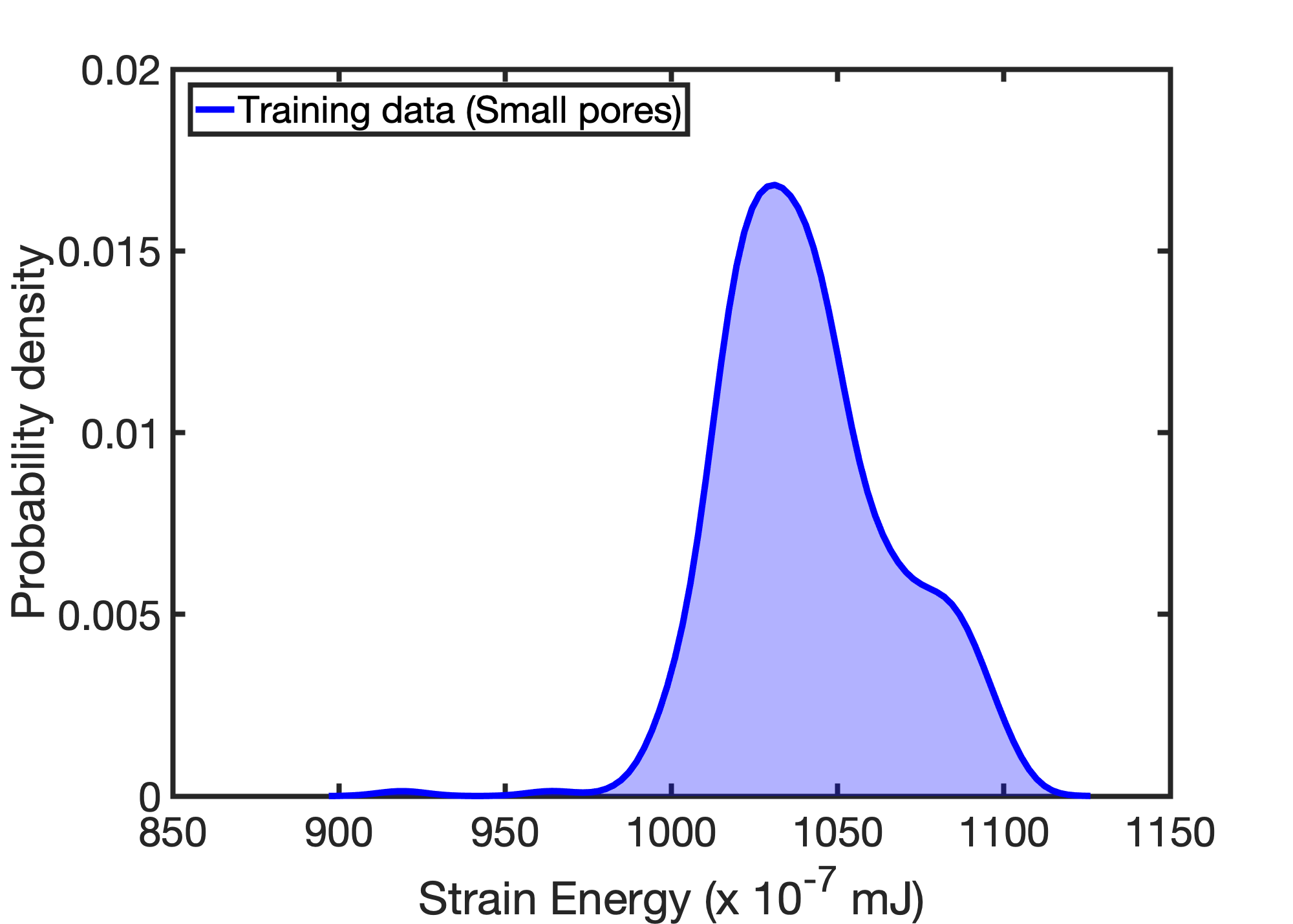}
    ~
    \includegraphics[trim = 0mm 0mm 0mm 0mm, clip, width=0.48\textwidth]{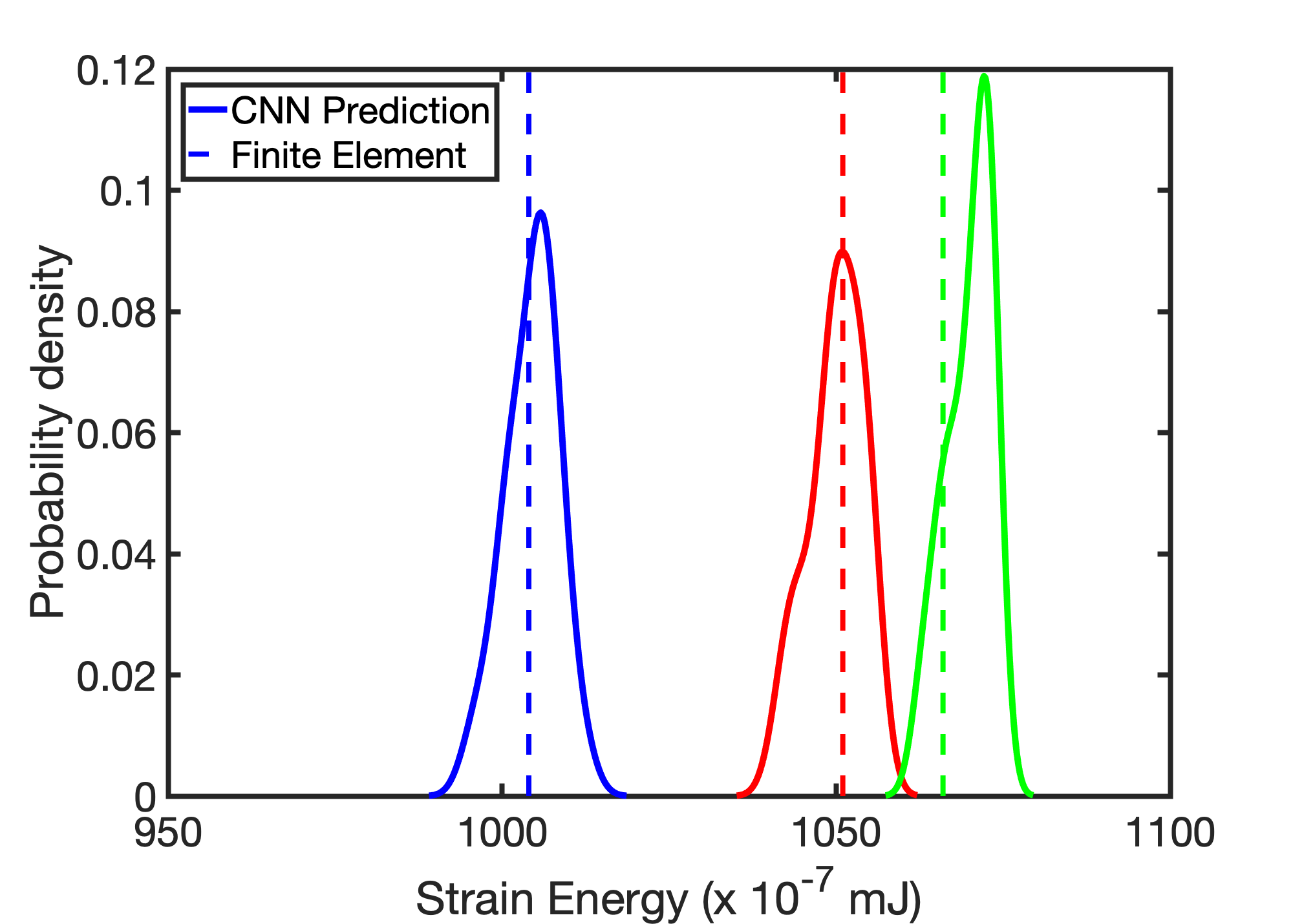}
    \caption{
    In-distribution prediction of BayesCNN for 
    \textit{small pores} microstructures (average pore size = 6$\mu$m):
    (left) Training dataset consist of 356 microstructure-strain energy pairs computed from finite element simulations; 
    (right) Comparison of BayesCNN-predicted strain energy values with corresponding finite element results for three representative microstructure samples with the same average pore sizes.
    }
    \label{fig:prediction_small}
    \vspace{-0.2in}
\end{figure} 

\begin{figure}[H]
    \centering
    \includegraphics[trim = 0mm 0mm 0mm 0mm, clip, width=0.48\textwidth]{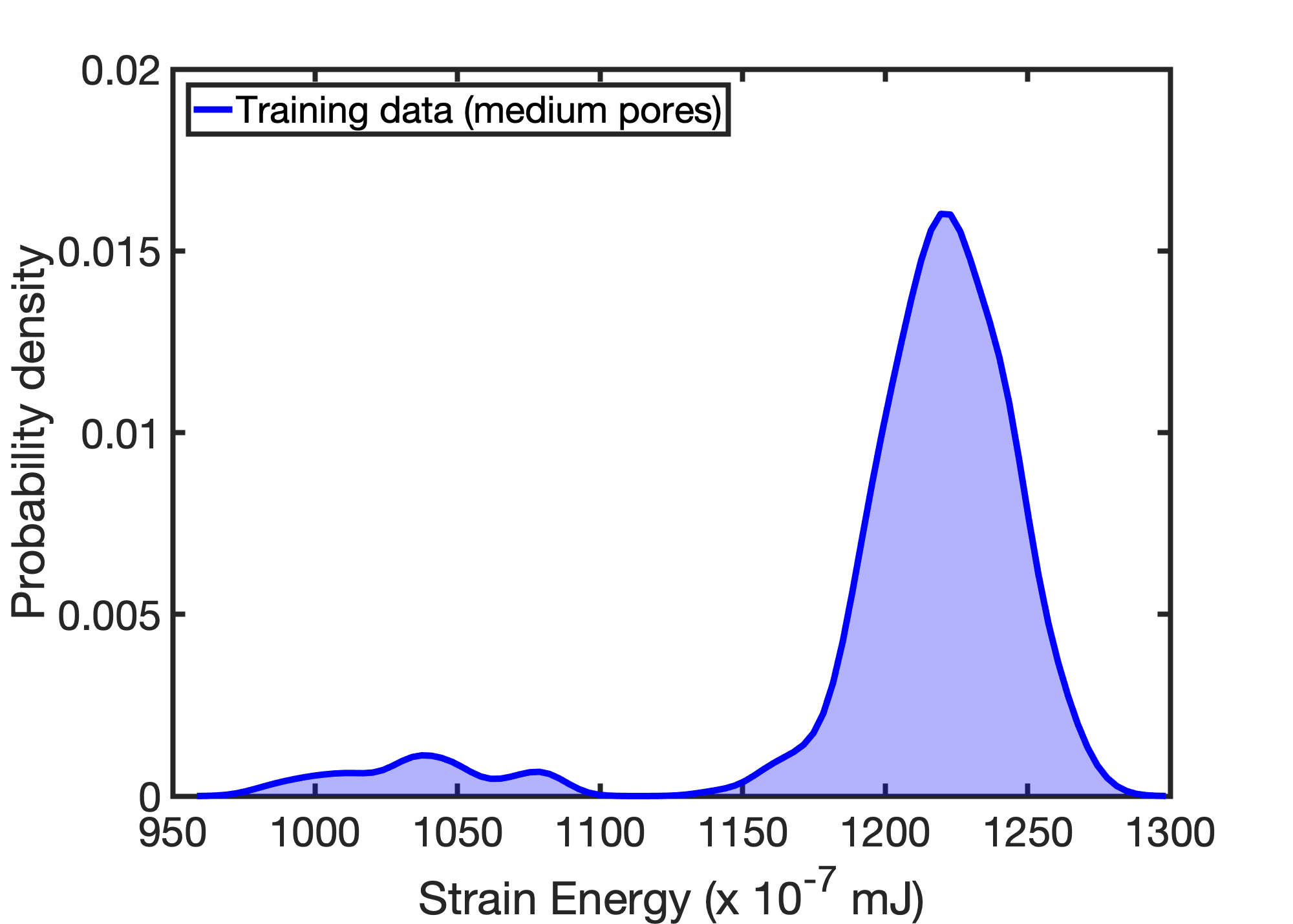}
    ~
    \includegraphics[trim = 0mm 0mm 0mm 0mm, clip, width=0.48\textwidth]{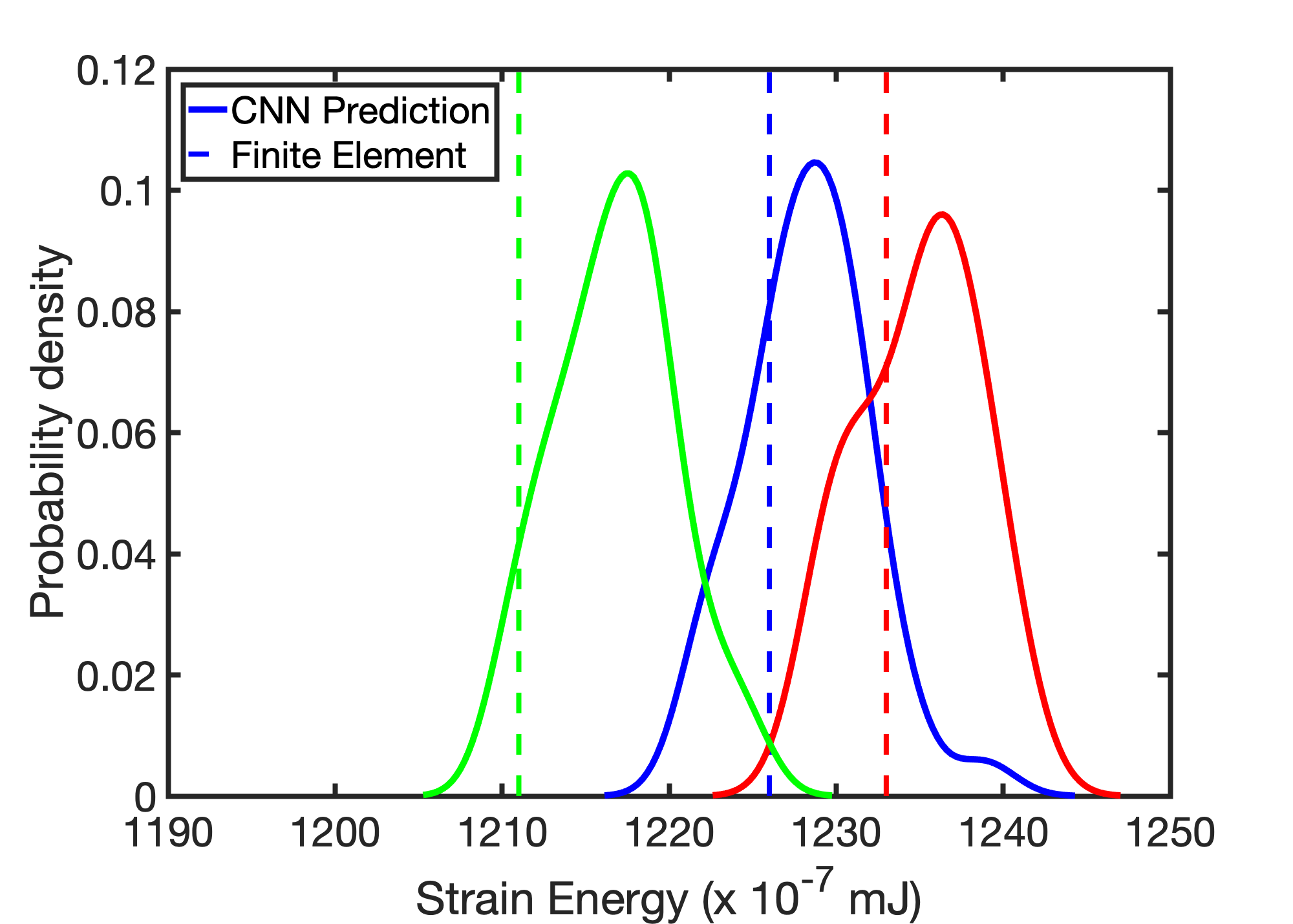}
    \caption{
    In-distribution prediction of BayesCNN for 
    \textit{medium pores} microstructures (average pore size = 18$\mu$m):
    (left) Training dataset consist of 356 microstructure-strain energy pairs computed from finite element simulations; 
    (right) Comparison of BayesCNN-predicted strain energy values with corresponding finite element results for three representative microstructure samples with the same average pore sizes.
    }
    \label{fig:prediction_medium}
    \vspace{-0.1in}
\end{figure}

\begin{figure}[H]
    \centering
    \includegraphics[trim = 0mm 0mm 0mm 0mm, clip, width=0.48\textwidth]{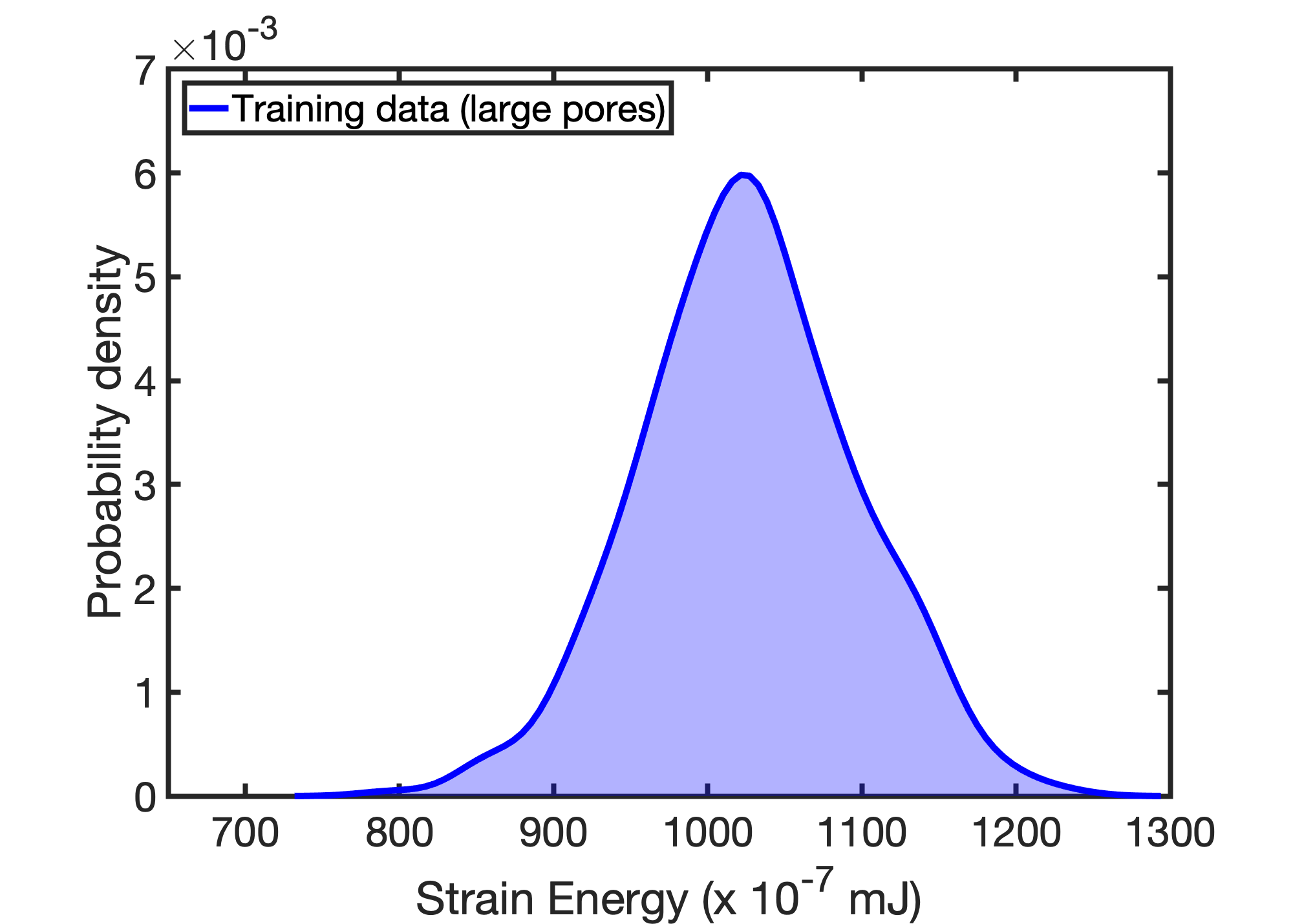}
    ~
    \includegraphics[trim = 0mm 0mm 0mm 0mm, clip, width=0.48\textwidth]{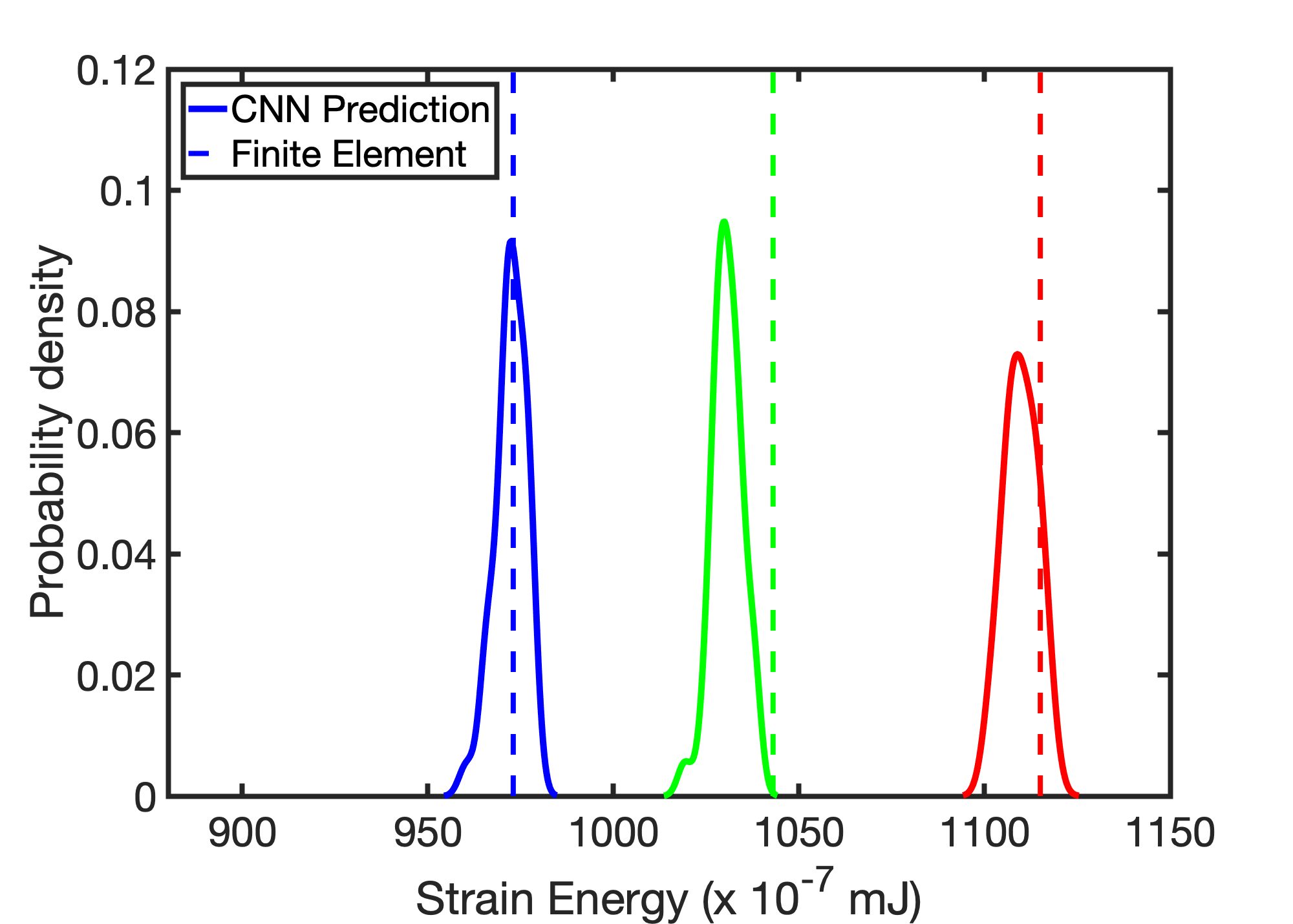}
    \caption{
    In-distribution prediction of BayesCNN for 
    \textit{large pores} microstructures (average pore size = 25$\mu$m):
    (left) Training dataset consist of 356 microstructure-strain energy pairs computed from finite element simulations; 
    (right) Comparison of BayesCNN-predicted strain energy values with corresponding finite element results for three representative microstructure samples with the same average pore sizes.
    }
    \label{fig:prediction_large}
    \vspace{-0.1in}
\end{figure}

\paragraph{BayesCNN prediction}
Figures \ref{fig:prediction_small}–\ref{fig:prediction_large} show the in-distribution prediction results of the BayesCNN model for the \textit{small pores}, \textit{medium pores}, and \textit{large pores} microstructure classes. Each figure includes the probability distribution of high-fidelity strain energy data for the corresponding class and the three representative microstructure samples, including the true strain energy value from the FE model, $Q^{FE} = \mathcal{F}_{FE}(\varphi, \bs\theta_{FE})$  and the BayesCNN-predicted probability distribution, $Q^{CNN} = \mathcal{G}_{CNN}(\varphi, \bs{w}_{CNN})$, obtained using 40 samples of weight parameters $\bs{w}_{CNN}$. In all cases, except the \blue{anomalous microstructural patterns} discussed in Figures \ref{fig:fe_small} and \ref{fig:fe_medium}, the true values consistently fall within the BayesCNN-predicted distributions, demonstrating the model's reliability for in-distribution predictions.

%

\begin{figure}[H]
    \centering
    % Second column
    \begin{minipage}[t]{0.5\textwidth}
        \centering
        \includegraphics[scale=0.107]{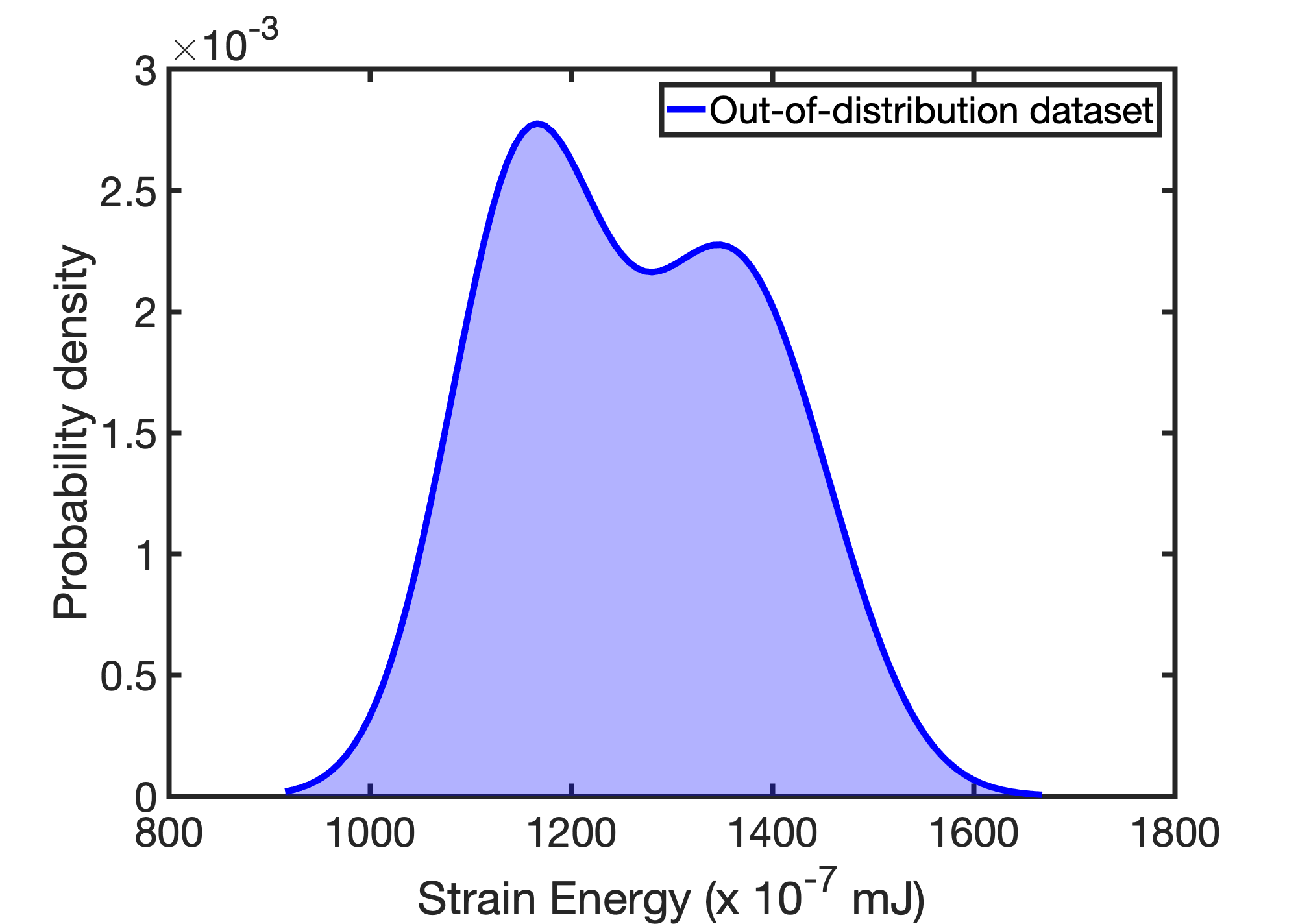}
    \end{minipage}%
    ~
    \begin{minipage}[t]{0.5\textwidth}
        \centering
        \includegraphics[scale=0.16]{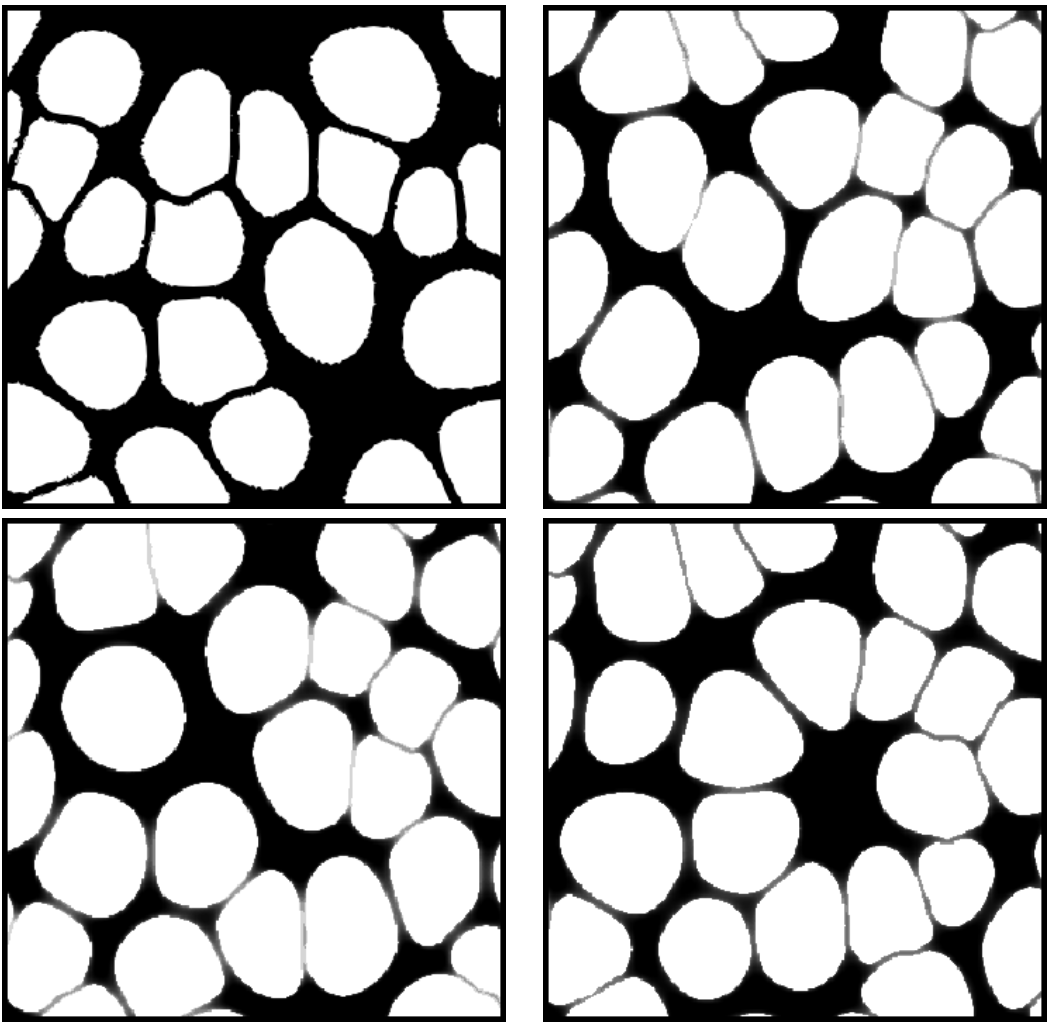}
        \includegraphics[scale=0.16]{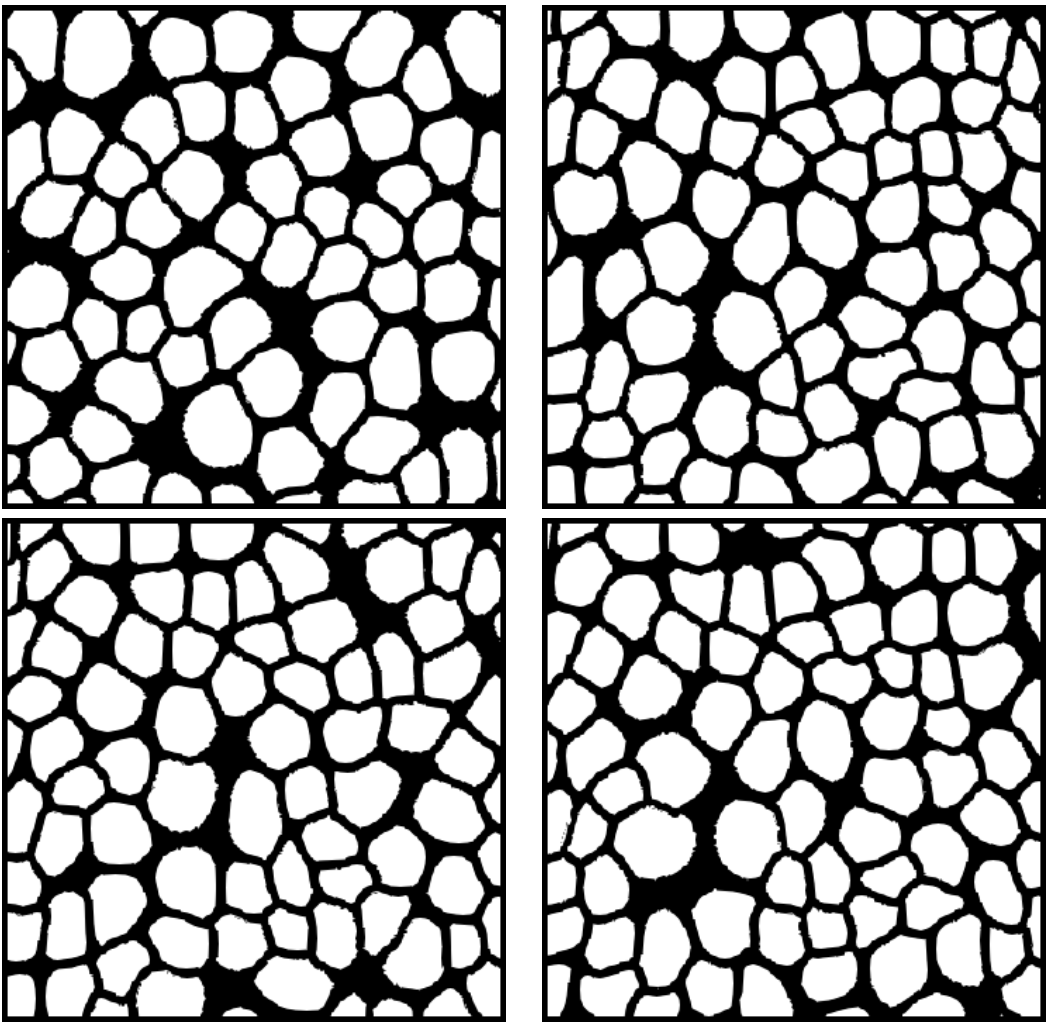}
    \end{minipage}%
    \\ (a) \hspace{2.5in} (b) \\
    \includegraphics[trim = 0mm 0mm 0mm 0mm, clip, width=0.48\textwidth]{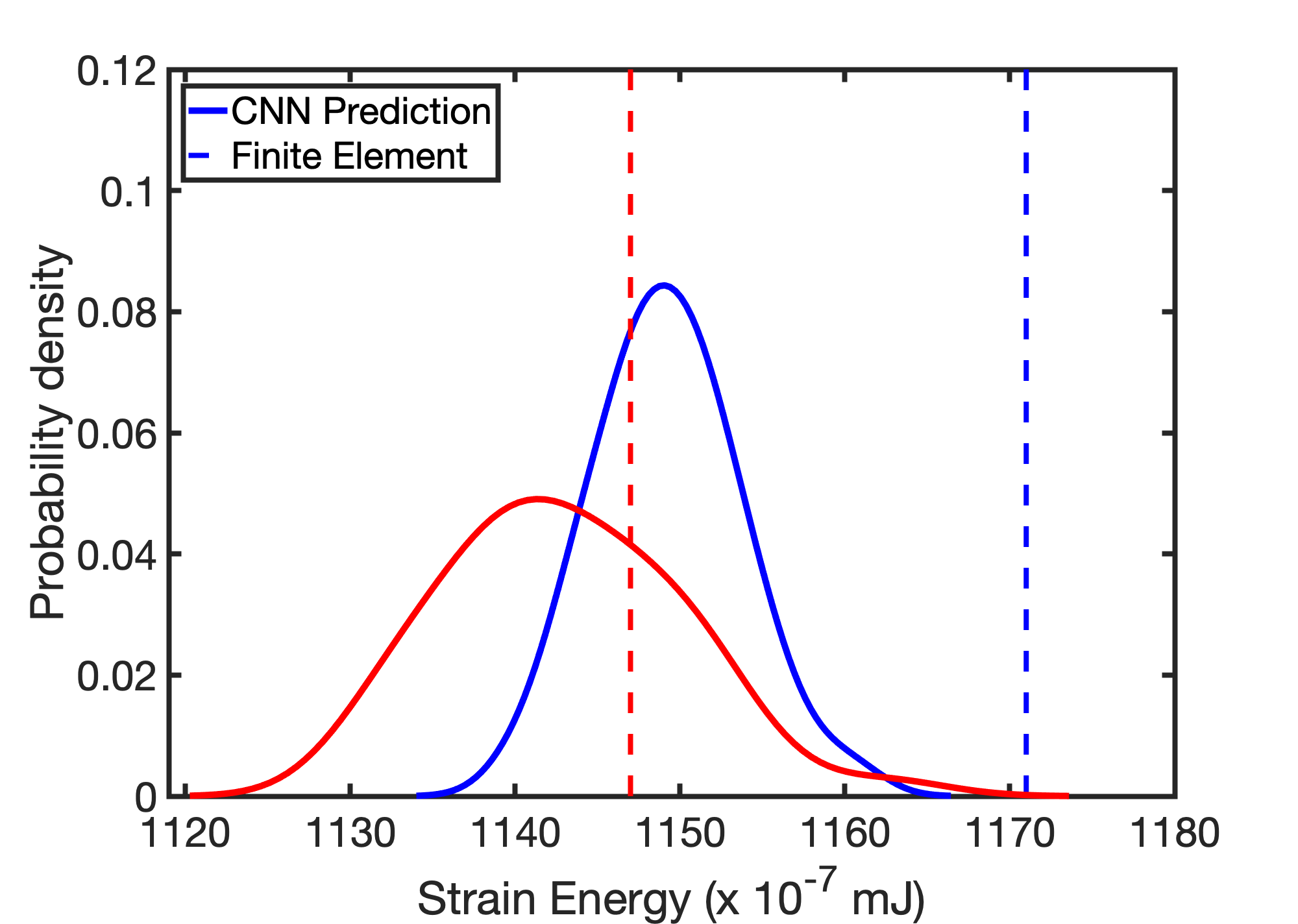}
    ~
    \includegraphics[trim = 0mm 0mm 0mm 0mm, clip, width=0.48\textwidth]{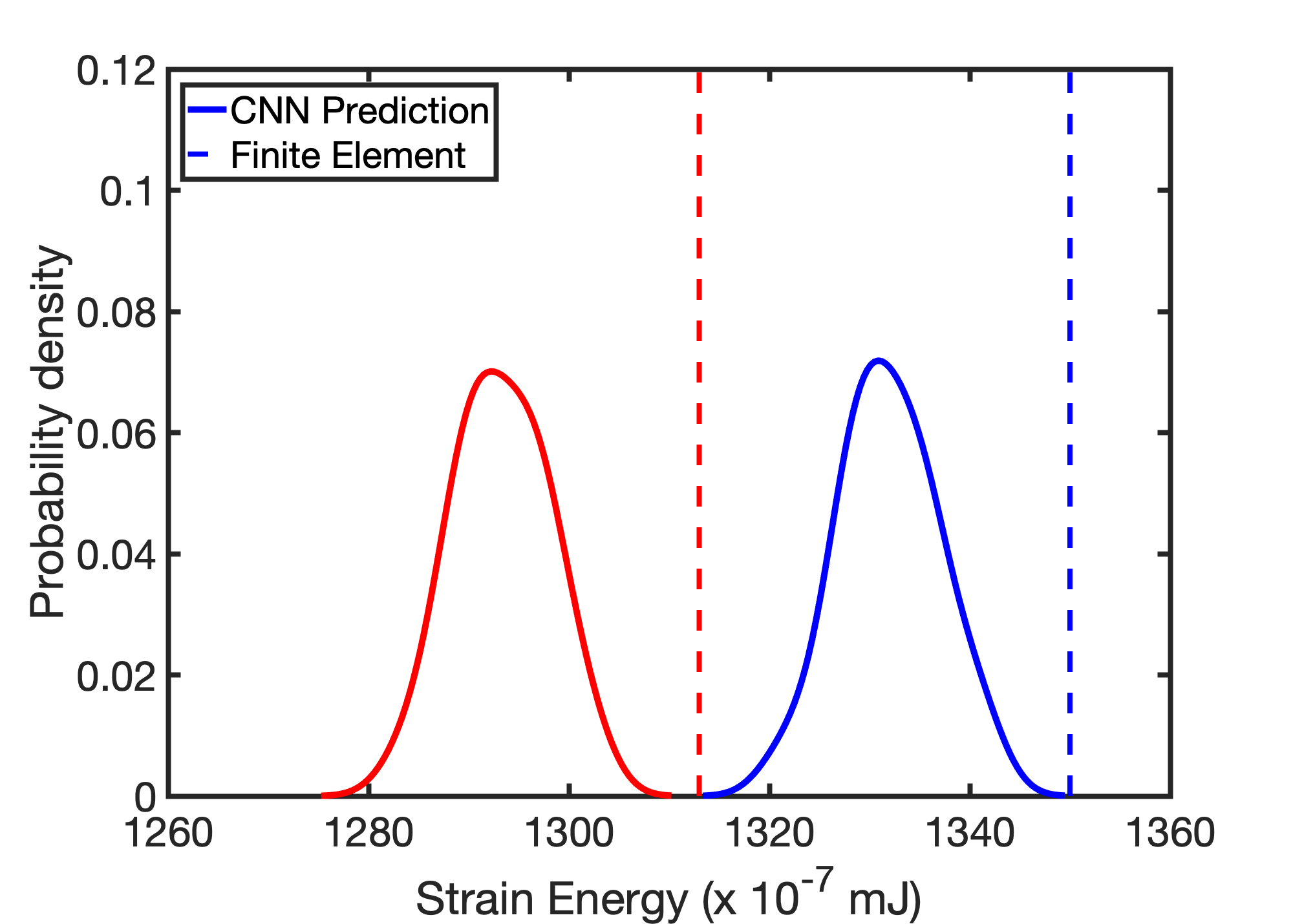}
      \\ (c) \hspace{2.5in} (d) \\
    \caption{
    Out-of-distribution predictions of BayesCNN: 
    (a) Probability distribution of strain energy in out-of-distribution dataset consisting of 20 aerogel microstructure samples with average pore sizes of 12 $\mu m$ (interpolating between \textit{small pores} and \textit{medium pores}) and 22 $\mu m$ (interpolating between \textit{medium pores} and \textit{large pores}).
    (b) Representative microstructure samples for the 12 $\mu m$ and 22 $\mu m$ pore size classes. 
    Comparison of BayesCNN-predicted strain energy with finite element simulation results for two representative out-of-distribution microstructures
    (c) 12 $\mu m$ pore size.
    (d) 22 $\mu m$ pore size.
    }
    \label{fig:outdis_prediction}
    \vspace{-0.05in}
\end{figure}

\begin{figure}[H]
    \centering
        \includegraphics[width=0.48\textwidth]{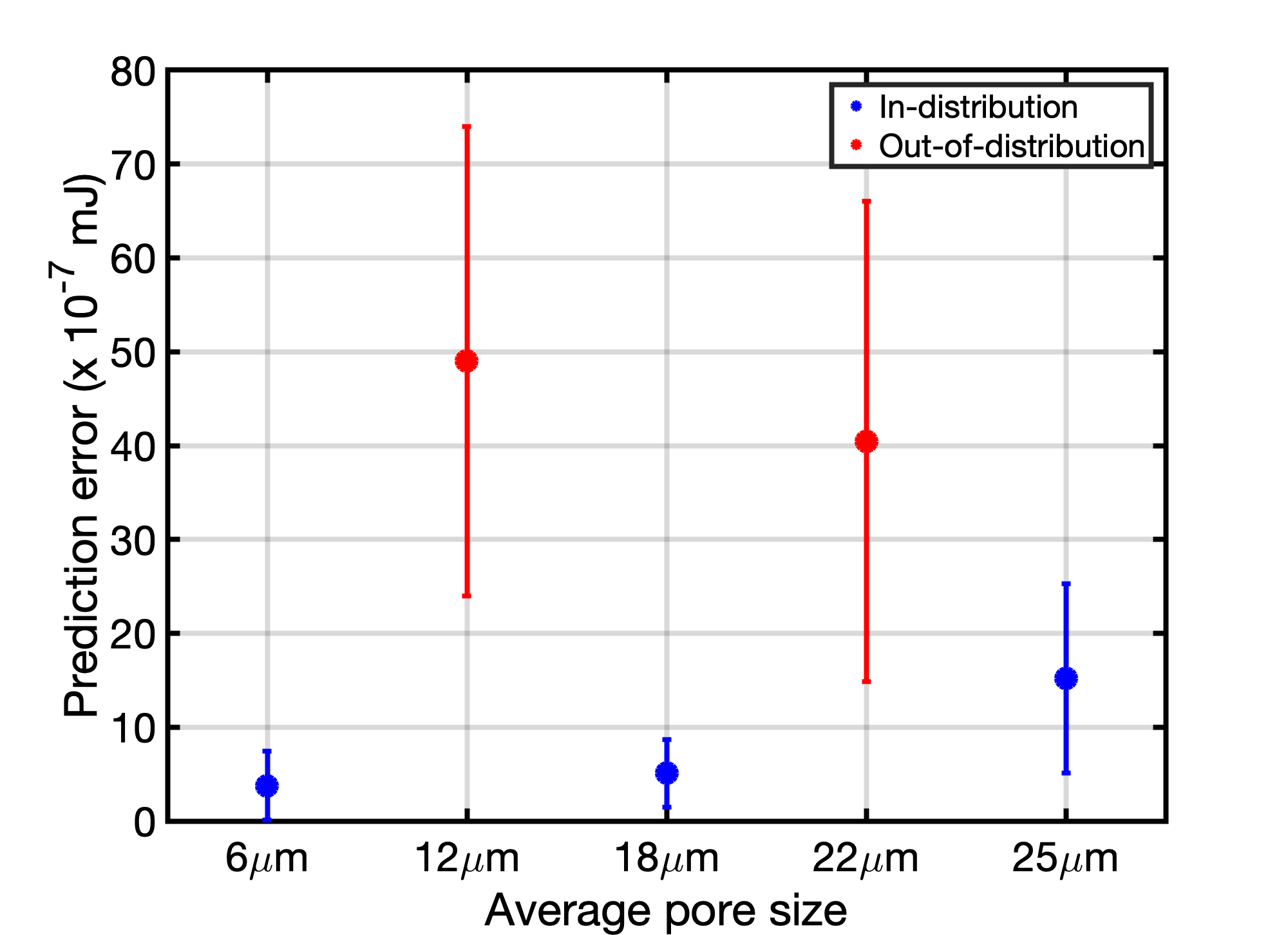}
        \includegraphics[width=0.48\textwidth]{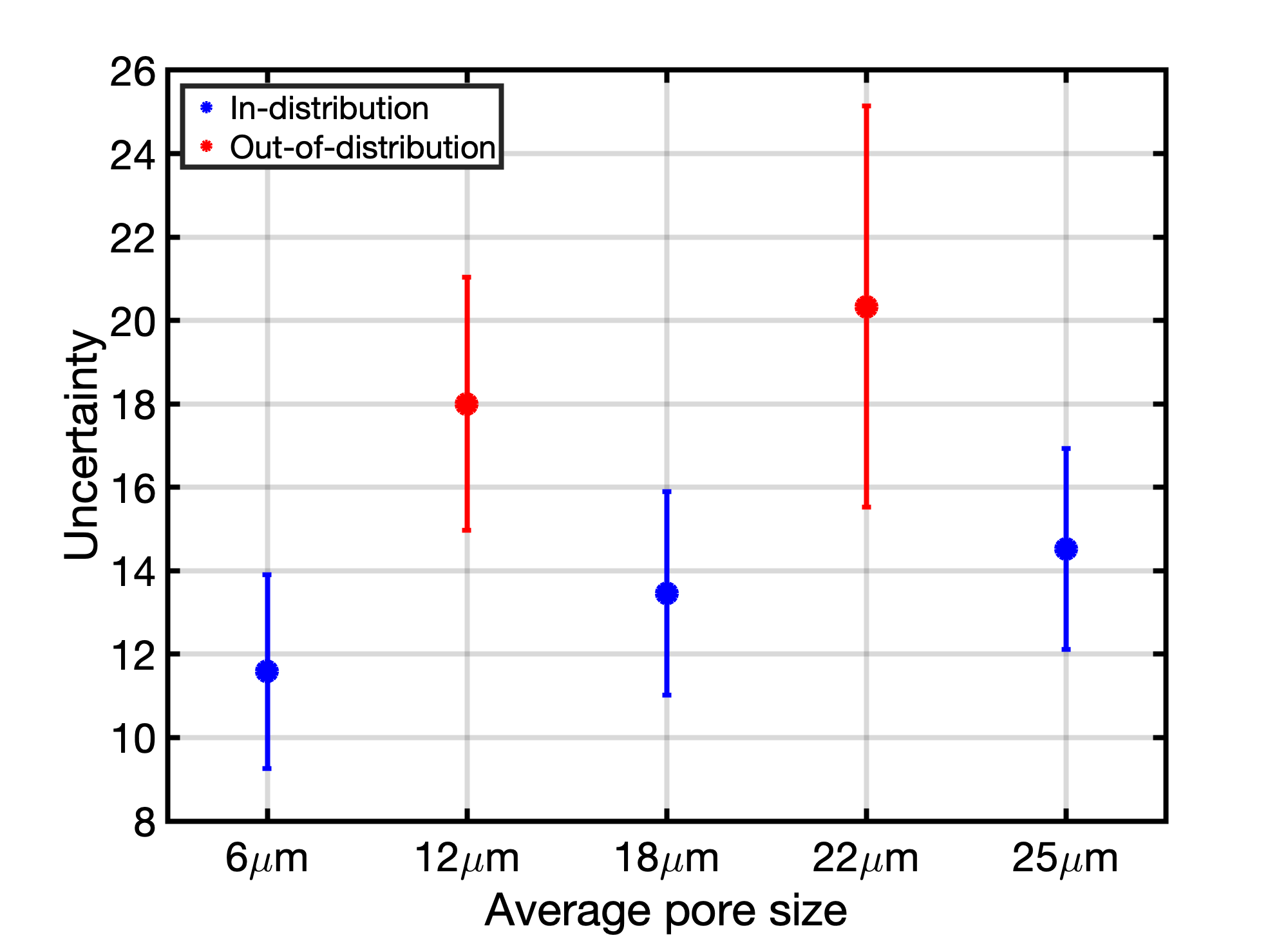}
    \caption{
    Prediction accuracy and uncertainty of the BayesCNN model evaluated on both in-distribution and out-of-distribution datasets using 30 samples for each average pore size.
    The \textit{prediction error} is defined as $\left| \mathbb{E}[Q^{CNN}] - Q^{FE} \right|$, and the \textit{uncertainty} is defined as $\mathbb{V}[Q^{CNN}]$.
    }
    \label{fig:errorbar}
    \vspace{-0.1in}
\end{figure}

%% Out-of-distribution
A key advantage of deep learning surrogate models for microstructure-property mapping in materials design and engineering lies in their accuracy and reliability for \textit{out-of-distribution} predictions. This refers to the model’s ability to generalize to new microstructural features beyond the classes represented in the training data.
Figure \ref{fig:outdis_prediction} compares the true strain energy, 
$Q^{FE} = \mathcal{F}_{FE}(\varphi, \bs\theta_{FE})$,
computed via the FE model, with the probability distribution of strain energy predicted by the BayesCNN surrogate model, $Q^{CNN} = \mathcal{G}_{CNN}(\varphi, \bs{w}_{CNN})$, for a few samples.
The accuracy and reliability of the BayesCNN predictions are further evaluated for both in-distribution and out-of-distribution dataset in Figure \ref{fig:errorbar}. In this figure, the \textit{prediction error} is defined as 
$\left| \mathbb{E}[Q^{CNN}] - Q^{FE} \right|$,
and the \textit{uncertainty} is represented by the variance of the BayesCNN predictions, $\mathbb{V}[Q^{CNN}]$.
For in-distribution data, both the prediction error and uncertainty are observed to increase with average pore size. This trend correlates with the variability in the training data distributions for each microstructure class, as illustrated in Figures \ref{fig:prediction_small}--\ref{fig:prediction_large}. Specifically, the training data for the small-pore class exhibit low variance in strain energy, leading to more accurate and confident BayesCNN predictions. In contrast, the higher range of strain energy training data in the large-pore class is reflected in increased prediction error and uncertainty, indicating that the BayesCNN effectively detects the broader variability in this regime.
These results indicate that the BayesCNN model successfully detects interpolation in pore size within the out-of-distribution samples, as reflected by lower accuracy (represented by prediction error $\left| \mathbb{E}[Q^{CNN}] - Q^{FE} \right|$) and reliability (represented by the variance $\mathbb{V}[Q^{CNN}]$) in its predictions for such cases. 
We acknowledge that interpolation in pore size alone does not fully characterize the complexity of microstructural variability affecting mechanical response. Incorporating more expressive topological or morphological descriptors, along with principled out-of-distribution detection frameworks, such as those described in \cite{li2025probing}, could offer a more robust assessment of surrogate model generalization.

\begin{figure}[H]
    \centering
        \includegraphics[width=0.49\textwidth]{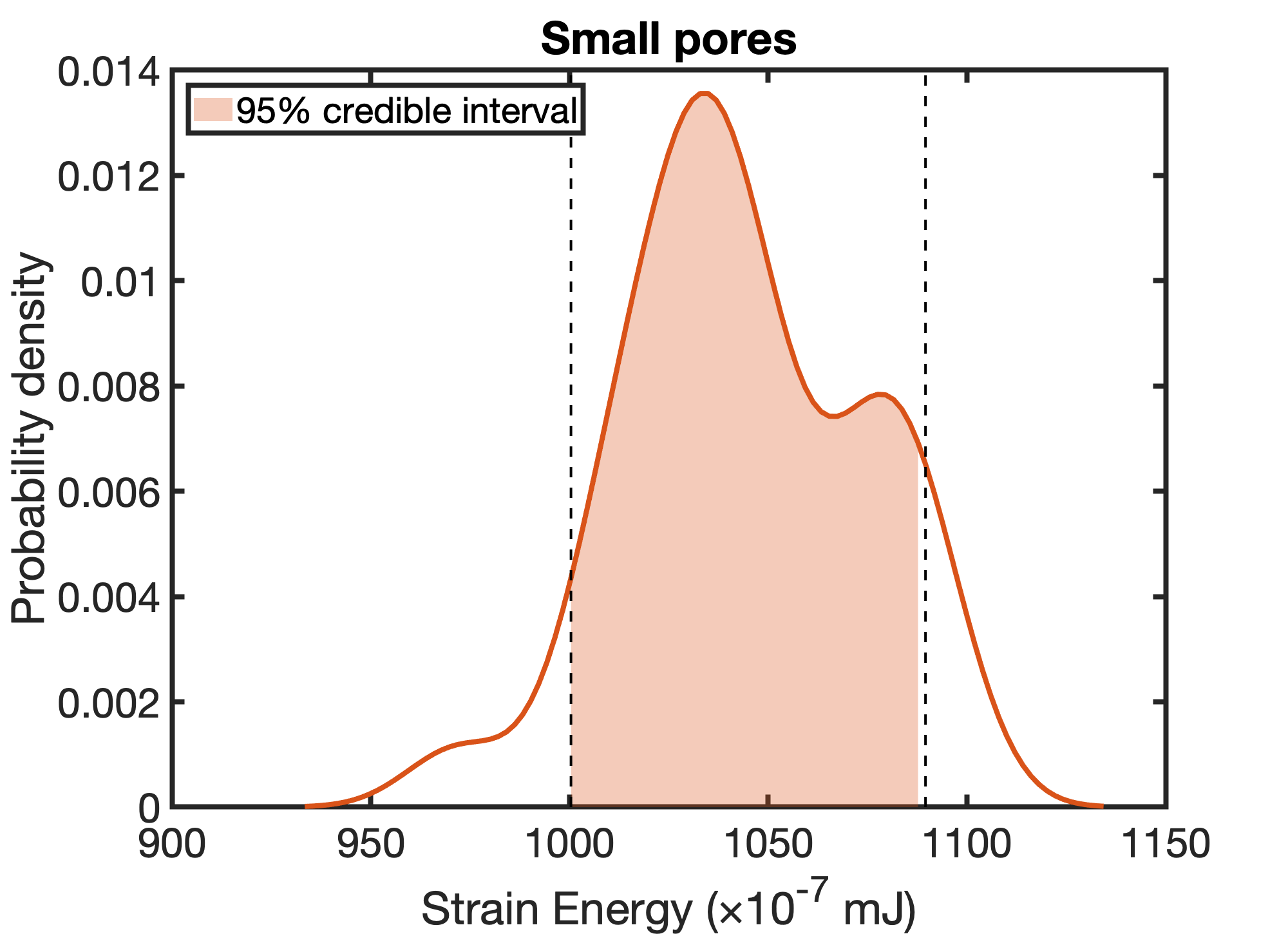}
        \includegraphics[width=0.49\textwidth]{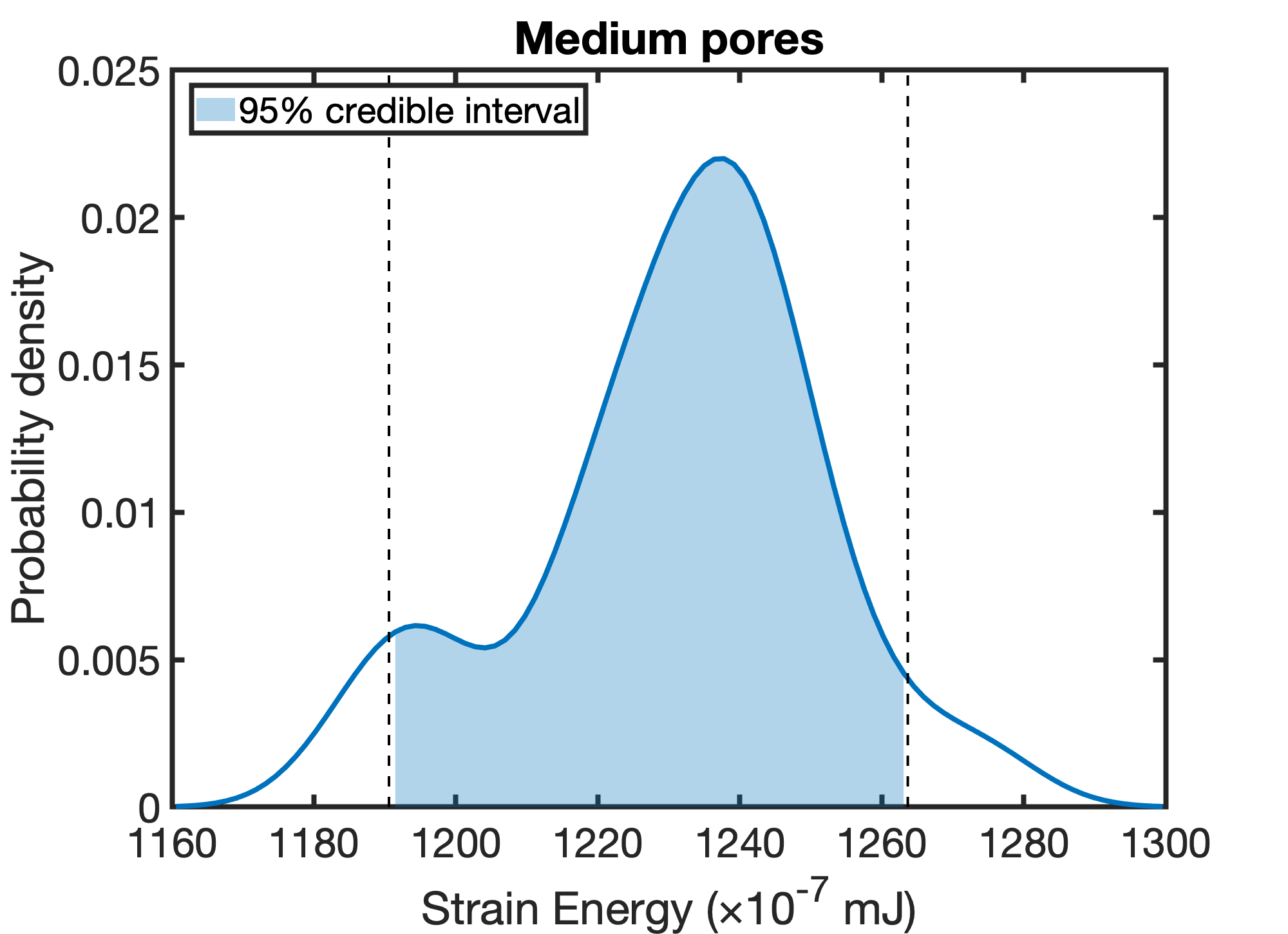}
        \\ (a) \hspace{2.5in} (b) \\
        \includegraphics[width=0.49\textwidth]{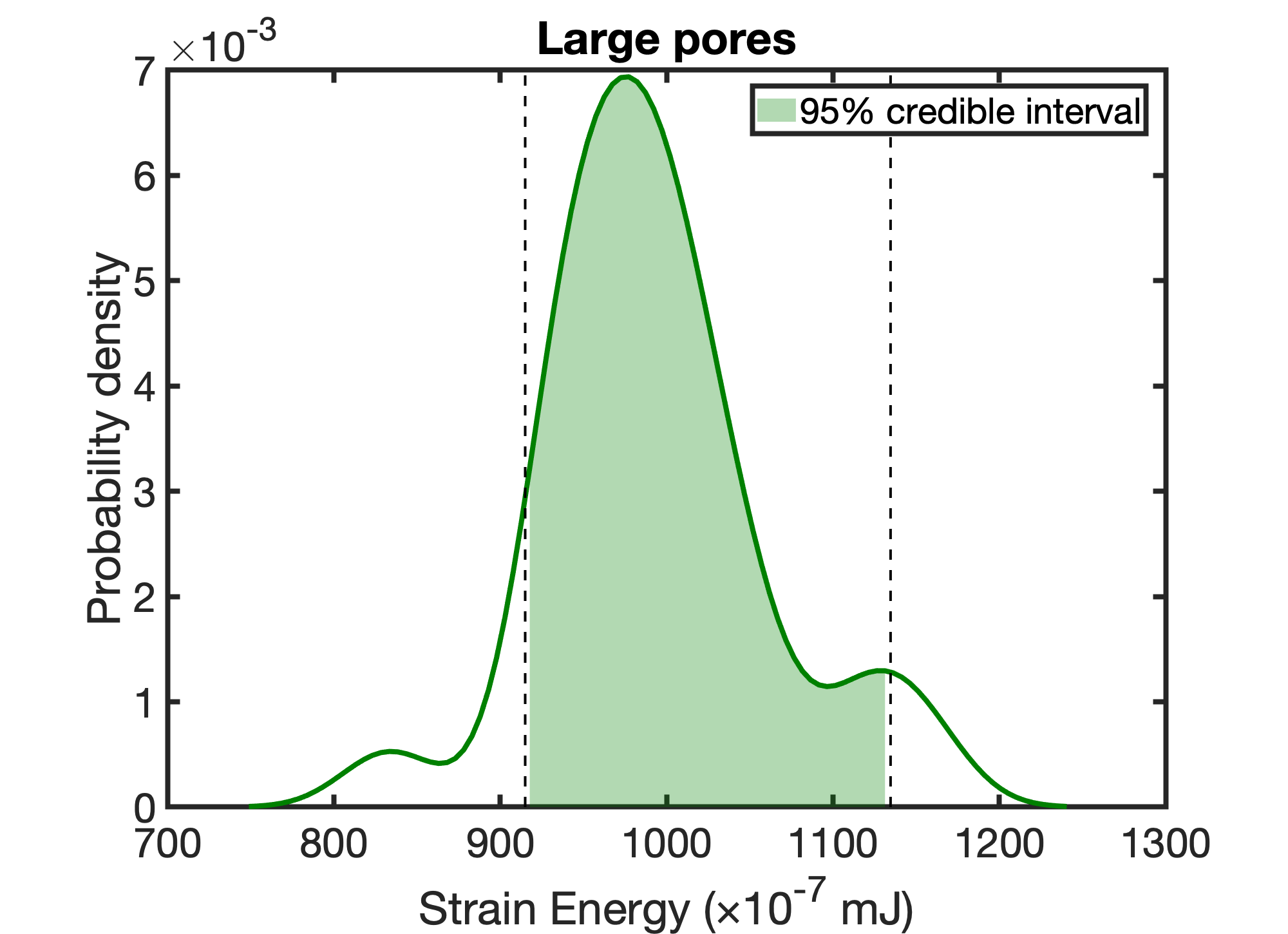}
             \\ (c) \\
    \caption{
    Uncertainty propagation from microstructure to mechanical response, shown as the predicted probability distributions of strain energy (QoI) for in-distribution samples corresponding to:
(a) small-pore,
(b) medium-pore, and
(c) large-pore microstructure classes.
    }
    \label{fig:uncertaintyprop}
    \vspace{-0.2in}
\end{figure}

Finally, leveraging the WGAN-GP and BayesCNN surrogate models, we perform uncertainty propagation analyses within the microstructure–property relationship, as illustrated in Figure \ref{fig:uncertaintyprop}. \blue{Given the demonstrated} accuracy and reliability of the BayesCNN predictions for in-distribution samples (see Figure \ref{fig:errorbar}), the results in this figure focus exclusively on the predicted probability distributions of strain energy 
$\pi(Q^{CNN})$ for each microstructure class. For each class, we report the posterior mean $\mathbb{E}[Q^{CNN}]$ and the 95\% credible interval $\mathcal{CI}_{95\%}[Q^{CNN}]$ of the predicted QoI, representing the uncertainty associated with microstructural variability.

\begin{eqnarray*}
    \rm{small \; pores:} & \mathbb{E}[Q^{CNN}] = 1046.9, & \mathcal{CI}_{95\%}[Q^{CNN}] = [1001.2, 1089.5]\\
    \rm{medium \; pores:} & \mathbb{E}[Q^{CNN}] = 1225.7, & \mathcal{CI}_{95\%}[Q^{CNN}] = [1173.6, 1263.6]\\
    \rm{large \; pores:} & \mathbb{E}[Q^{CNN}] = 1006.15, & \mathcal{CI}_{95\%}[Q^{CNN}] = [915, 1142.7] 
\end{eqnarray*}

%========================================================
% Conclusions
%========================================================
\section{Conclusions}\label{sec:conclusions}

%-------------------------------------------------------
% 1. Summarize the key methods and results (two paragraphs)
%-------------------------------------------------------
This study introduces an integrated deep learning surrogate modeling framework for efficient uncertainty propagation across the synthesis, microstructure, and properties of ceramic aerogel porous materials. The stochastic microstructure surrogate employs a scalable Wasserstein Generative Adversarial Network with Gradient Penalty (WGAN-GP) to generate ensembles of morphologically consistent microstructures over arbitrarily large domains, trained on smaller domain size samples from Lattice Boltzmann simulations of the foaming process. For property prediction, a Bayesian Convolutional Neural Network (BayesCNN) is developed to estimate the probability distribution of weight parameters from finite element training data, enables not only prediction of the strain energy but also quantification of the associated uncertainty, providing a measure of confidence in the surrogate model’s predictions.
Numerical results demonstrate that the WGAN-GP effectively generates microstructures consistent with the morphology of training data (in terms of pore size distribution and two-point correlation) across larger domains. The BayesCNN demonstrates accurate mapping from microstructure to strain energy for in-distribution samples, with uncertainty estimates that reflect the level of confidence in prediction. Notably, both prediction error and uncertainty (i.e., variance) increase with average pore size, consistent with the broader distribution of mechanical responses in larger-pore microstructures. However, the model exhibits reduced predictive accuracy and confidence for microstructures with pore sizes interpolating between the training classes.

%-------------------------------------------------------
% 2. Limitations of the study and possible extensions (one paragraph)
%-------------------------------------------------------

Future work will focus on enhancing the surrogate modeling framework by incorporating aerogel synthesis parameters as additional inputs to the WGAN-GP, enabling the discovery of novel morphologies that interpolate between training data for specific synthesis parameter combinations, as demonstrated in \cite{chun2020deep}. 
The predictive capability of the BayesCNN surrogate will be further evaluated on a broader set of 3D aerogel microstructures with diverse pore size distributions. 
\blue{
Recent work has demonstrated the effectiveness of advanced CNN architectures for predicting material properties from complex 3D porous microstructures (e.g., \cite{RAO2020109850,YANG2018278}). Extending our framework to such 3D settings will be a natural next step. In particular, we are interested in applying Bayesian training to these 3D models to not only assess predictive accuracy but also evaluate the associated confidence levels in model predictions. In parallel, we will extend the current framework to incorporate nonlinear and nonlocal finite element simulations with micromechanical damage models, enabling more accurate predictions of deformation behavior beyond the linear elastic regime. Such size-dependent models will account for the influence of solid wall thickness and  failure mechanisms critical to the mechanical strength of ceramic aerogels.} To improve BayesCNN reliability, architecture and hyperparameter selection will be guided by a hierarchical Bayesian inference framework \cite{singh2024opal}, enabling identification of the most plausible model given limited training data, supporting more trustworthy surrogate modeling.
\blue{
An additional direction to improve the predictive capability of surrogate models is the incorporation of governing physical laws into the training process through physics-informed CNNs, e.g., \cite{zhang2024weak, rezaei2025finite, liu2024multi}, that has been shown to enhance extrapolation behavior and reduce data requirements. However, extending this approach within a Bayesian framework introduces new challenges. Specifically, it requires careful construction of a composite likelihood that balances data fidelity with residual physics violations under uncertainty. Hyperparameter tuning becomes particularly critical to ensure the model appropriately weighs physical constraints without compromising fit to available data, e.g., \cite{yang2021b, dabrowski2023bayesian}.
}
The FSLBM simulations will also be extended to more faithfully represent the ceramic aerogel foaming process. This includes integrating a homogeneous nucleation model \cite{wagner1984measurements} to probabilistically determine initial bubble configurations based on synthesis parameters. The nucleation model incorporates two key parameters: surface tension and critical bubble radius. Surface tension will be modeled as a decreasing function of CTAB concentration, accounting for surfactant-induced stabilization at bubble interfaces. Higher CTAB concentrations enhance disjoining pressure, promoting bubble generation and suppressing coalescence \cite{Ren2019hierarchical}. In parallel, urea concentration, affecting bubble formation through thermal hydrolysis, will be used to inversely scale the critical bubble radius, as it governs the pressure differential driving nucleation.
Finally, while the current microstructure-property surrogate relies on standard CNN architectures, future work will explore more advanced deep learning models \cite{RASHID2022105452, EIDEL2023115741, mianroodi2022lossless, gollapalli2025design, templeton2024expediting} with superior generalization capabilities. Embedding these models within a Bayesian inference framework will be investigated to evaluate their effectiveness for uncertainty-aware surrogate modeling of complex porous materials such as ceramic aerogels.

%-------------------------------------------------------
% 3. Summarize the main take-home points from the study
%-------------------------------------------------------
This study addresses key computational challenges in developing reliable deep learning surrogate models for synthesis-microstructure-property relationships. The proposed framework, along with its extensions, establishes a foundation for uncertainty quantification in data-driven surrogate modeling, enabling broad applications in computational materials science, including uncertainty propagation, materials design, and optimization.

%++++++++++++++++++++++++++++++++++++++++++++++++++++++++++++++++++++++++
\section*{Data availability}
\noindent
The dataset and python code pertinent to this paper is available at
\href{https://github.com/pce-lab/BayesCNNSurrogate}{https://github.com/pce-lab/BayesCNNSurrogate}.

%++++++++++++++++++++++++++++++++++++++++++++++++++++++++++++++++++++++++
\section*{Acknowledgments}
\noindent
MAI and DF extend their gratitude for the financial support of this work 
from the Research Foundation of The State University of New York (SUNY) under grant number 1191358.
%PKS and DF also appreciate the support from the U.S. National Science Foundation (NSF) CAREER Award CMMI-2143662.
Additionally, the authors would like to acknowledge the support provided by the Center for Computational Research at the University at Buffalo.

%========================================================================
% Appendix
\appendix
\section{Appendix: Deep learning architectures}
\begin{table}[h!]
    \centering
    \resizebox{0.8\linewidth}{!}{%
    \begin{tabular}{c c c c c c c}
         \hline
         Layer & Dimension & Kernel & Stride & Padding & Batch Normalization & Activation Function  \\ \hline \\\textbf{Generator}\\
         \\Layer 1 & -1 $\times$ 8 $\times$ 8 $\times$ 512 & 4 $\times$ 4 & 2 $\times$ 2 & 1 $\times$ 1 & Yes & ReLU\\
         \\Layer 2 & -1 $\times$ 16 $\times$ 16 $\times$ 256 & 4 $\times$ 4 & 2 $\times$ 2 & 1 $\times$ 1 & Yes & ReLU\\
         \\Layer 3 & -1 $\times$ 32 $\times$ 32 $\times$ 128 & 4 $\times$ 4 & 2 $\times$ 2 & 1 $\times$ 1 & Yes & ReLU\\
         \\Layer 4 & -1 $\times$ 64 $\times$ 64 $\times$ 64 & 4 $\times$ 4 & 2 $\times$ 2 & 1 $\times$ 1 & Yes & ReLU\\
         \\Layer 5 & -1 $\times$ 128 $\times$ 128 $\times$ 1 & 4 $\times$ 4 & 2 $\times$ 2 & 1 $\times$ 1 & No & Tanh\\
         \\
         \textbf{Discriminator}\\
         \\Layer 1 & -1 $\times$ 64 $\times$ 64 $\times$ 64 & 4 $\times$ 4 & 2 $\times$ 2 & 1 $\times$ 1 & Yes & LeakyReLU\\
         \\Layer 2 & -1 $\times$ 32 $\times$ 32 $\times$ 128 & 4 $\times$ 4 & 2 $\times$ 2 & 1 $\times$ 1 & Yes & LeakyReLU\\
         \\Layer 3 & -1 $\times$ 16 $\times$ 16 $\times$ 256 & 4 $\times$ 4 & 2 $\times$ 2 & 1 $\times$ 1 & Yes & LeakyReLU\\
         \\Layer 4 & -1 $\times$ 8 $\times$ 8 $\times$ 512 & 4 $\times$ 4 & 2 $\times$ 2 & 1 $\times$ 1 & Yes & LeakyReLU\\
         \\Layer 5 & -1 $\times$ 1 $\times$ 1 $\times$ 1 & 4 $\times$ 4 & 2 $\times$ 2 & 1 $\times$ 1 & No & -\\
         \hline
    \end{tabular}}
    \caption{Architecture of WGAN-GP with the output dimension from each layer.}
    \label{tab:gan_arch}
\end{table}
\begin{table}[H]
    \centering
    \resizebox{0.8\linewidth}{!}{%
    \begin{tabular}{l c c c c c c}
    \hline
    \textbf{Layer} & \textbf{Type} & \textbf{Kernel} & \textbf{Stride} & \textbf{Input Shape} & \textbf{Output Shape} & \textbf{Parameters} \\
    \hline
    Input & - & - & - & $3 \times 168 \times 168$ & $3 \times 168 \times 168$ & 0 \\[6pt]

    Conv1 & BBBConv2d & $5 \times 5$ & $1 \times 1$ & $3 \times 168 \times 168$ & $6 \times 164 \times 164$ & $456$ \\[6pt]

    Act1 & Softplus & - & - & $6 \times 164 \times 164$ & $6 \times 164 \times 164$ & 0 \\[6pt]

    Pool1 & MaxPool & $2 \times 2$ & $2 \times 2$ & $6 \times 164 \times 164$ & $6 \times 82 \times 82$ & 0 \\[6pt]

    Conv2 & BBBConv2d & $5 \times 5$ & $1 \times 1$ & $6 \times 82 \times 82$ & $16 \times 78 \times 78$ & $2{,}416$ \\[6pt]

    Act2 & Softplus & - & - & $16 \times 78 \times 78$ & $16 \times 78 \times 78$ & 0 \\[6pt]

    Pool2 & MaxPool & $2 \times 2$ & $2 \times 2$ & $16 \times 78 \times 78$ & $16 \times 39 \times 39$ & 0 \\[6pt]

    Flatten & Flatten & - & - & $16 \times 39 \times 39$ & $24{,}336$ & 0 \\[6pt]

    FC1 & BBBLinear & - & - & $24{,}336$ & $120$ & $2{,}920{,}440$ \\[6pt]

    Act3 & Softplus & - & - & $120$ & $120$ & 0 \\[6pt]

    FC2 & BBBLinear & - & - & $120$ & $84$ & $10{,}164$ \\[6pt]

    Act4 & Softplus & - & - & $84$ & $84$ & 0 \\[6pt]

    Regression & BBBLinear & - & - & $84$ & $1$ & $85$ \\[6pt]

    \hline
    \end{tabular}
    }    
    \caption{Architecture of the BayesCNN model and total number of parameters.}
    \label{tab:cnn_arch}
\end{table}
%
% \label{sec:sample:appendix}
% \blue{
% Lorem ipsum dolor sit amet, consectetur adipiscing elit, sed do eiusmod tempor section \ref{sec:sample1} incididunt ut labore et dolore magna aliqua. Ut enim ad minim veniam, quis nostrud exercitation ullamco laboris nisi ut aliquip ex ea commodo consequat. Duis aute irure dolor in reprehenderit in voluptate velit esse cillum dolore eu fugiat nulla pariatur. Excepteur sint occaecat cupidatat non proident, sunt in culpa qui officia deserunt mollit anim id est laborum.
% }

%========================================================================
% Bibliography
%========================================================================

 \bibliographystyle{elsarticle-num} 
 \bibliography{refs, references}

\end{document}